\documentclass[aps,preprint,floats,epsf,epsfig,nofootinbib,letter]{revtex4}
\usepackage{epsfig}
\usepackage{graphicx}
\usepackage{dcolumn}
\usepackage{bm}
\usepackage{subfigure}
\usepackage{hyperref}                 
\usepackage{amsmath}

\begin{document}
\def\be{\begin{eqnarray}}
\def\en{\end{eqnarray}}
\def\non{\nonumber}
\def\la{\langle}
\def\ra{\rangle}
\def\Br{{\mathcal B}}
\def\A{{\mathcal A}}
\def\B{{\cal B}}
\def\D{{\cal D}}
\def\Bbar{\overline{\cal B}}
\def\bfB{{\rm\bf B}}
\def\bfBB{{\rm\bf B}\overline{\rm\bf B}}
\def\BB{{{\cal B} \overline {\cal B}}}
\def\BD{{{\cal B} \overline {\cal D}}}
\def\DB{{{\cal D} \overline {\cal B}}}
\def\DD{{{\cal D} \overline {\cal D}}}
\def\sq{\sqrt}


\title{Charmless Semileptonic Baryonic $B_{u,d,s}$ Decays}

\author{Chun-Khiang Chua}
\affiliation{Department of Physics and Center for High Energy Physics,
Chung Yuan Christian University,
Chung-Li, Taiwan 320, Republic of China}

\date{\today}

\begin{abstract}
We study $\overline B_q\to {{\rm\bf B}\overline{\rm\bf B}}' l \bar\nu$ and $\overline B_q\to {{\rm\bf B}\overline{\rm\bf B}}' \nu \bar\nu$ decays with all low lying octet (${\cal B}$) and decuplet (${\cal D}$) baryons using a topological amplitude approach. In tree induced $\overline B_q\to {{\rm\bf B}\overline{\rm\bf B}}' l \bar\nu$ decay modes, we need two tree amplitudes and one annihilation amplitude in $\overline B_q\to{\cal B}\overline{\cal B}' l\bar\nu$ decays, one tree amplitude in $\overline B_q\to{\cal B}\overline{\cal D} l\bar\nu$ decays, one tree amplitude in $\overline B_q\to{\cal D}\overline{\cal B} l\bar\nu$ decays and one tree amplitude and one annihilation amplitude in $\overline B_q\to{\cal D}\overline{\cal D}' l\bar\nu$ decays. In loop induced $\overline B_q\to {{\rm\bf B}\overline{\rm\bf B}}' \nu \bar\nu$  decay modes, similar numbers of penguin-box and penguin-box-annihilation amplitudes are needed. As the numbers of independent topological amplitudes are highly limited, there are plenty of relations on these semileptonic baryonic $B_q$ decay amplitudes. Furthermore, the loop topological amplitudes and tree topological amplitudes have simple relations, as their ratios are fixed by known CKM factors and loop functions. It is observed that the $B^-\to p\bar p \mu^-\bar\nu$ differential rate exhibits threshold enhancement, which is expected to hold in all other semileptonic baryonic modes. The threshold enhancement effectively squeezes the phase space toward the threshold region and leads to very large SU(3) breaking effects in the decay rates. They are estimated using the measured $B^-\to p\bar p \mu^-\bar\nu$ differential rate and model calculations. From the model calculations, we find that branching ratios of non-annihilation $\overline B_{q}\to  {{\rm\bf B}\overline{\rm\bf B}}' l \bar \nu$ modes are of the orders of $10^{-9}\sim 10^{-6}$, while branching ratios of non-penguin-box-annihilation $\overline B_{q}\to  {{\rm\bf B}\overline{\rm\bf B}}' \nu \bar \nu$ modes are of the orders of $10^{-12}\sim 10^{-8}$. Modes with relatively unsuppressed rates and good detectability are identified. These modes can be searched experimentally in near future and the rate estimations can be improved when more modes are discovered. Ratios of rates of some loop induced $\overline B_q\to {{\rm\bf B}\overline{\rm\bf B}}' \nu \bar\nu$ decays and tree induced $\overline B_q\to {{\rm\bf B}\overline{\rm\bf B}}' l \bar\nu$ decays are predicted and can be checked experimentally. They can be tests of the SM.  
Some implications on $\overline B_{q}\to \bfBB' l^+ l^-$ decays are also given.

\end{abstract}


\maketitle

\tableofcontents

\vfill\eject

\section{Introduction}

Recently, there are some experimental activities on $\overline B\to \bfBB' l\bar\nu$ and $\overline B\to \bfBB'\nu\bar\nu$ decays, where $\bfBB'$ are baryon anti-baryon pairs \cite{CLEO:2003bit, Belle:2013uqr,LHCb:2019cgl,PDG,BaBar:2019awu}.
The present experimental results are summarized in Table~\ref{tab: expt}.
In particular, the branching ratio of $B^-\to p\bar p \mu^-\bar\nu_\mu$ decay is measured to be $(5.27^{+0.23}_{-0.24}\pm 0.21\pm 0.15)\times 10^{-6}$
by LHCb~\cite{LHCb:2019cgl} and $Br(B^-\to p\bar p l\bar\nu)=(5.8^{+2.6}_{-2.3})\times 10^{-6}$ by Belle \cite{Belle:2013uqr} (see also \cite{PDG}),
while only upper limit of $Br(B^-\to \Lambda \bar p\nu\bar\nu)<3.0\times 10^{-5}$ was reported by BaBar~\cite{BaBar:2019awu}.

Theoretically, the branching ratios of $\overline B\to \bfBB' l\bar\nu$ decays were estimated and predicted to be of the order of $10^{-6}$ to $10^{-4}$ \cite{Hou:2000bz,Geng:2011tr}. 
Some recent studies are devoted to understand the rate of the $B^-\to p\bar p l\bar\nu$ decay \cite{Geng:2021sdl, Hsiao:2022uzx} as the measured rate is roughly 20 times smaller than a previous theoretical prediction \cite{Geng:2011tr}, while the shape of the predicted differential rate using QCD counting rules agrees well with data, which exhibits  threshold enhancement~\cite{Geng:2011tr, LHCb:2019cgl}.
In this work we will employ the approach of refs. \cite{Chua:2003it,Chua:2013zga,Chua:2016aqy,Chua:2022wmr}, 
which was used to study two-body baryonic $B$ decays, $\overline B\to\bfBB'$, making use of the well established topological amplitude formalism~\cite{Zeppenfeld:1980ex,Chau:tk,Chau:1990ay,Gronau:1994rj,Gronau:1995hn,Chiang:2004nm,Cheng:2014rfa,Savage:ub,He:2018php,He:2018joe,Wang:2020gmn}.
The decay amplitudes of $\overline B\to \bfBB' l\bar\nu$ and $\overline B\to \bfBB'\nu\bar\nu$ decays with all low lying octet (${\cal B}$) and decuplet (${\cal D}$) baryons will be decomposed into combinations of several topological amplitudes.
As the numbers of topological amplitudes are highly limited, there are many relations of decay amplitudes.

\begin{figure}[tb]
\centering
  \includegraphics[width=0.5\textwidth]{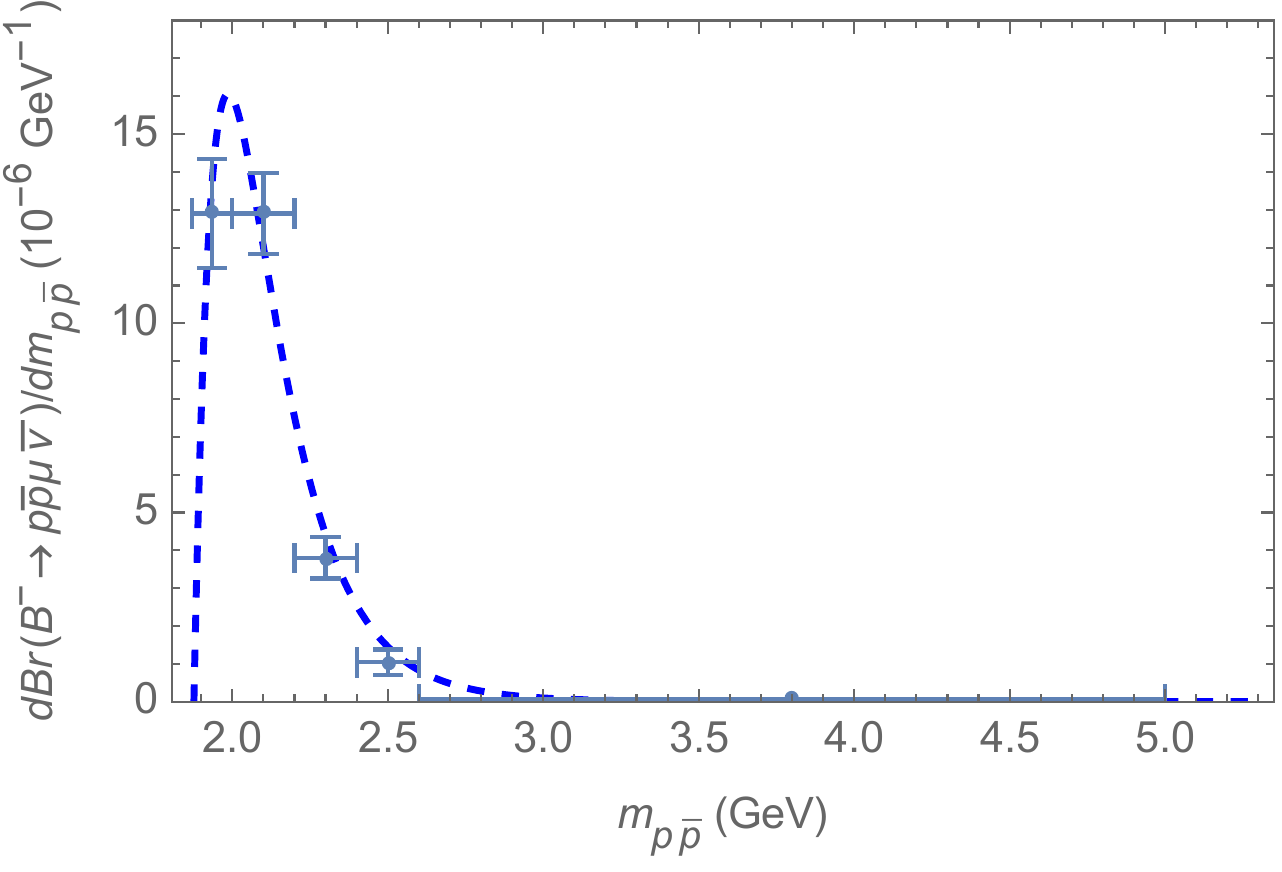}
\caption{
The experimental differential rate $dBr/dm_{p\bar p}$ of $B^-\to p\bar p \mu^-\bar\nu$ decay from LHCb~\cite{LHCb:2019cgl} exhibits threshold enhancement.
The threshold enhancement effectively squeezes the phase space toward the threshold region.
}
 \label{fig: dBBdm0}
\end{figure}

It is well known that a decay rate strongly depends on the masses of the final state particles when the decay is just above the threshold. 
The rates may vary in orders of magnitudes even if the amplitudes are of similar sizes. 
One normally does not expect such behavior in $B_q$ decays when large phase spaces are available.
From the experimental differential rate $dBr/dm_{p\bar p}$ of $B^-\to p\bar p \mu^-\bar\nu$ decay from LHCb~\cite{LHCb:2019cgl} as shown in 
Fig.~\ref{fig: dBBdm0}, one can easily see that the spectrum exhibits prominent threshold enhancement, which is a comment feature in three or more body baryonic $B_q$ decays~\cite{Hou:2000bz, Chua:2001vh, Cheng:2001tr, Chua:2002wn, Chua:2002yd, Geng:2011tr, Huang:2021qld, Geng:2021sdl, Hsiao:2022uzx}. 
Threshold enhancement is expected to hold in all other semileptonic baryonic modes considered in this work as well.
The threshold enhancement effectively squeezes the phase space to the threshold region, see Fig.~\ref{fig: dBBdm0}, and thus mimics the decay just above threshold situation.
Consequently, it amplifies the effects of SU(3) breaking in final state baryon masses and
can lead to very large SU(3) breaking effects in the decay rates.
The SU(3) breakings in the decay rates from threshold enhancements will be estimated using the measured $B^-\to p\bar p \mu^-\bar\nu$ differential rate and model calculations with available theoretical inputs from refs. \cite{Geng:2021sdl, Hsiao:2022uzx}, which can reproduce the measured $B^-\to p\bar p \mu^-\bar\nu$ differential rate.

\begin{table}[t!]
\caption{\label{tab: expt}
Experimental results of $\B^-\to\bfBB' l\bar\nu$ and $\bfBB'\nu\bar\nu$ branching ratios. The upper limits are at 90\% confidence level.
}
\footnotesize{
\begin{ruledtabular}
\begin{tabular}{lcccr}
Mode
          & Branching ratio
          & References
          \\
\hline $B^-\to p\bar p e^-\bar \nu_e$
          & $(5.8\pm 3.7\pm 3.6)\times 10^{-4}$ $(<5.2\times 10^{-3})$
          & CLEO~\cite{CLEO:2003bit}
          \\
          & $(8.2^{+3.7}_{-3.2}\pm 0.6)\times 10^{-6}$
          & Belle~\cite{Belle:2013uqr}
          \\
          &  $(8.2^{+4.0}_{-3.3})\times 10^{-6}$
          &  PDG~\cite{PDG}
          \\  
$B^-\to p\bar p \mu^-\bar \nu_\mu$
          & $(3.1^{+3.1}_{-2.4}\pm 0.7)\times 10^{-6}$
          & Belle~\cite{Belle:2013uqr}
          \\
          & $(5.27^{+0.23}_{-0.24}\pm 0.21\pm 0.15)\times 10^{-6}$
          & LHCb~\cite{LHCb:2019cgl}
          \\
          &  $(5.32\pm 0.34 )\times 10^{-6}$
          &  PDG~\cite{PDG}
          \\  
$B^-\to p\bar p l \bar \nu\, (l=e,\mu)$
          &  $(5.8^{+2.6}_{-2.3})\times 10^{-6}$
          &  Belle\cite{Belle:2013uqr}, PDG~\cite{PDG}
          \\             
\hline $B^-\to \Lambda \bar p \nu\bar \nu$
          & $(0.4\pm 1.1\pm 0.6)\times 10^{-5}$ $(<3.0\times 10^{-5})$
          &  BaBar ~\cite{BaBar:2019awu}                                  
\end{tabular}
\end{ruledtabular}
}
\end{table}

We will try to identify modes with relatively unsuppressed rates and good detectability.
The estimation on rates can be improved when more modes are discovered.
Recently hints of new physics effects in rare $B$ decays are accumulating, 
see, for example, \cite{Altmannshofer:2021qrr,Cornella:2021sby, Geng:2021nhg}.
Given the present situation and the fact that $\overline B\to \bfBB' l\bar\nu$ decays are tree induced decay modes, while $\overline B\to \bfBB' \nu\bar\nu$ decays are loop induced decay modes, 
it will be interesting and useful to identify $\overline B\to \bfBB'\nu\bar\nu$ and $\overline B\to \bfBB' l\bar\nu$ decay modes which have good detectability. 
Their rate ratios, especially, those insensitive to the modeling of SU(3) breaking from threshold enhancement, can be tests of the Standard Model (SM). 

The layout of this paper is as following. 
We give the formalism for decomposing amplitudes in terms of topological amplitudes and modeling of the topological amplitudes in Sec.~II.
In Sec.~III, results on decay amplitudes in term of topological amplitudes, relations of decay amplitudes and decay rates are provided.  
Conclusion and discussions are given in Sec.~IV, where some comments on $\overline B_{q}\to \bfBB' l^+ l^-$ decays will also be given.
Appendix A concerning the transition matrix elements in the asymptotic limit and Appendix B with some useful formulas for calculating 4-body decay rates are added at the end of the paper.

\section{Formalism}

\subsection{Topological amplitudes}

The decay amplitudes of $\overline B_q\to {\bf B}\overline {\bf B}' l\bar \nu$ and $\overline B_q\to {\bf B}\overline {\bf B}' \nu\bar \nu$ decays are given by~\cite{Geng:2011tr,Geng:2012qn}
\be
A(\overline B_q\to {\bf B}\overline {\bf B}' l\bar \nu)&=&\frac{G_F}{\sqrt2} V_{ub} \la {\bf B}\overline {\bf B}'|\bar u_L\gamma_\mu b_L|\overline B_q\ra 
\bar l_L \gamma^\mu \nu_L,
\non\\
A(\overline B_q\to {\bf B}\overline {\bf B}' \nu\bar \nu)&=&\frac{G_F}{\sqrt2} \frac{\alpha_{\rm em}}{2\pi\sin^2\theta_W}
V^*_{ts} V_{tb}   D(m_t^2/m_W^2) \la {\bf B}\overline {\bf B}'|\bar s_L\gamma_\mu b_L|\overline B_q\ra 
\bar \nu_L \gamma^\mu \nu_L,
\label{eq: AA}
\en
where $V_{ub}$, $V_{ts}$ and $V_{tb}$ are Cabibbo-Kobayashi-Maskawa (CKM) matrix elements,
$D(x)$, $D_0(x)$ and $D_1(x)$ are loop functions with~\cite{loopfunctions}
\be
D(x)&=& D_0(x)+\frac{\alpha_s}{4\pi} D_1(x),
\non\\
D_0(x)&=& \frac{x}{8} \bigg(-\frac{2+x}{1-x}+\frac{3x-6}{(1-x)^2}\ln x\bigg),
\non\\
D_1(x)&=& 
-\frac{23 x+5 x^2-4 x^3}{3(1-x)^2}
+\frac{x-11 x^2+x^3+x^4}{(1-x)^3}\ln x
+\frac{8x+4x^2+x^3-x^4}{2(1-x)^3}\ln^2 x
\non\\
&&-\frac{4 x-x^3}{(1-x)^2} Li_2 (1-x)+8 x \frac{d D_0(x)}{dx} \ln \frac{\mu^2}{m_W^2}.
\en
Note that the $\overline B_q\to {\bf B}\overline {\bf B}' l\bar \nu$ decay is governed by the matrix element, $\la {\bf B}\overline {\bf B}'|\bar u_L\gamma_\mu b_L|\overline B_q\ra$, while the $\overline B_q\to {\bf B}\overline {\bf B}' \nu\bar \nu$ decay is governed by the matrix element, $ \la {\bf B}\overline {\bf B}'|\bar s_L\gamma_\mu b_L|\overline B_q\ra$.
These two matrix elements are difficult to calculate as they involve baryon pairs ${\bf B}\overline {\bf B}'$ in the final state.
Nevertheless they are related by interchanging $u$ and $s$ and, hence, can be related by SU(3) transformations.


\begin{figure}[t]
\centering
 \subfigure[]{
  \includegraphics[width=0.35\textwidth]{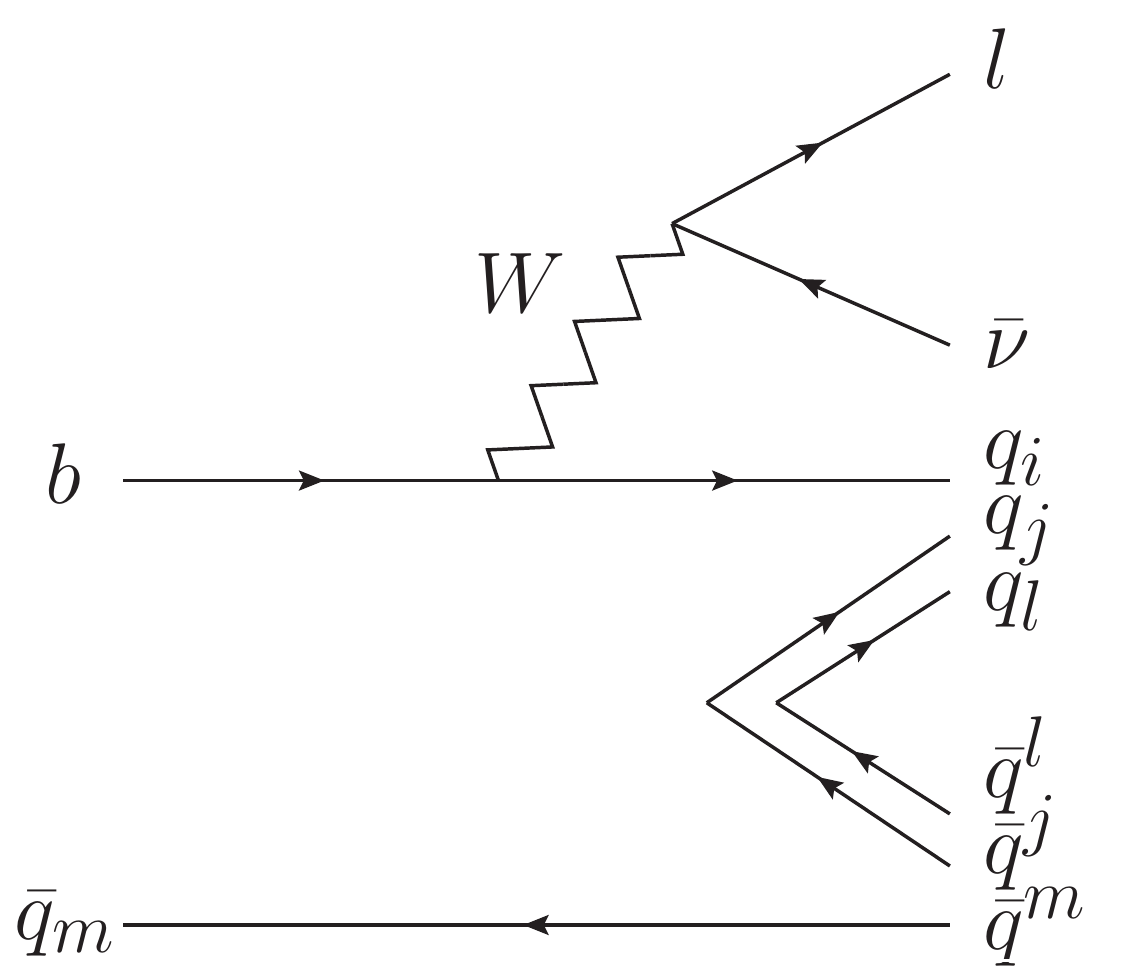}
}
\hspace{12pt}
\subfigure[]{
  \includegraphics[width=0.35\textwidth]{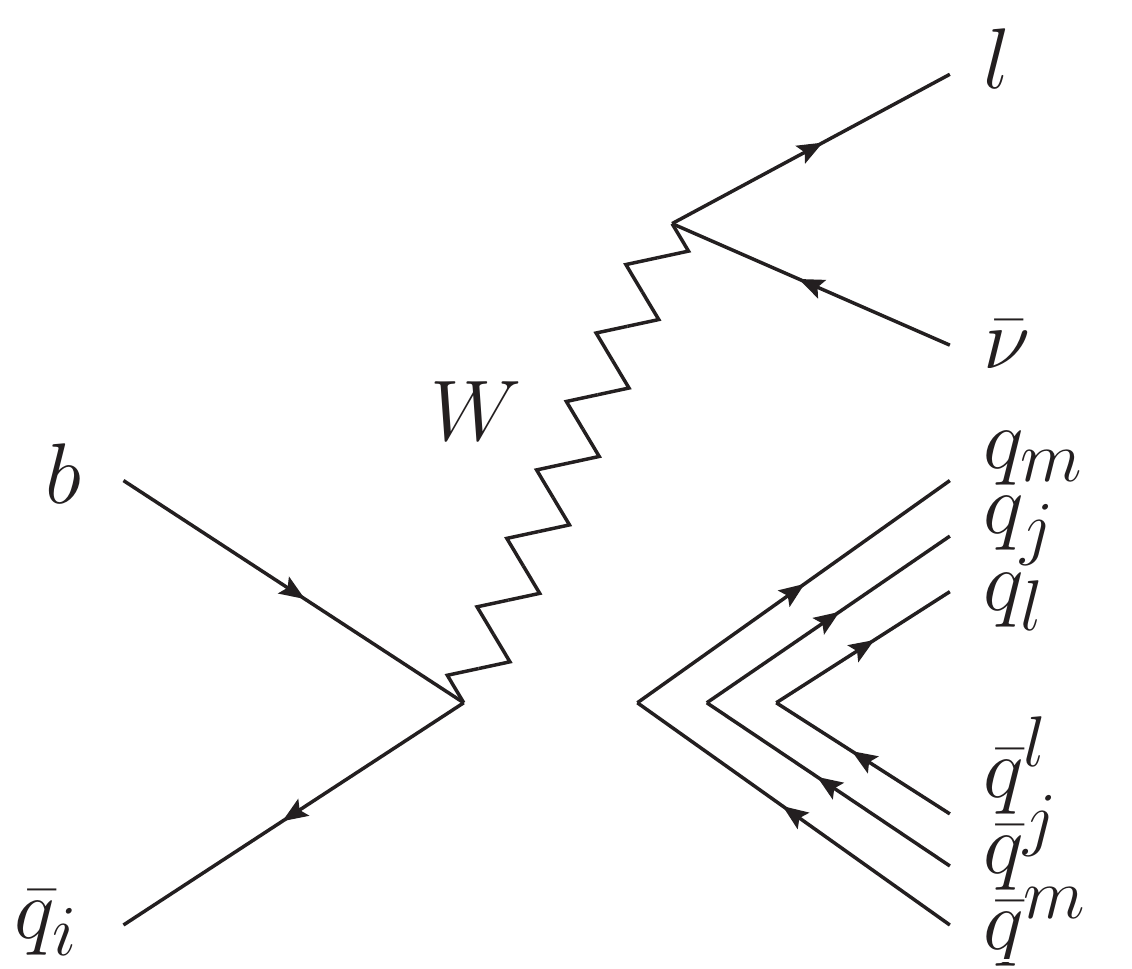}
}
\\
 \subfigure[]{
  \includegraphics[width=0.35\textwidth]{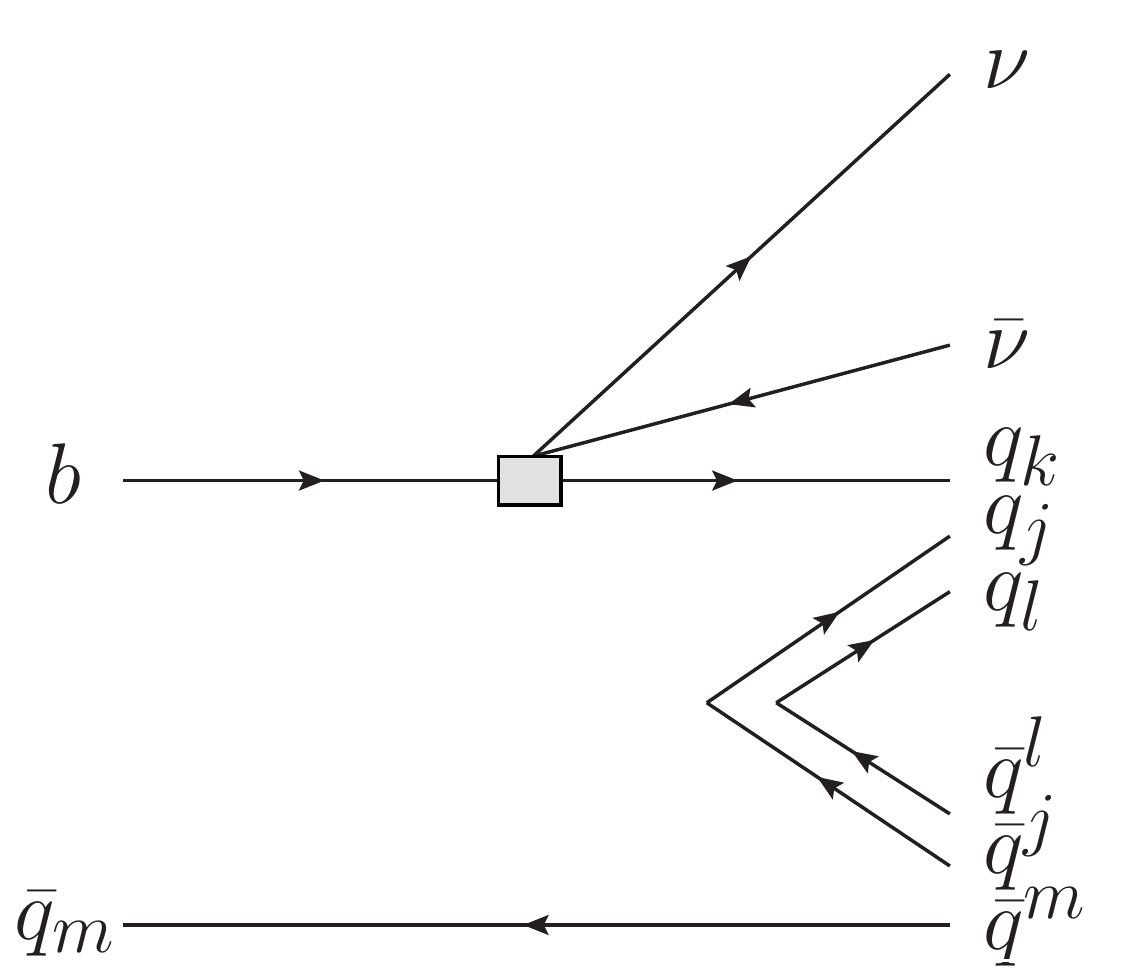}
}
\hspace{12pt}
\subfigure[]{
  \includegraphics[width=0.35\textwidth]{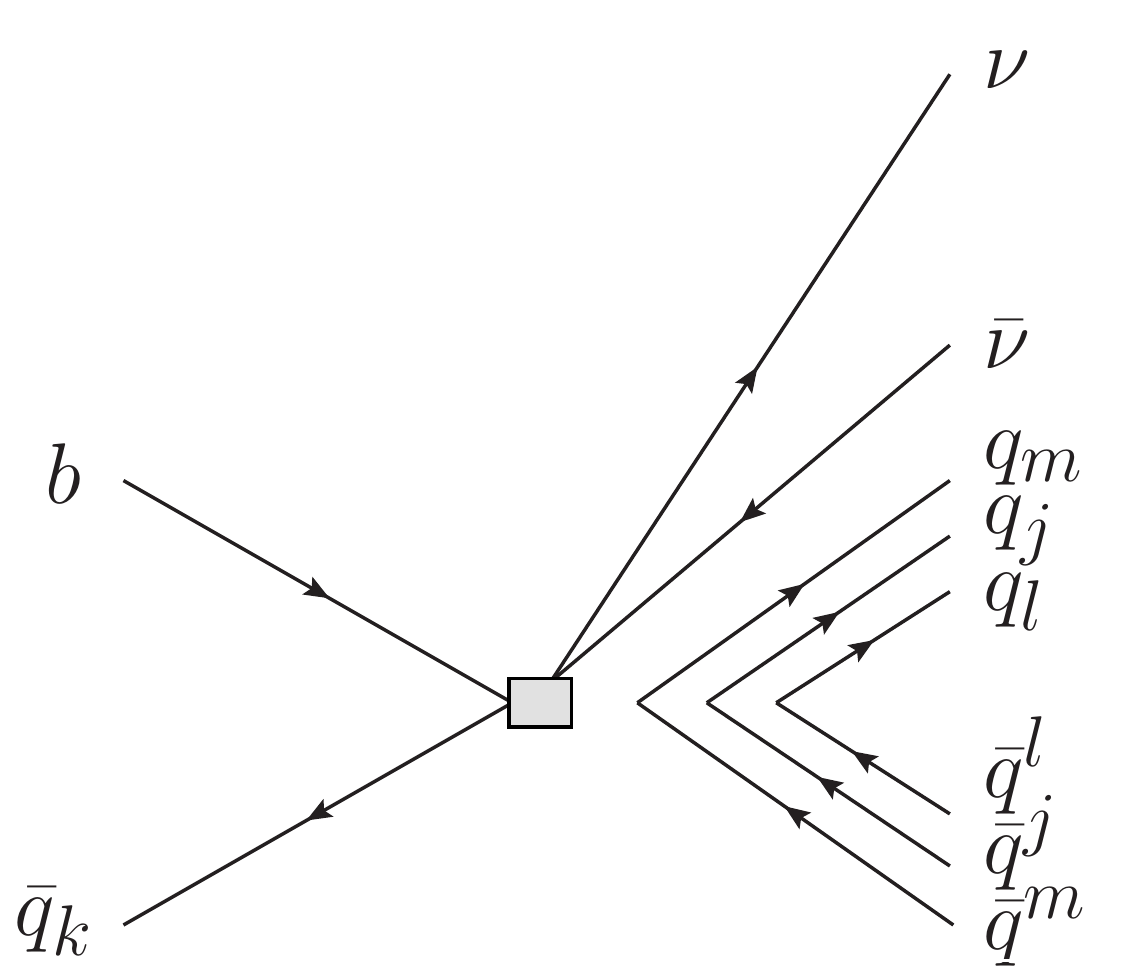}
}
\caption{Topological diagrams of 
  (a) $T$ (tree) and (b) $A$ (annihilation) amplitudes in $\overline B_q\to\bfBB' l\bar\nu$ decays and (c) $PB$ (penguin and box) and (d) $PBA$ (penguin-box annihilation) amplitudes in $\overline B_q\to\bfBB' \nu\bar\nu$ decays.
  These are flavor flow diagrams, where gluon lines are not shown. 
} \label{fig:TA}
\end{figure}

It is known that topological amplitude approach is related to SU(3) approach~\cite{Zeppenfeld:1980ex,Gronau:1994rj,Savage:ub}.
We follow the approach similar to the one employed in the study of $\overline B_q\to {\bf B}\overline {\bf B}'$ decays \cite{Chua:2003it,Chua:2013zga,Chua:2016aqy,Chua:2022wmr} to decompose $\overline B_q\to {\bf B}\overline {\bf B}' l\bar\nu$ and ${\bf B}\overline {\bf B}' \nu\bar\nu$ decay amplitudes
into topological amplitudes.

From Eq. (\ref{eq: AA}), we see that the 
Hamiltonian governing 
$\overline B_q\to {\bf B}\overline {\bf B}' l\bar\nu$ decays
has the following flavor structure,
\begin{eqnarray}
(\bar u b )
 =H^i_T (\bar q_i b),
\en
with
\be
H^{1}_T=1,
\quad
{\rm otherwise}\,\,\,H^{i}_T=0,
\end{eqnarray}
where we take $q_{1,2,3}=u,d,s$ as usual.
Similarly, the 
Hamiltonian governing 
$\overline B_q\to{\bf B}\overline {\bf B}' \nu\bar\nu$ decays
has the following flavor structure,
\begin{eqnarray}
(\bar s b) 
 =H^k_{PB} (\bar q_k b), 
\en
with
\be
H^3_{PB}=1, 
\quad
{\rm otherwise}\,\,\,H^k_{PB}=0.
\end{eqnarray}
These $H_T$ and $H_{PB}$ will be used as spurion fields in the following constructions of effective Hamiltonian, $H_{\rm eff}$.

We shall start with $\overline B_q\to \DD' l\bar\nu$ decays with $\D$ the low-lying decuplet baryon.
The flavor flow diagram for a $\overline B_q\to \bfBB' l\bar\nu$ decay  is given in Fig.~\ref{fig:TA}.
Note that in the case of a $\overline B_q\to \DD' l\bar\nu$ decay, 
the $q_i q_j q_l$ and $\bar q^l \bar q^j \bar q^m$ flavors as
shown in Fig.~\ref{fig:TA} correspond to the following fields,
\be
q_i q_j q_l \to \overline {\cal D}_{ijl},
\quad
\bar q^l \bar q^j \bar q^m\to {\cal D}^{jlm},
\en
in the Hamiltonian, respectively,
where ${\cal D}^{jlm}$ denotes
the familiar decuplet field, and, explicitly, we have
${\cal D}^{111}=\Delta^{++}$,
${\cal D}^{112}=\Delta^{+}/\sqrt3$, 
${\cal D}^{122}=\Delta^0/\sq3$,
${\cal D}^{222}=\Delta^-$, 
${\cal D}^{113}=\Sigma^{*-}/\sq3$,
${\cal D}^{123}=\Sigma^{*0}/\sq6$, 
${\cal D}^{223}=\Sigma^{*-}/\sq3$,
${\cal D}^{133}=\Xi^{*0}/\sq3$, 
${\cal D}^{233}=\Xi^{*-}/\sq3$
and ${\cal D}^{333}=\Omega^-$
(see, for example~\cite{text}).
By using the above correspondent rule,
we obtain the following effective Hamiltonian for $\overline B_q\to \DD' l\bar\nu$ decays,
\begin{eqnarray}
H_{\rm eff}(\overline B_q\to\DD l\bar\nu)
&=& 6\,T_{{\cal D}\overline{\cal D}}\,\overline B_m H^{i}_T \overline {\cal D}_{ijl} {\cal D}^{ljm}
+ A_{{\cal D}\overline{\cal D}}\,\overline B_i H^{i}_T \overline {\cal D}_{mjl} {\cal D}^{ljm},
\label{eq: DDT}
\end{eqnarray}
with ${\overline B}_m=\left(
B^- ,\overline B {}^0, \overline B {}^0_s\right)$.
Without lost of generality,
the pre-factors are assigned for
latter purpose.

For the $\overline B_q \to{\cal D}\overline{\cal B}l \bar\nu$ decays, 
we note that the anti-octet final state is produced by the
${\cal B}^j_k$ field with~\cite{text}
\begin{eqnarray}
{\mathcal B}= \left(
\begin{array}{ccc}
{{\Sigma^0}\over\sqrt2}+{{\Lambda}\over\sqrt6}
       &{\Sigma^+}
       &{p}
       \\
{\Sigma^-}
       &-{{\Sigma^0}\over\sqrt2}+{{\Lambda}\over\sqrt6}
       &{n}
       \\
{\Xi^-}
       &{\Xi^0}
       &-\sqrt{2\over3}{\Lambda}
\end{array}
\right),
\label{eq: octet}
\end{eqnarray}
where ${\cal B}^j_k$ 
has the following flavor structure $q^j q^a
q^b \epsilon_{abk}-\frac{1}{3}\,\delta^j_k q^c q^a
q^b$~\cite{text}. To match the flavor of $\bar q^l\bar q^j\bar q^m$ in the final state as shown in Fig.~\ref{fig:TA},
we use
\begin{eqnarray}
\bar q^l\bar q^j\bar q^m\to \epsilon^{ljb} {\cal B}^m_b,\,\
                              \epsilon^{lbm} {\cal B}^j_b,\,\
                              \epsilon^{bjm} {\cal B}^l_b,
\label{eq: qqq}
\end{eqnarray}
which are, however, not totally independent, as
it can be easily shown that they are subjected to the following relation,
\be
\epsilon^{ljb} {\cal B}^m_b+
               \epsilon^{lbm} {\cal B}^j_b+
               \epsilon^{bjm} {\cal B}^l_b=0.
\en
Hence we only need two of the terms in the right-hand-side of Eq. (\ref{eq: qqq}), and, without loss of generality, the first two terms are chosen.
The effective Hamiltonian of the $\overline B\to{\cal D}\overline{\cal B}l \bar\nu$ decays can be obtained by
replacing ${\cal D}^{ljm}$ in Eq.~(\ref{eq: DDT}) by 
$({\cal B}_1)^{ljm}\equiv\epsilon^{ljb}{\cal B}^m_b$ and 
$({\cal B}_2)^{ljm}\equiv\epsilon^{bjm} {\cal B}^l_b$, 
and, consequently, we have
\begin{eqnarray}
H_{\rm eff}(\overline B_q\to\DB l\bar\nu)&=& \sqrt{6}\,T_{1\DB}\,
                        \overline B_m H^{i}_T \overline{\cal D}_{ijl}
                        \epsilon^{ljb} {\cal B}^m_b
               +\sqrt{6}\,T_{2\DB}\,
                        \overline B_m H^{i}_T \overline{\cal D}_{ijl}
                        \epsilon^{bjm} {\cal B}^l_b
\non\\
&&+\sqrt{6}\,A_{1\DB}\,
                        \overline B_i H^{i}_T \overline{\cal D}_{mjl}
                        \epsilon^{ljb} {\cal B}^m_b
               +\sqrt{6}\,A_{2\DB}\,
                        \overline B_i H^{i}_T \overline{\cal D}_{mjl}
                        \epsilon^{bjm} {\cal B}^l_b,
\label{eq: DBT0}
\end{eqnarray}
where some pre-factors are introduced without lost of generality.
Note that the $T_{1\DB}$, $A_{1\DB}$ and $A_{2\DB}$ terms are vanishing and we only have
\be
H_{\rm eff}(\overline B_q\to\DB l\bar\nu)=\sqrt{6}\,T_{\DB}\,
                        \overline B_m H^{i}_T \overline{\cal D}_{ijl}
                        \epsilon^{bjm} {\cal B}^l_b,
\label{eq: DBT}
\en
with $T_{2\DB}$ relabeled to $T_{\DB}$.

Similarly for $\overline B\to \BD l\bar\nu$ decays, 
the
$q_i q_k q_l$ flavor in the final state corresponds to $\epsilon_{ika} \overline {\cal B}^a_l$,
$\epsilon_{ial} \overline {\cal B}^a_k$ and $\epsilon_{akl} \overline {\cal B}^a_i$, while the last one is redundant,
since it can be expressed by the formers using the following relation,
$\epsilon_{ika} \overline {\cal B}^a_l+\epsilon_{ial} \overline
{\cal B}^a_k+\epsilon_{akl} \overline {\cal
B}^a_i=0$.
Hence we replace the $\overline{\cal D}_{ijl}$ in Eq.~(\ref{eq: DDT}) by
$(\overline {\cal B}_1)_{ijl}\equiv \epsilon_{ija}\overline {\cal B}^a_l$ and
$(\overline {\cal B}_2)_{ijl}\equiv\epsilon_{ajl}\overline {\cal B}^a_i$
and
obtain
\begin{eqnarray}
H_{\rm eff}(\overline B_q\to\BD l\bar\nu)&=& -\sqrt{6}\,T_{1\BD}\,
                        \overline B_m H^{i}_T
                        \epsilon_{ija}\overline {\cal B}^a_l
                        {\cal D}^{ljm}
               -\sqrt{6}\,T_{2\BD}\,
                        \overline B_m H^{i}
                        \epsilon_{ajl}\overline {\cal B}^a_i
                       {\cal D}^{ljm}
\non\\
&&-\sqrt{6}\,A_{1\BD}\,
                        \overline B_i H^{i}_T
                        \epsilon_{mja}\overline {\cal B}^a_l
                        {\cal D}^{ljm}
               -\sqrt{6}\,A_{2\BD}\,
                        \overline B_i H^{i}
                        \epsilon_{ajl}\overline {\cal B}^a_m
                       {\cal D}^{ljm}
\non\\                       
&=& -\sqrt{6}\,T_{\BD}\,
                        \overline B_m H^{i}_T
                        \epsilon_{ija}\overline {\cal B}^a_l
                        {\cal D}^{ljm}
\label{eq: BDT}
\end{eqnarray}
where the $T_{2\BD}$, $A_{1\BD}$ and $A_{2\BD}$ terms in the equation are vanishing as 
$\epsilon_{ajl} {\cal D}^{ljm}=\epsilon_{ajl}{\cal D}^{ljm}=0$, 
and $T_{1\BD}$ is relabeled to $T_{\BD}$ in the last step.

To obtain the effective Hamiltonian of $\overline B_q\to {\cal B}\overline {\cal B}l\bar\nu$ decays, we first replace 
$\overline{\cal D}_{ijl}$ and ${\cal D}^{ljm}$ in Eq.~(\ref{eq: DDT}) by 
$(\overline{\cal B}_1)_{ijl}\equiv\epsilon_{ija}\overline {\cal B}^a_l$, $(\overline{\cal B}_2)_{ijl}\equiv\epsilon_{akl}\overline {\cal B}^a_i$ 
and
$({\cal B}_1)^{lim}\equiv\epsilon^{ljb} {\cal B}^m_b$, $({\cal B}_2)^{lim}\equiv\epsilon^{bjm} {\cal B}^l_b$,
respectively, and obtain
\be
H_{\rm eff}(\overline B_q\to\BB l\bar\nu)
&=&
-T_{11\BB}\overline B_m H^{i}_T (\overline {\cal B}_1)_{ijl}({\cal B}_1)^{ljm}
 -T_{12\BB}\overline B_m H^{i}_T (\overline {\cal B}_1)_{ijl}({\cal B}_2)^{ljm}
\non\\
&& 
 -T_{21\BB}\overline B_m H^{i}_T (\overline {\cal B}_2)_{ijl}({\cal B}_1)^{ljm}
 -T_{22\BB}\overline B_m H^{i}_T (\overline {\cal B}_2)_{ijl}({\cal B}_2)^{ljm}
\non\\
&&
-A_{11\BB}\overline B_i H^{i}_T (\overline {\cal B}_1)_{mjl}({\cal B}_1)^{ljm}
 -A_{12\BB}\overline B_i H^{i}_T (\overline {\cal B}_1)_{mjl}({\cal B}_2)^{ljm}
\non\\
&& 
 -A_{21\BB}\overline B_i H^{i}_T (\overline {\cal B}_2)_{mjl}({\cal B}_1)^{ljm}
 -A_{22\BB}\overline B_i H^{i}_T (\overline {\cal B}_2)_{mjl}({\cal B}_2)^{ljm}.
\en
Using the following identity
\be
-2(\overline {\cal B}_1)_{ijl} ({\cal B}_1)^{ljm}
&=&(\overline {\cal B}_2)_{ijl} ({\cal B}_1)^{ljm}
=-2(\overline {\cal B}_2)_{ijl} ({\cal B}_2)^{ljm},
\non\\
-2(\overline {\cal B}_1)_{mjl}({\cal B}_1)^{ljm}
&=&(\overline {\cal B}_1)_{mjl}({\cal B}_2)^{ljm}
=(\overline {\cal B}_2)_{mjl}({\cal B}_1)^{ljm}
=-2(\overline {\cal B}_2)_{mjl}({\cal B}_2)^{ljm}, 
\en
the above Hamiltonian can be expressed as
\be
H_{\rm eff}(\overline B_q\to\BB l\bar\nu)&=&
(-T_{11\BB}+2T_{21\BB}-T_{22\BB})\overline B_m H^{i}_T (\overline {\cal B}_1)_{ijl}({\cal B}_1)^{ljm}
 \non\\
&& -T_{12\BB}\overline B_m H^{i}_T (\overline {\cal B}_1)_{ijl}({\cal B}_2)^{ljm}
\non\\
&&
+(A_{11\BB}-2A_{12\BB}-2A_{21\BB}+A_{22\BB}) \overline B_i H^{i}_T (\overline {\cal B}_1)_{mjl}({\cal B}_1)^{ljm}
\non\\
&=&
-T_{1\BB}\overline B_m H^{i}_T \epsilon_{ija}\overline {\cal B}^a_l \epsilon^{bjm} {\cal B}^l_b
+T_{2\BB}\overline B_m H^{i}_T \epsilon_{ija}\overline {\cal B}^a_l \epsilon^{ljb} {\cal B}^m_b
\non\\
&&+A_{\BB}\overline B_i H^{i}_T \epsilon_{mja}\overline {\cal B}^a_l \epsilon^{ljb} {\cal B}^m_b,
\label{eq: BBT}
 \en
where the topological amplitudes are redefined as following
\be
T_{1\BB}&\equiv& T_{12\BB},
\non\\
T_{2\BB}&\equiv& -T_{11\BB}+2T_{21\BB}-T_{22\BB},
\non\\
A_{\BB}&\equiv& A_{11\BB}-2A_{12\BB}-2A_{21\BB}+A_{22\BB}.
\en
With this all effective Hamiltonians of $\overline B_q\to {\bf B} \overline{\bf B}' l\bar\nu$ decays with low-lying octet and decuplet baryons are obtained.

The effective Hamiltonian of the $\overline B_q\to {\bf B} \overline{\bf B}'\nu\bar\nu$ decays can be obtained similarly. We simply give the results
in the following equation,
\begin{eqnarray}
H_{\rm eff}(\overline B_q\to\DD\nu\bar\nu)&=& 
              6PB_{{\cal D}\overline{\cal D}}\,\overline B_m H^k_{PB}
             \overline {\cal D}_{kjl} {\cal D}^{ljm}
             +PBA_{{\cal D}\overline{\cal D}}\,\overline B_k H^k_{PB}
             \overline {\cal D}_{mjl} {\cal D}^{ljm},
\non\\
H_{\rm eff}(\overline B_q\to\DB\nu\bar\nu)&=&
           \sqrt{6} PB_\DB\,
                        \overline B_m H^k_{PB}\overline{\cal D}_{kjl} 
                        \epsilon^{bjm}{\cal B}^l_{b},
\non\\
H_{\rm eff}(\overline B_q\to\BD\nu\bar\nu)&=& 
           -\sqrt{6} PB_\BD\,
                        \overline B_m H^k_{PB}
                        \epsilon_{kja} \overline {\cal B}^a_{l}
                       {\cal D}^{ljm},
\non\\
H_{\rm eff}(\overline B_q\to\BB\nu\bar\nu)&=&
           -PB_{1\BB}\,
                        \overline B_m H^k_{PB}
                        \epsilon_{kja} \overline {\cal B}^a_{l}
                        \epsilon^{bjm} {\cal B}^l_{b}
           +PB_{2\BB}\,
                        \overline B_m H^k_{PB}
                        \epsilon_{kja} \overline {\cal B}^a_{l}
                        \epsilon^{ljb} {\cal B}^m_{b}                
\non\\
&&+          PBA_{\BB}\,
                        \overline B_k H^k_{PB}
                        \epsilon_{mja} \overline {\cal B}^a_{l}
                        \epsilon^{ljb} {\cal B}^m_{b} .             
\label{eq: TPnunu}
\end{eqnarray}

In summary the effective Hamiltonians of $\overline B_q\to \bfBB' l\bar\nu$ and $\bfBB' \nu\bar\nu$ decays for low-lying octet and decuplet baryons are obtained and are shown in
Eqs. (\ref{eq: DDT}), (\ref{eq: DBT}), (\ref{eq: BDT}), (\ref{eq: BBT}) and (\ref{eq: TPnunu}).   
The decay amplitudes can be obtained readily by using these effective Hamiltonians.
The results of decay amplitudes in terms of these topological amplitudes and relations on the amplitudes will be given explicitly in the next section.

Before we end this section it is important to note that, as shown in Eq. (\ref{eq: AA}), the topological amplitudes $PB$ and $T$ and the topological amplitudes $PBA$ and $A$ should be related in the following manner,
\be
\zeta&\equiv& 
\frac{PB_{i\BB}}{T_{i\BB}}
=\frac{PBA_\BB}{A_\BB}
=\frac{PB_\BD}{T_\BD}
=\frac{PB_\DB}{T_\DB}
=\frac{PB_\DD}{T_\DD}
=\frac{PBA_\DD}{A_\DD}
\non\\
&=&\frac{\alpha_{\rm em}}{2\pi\sin^2\theta_W}
\frac{V_{ts}^*V_{tb}}{V_{ub}}  D(m_t^2/m_W^2), 
\label{eq: zeta}
\en
where numerically we use $|V_{ub}|=0.0036$ and have $\zeta=  -0.037 e^{i\phi_3}$, with $\phi_3=(65.5^{+1.3}_{-1.1})^\circ$ one of the unitary angle in the CKM matrix \cite{CKMfitter}.

\subsection{Modeling the topological amplitudes}\label{sec: model calculations}

In addition to the above decompositions of amplitudes in terms of topological amplitudes,
it will be useful to have some numerical results on rates. We will use the available theoretical inputs from refs~\cite{Geng:2021sdl} and \cite{ Hsiao:2022uzx} in our modeling of the topological amplitudes and we denote them as Model 1 and Model 2, respectively.
They are used as illustration and can be improved when more data are available.

In general the topological amplitudes $T_{1\BB}$, $T_{2\BB}$ and $A_\BB$ in $\overline B_q\to\BB' l\bar\nu$ decays can be expressed as
\be
T_{i\BB}
&=&
i\frac{G_F}{\sqrt2} V_{ub} 
\bar l_L \gamma^\mu \nu_L
\bar u(p_{{\cal B}})
\{
[g^{(i)}_1 \gamma_\mu
+ig^{(i)}_2 \sigma_{\mu\nu} q^\nu
+g^{(i)}_3 q_\mu
+g^{(i)}_4 (p_{{\cal B}}+p_{\overline{\cal B}^\prime})_\mu
+g^{(i)}_5 (p_{{\cal B}}-p_{\overline{\cal B}^\prime})_\mu
 ]\gamma_5
\non\\
&&
-[f^{(i)}_1 \gamma_\mu 
+if^{(i)}_2 \sigma_{\mu\nu} q^\nu
+f^{(i)}_3 q_\mu 
+f^{(i)}_4 (p_{{\cal B}}+p_{\overline{\cal B}^\prime})_\mu
+f^{(i)}_5  (p_{{\cal B}}-p_{\overline{\cal B}^\prime})_\mu]\}
v_R(p_{\overline{\cal B}^\prime}),
\non\\
A_{\BB}
&=&
i\frac{G_F}{\sqrt2} V_{ub} 
\bar l_L \gamma^\mu \nu_L
\bar u(p_{{\cal B}})
\{
[g^{(a)}_1 \gamma_\mu
+ig^{(a)}_2 \sigma_{\mu\nu} q^\nu
+g^{(a)}_3 q_\mu
+g^{(a)}_4 (p_{{\cal B}}+p_{\overline{\cal B}^\prime})_\mu
+g^{(a)}_5 (p_{{\cal B}}-p_{\overline{\cal B}^\prime})_\mu
 ]\gamma_5
\non\\
&&
-[f^{(a)}_1 \gamma_\mu 
+if^{(a)}_2 \sigma_{\mu\nu} q^\nu
+f^{(a)}_3 q_\mu 
+f^{(a)}_4 (p_{{\cal B}}+p_{\overline{\cal B}^\prime})_\mu
+f^{(a)}_5  (p_{{\cal B}}-p_{\overline{\cal B}^\prime})_\mu]\}
v_R(p_{\overline{\cal B}^\prime}),
\label{eq: Ti}
\en
with $q\equiv p_{B_q}-p_{{\rm\bf B}}-p_{\overline{\rm\bf B}^\prime}$, $i=1,2$, $j=1,\dots,5$, and $f^{(i)}_j$, $g^{(i)}_j$, $f^{(a)}_j$ and $g^{(a)}_j$ denoting form factors.
Similarly the topological amplitudes of $\overline B_q\to \BD l\bar \nu$ and $\overline B_q\to \DB l\bar \nu$ decays can be expressed as
 \be
T_{\BD}
&=&
i\frac{G_F}{\sqrt2} V_{ub} 
\bar l_L \gamma_\mu \nu_L
\non\\
&&
\times
\bar u(p_{{\cal D}},\lambda_{\cal B})
\Big\{
 \Big[
 g'_1 p_{{\cal B}\nu} \gamma_\mu
 +i g'_2 \sigma_{\mu\rho} p_{{\cal B}\nu} q^\rho
 +g'_3 p_{{\cal B}\nu} q_\mu
 +g'_4 p_{{\cal B}\nu}p_{{\cal B}\mu}
 +g'_5 g_{\nu\mu}
  +g'_6 q_\nu \gamma_\mu
 \non\\
 &&\quad
 +i g'_7 \sigma_{\mu\rho} q_\nu q^\rho
 +g'_8 q_\nu q_\mu
 +g'_9 q_\nu p_{{\cal B}\mu}
 \Big]
 \gamma_5
 -\Big[
 f'_1 p_{{\cal B}\nu} \gamma_\mu
 +i f'_2 \sigma_{\mu\rho} p_{{\cal B}\nu} q^\rho
 +f'_3 p_{{\cal B}\nu} q_\mu
 \non\\
 &&\quad
 +f'_4 p_{{\cal B}\nu}p_{{\cal B}\mu}
 +f'_5 g_{\nu\mu}
 +f'_6 q_\nu \gamma_\mu
 +i f'_7 \sigma_{\mu\rho} q_\nu q^\rho
 +f'_8 q_\nu q_\mu
 +f'_9q_\nu p_{{\cal B}\mu}
 \Big]
 \Big\}
 v^\nu(p_{\overline{\cal D}},\lambda_{\overline{\cal D}}),
\label{eq: T BD}
\en
and
\be
T_{\DB}
&=&
i\frac{G_F}{\sqrt2} V_{ub} 
\bar l_L \gamma_\mu \nu_L
\non\\
&&\times
\bar u^\nu(p_{{\cal D}},\lambda_{\cal D})
\Big\{
 \Big[
 g''_1 p_{\overline{\cal B}\nu} \gamma_\mu
 +i g''_2 \sigma_{\mu\rho} p_{\overline{\cal B}\nu} q^\rho
 +g''_3 p_{\overline{\cal B}\nu} q_\mu
 +g''_4 p_{\overline{\cal B}\nu}p_{\overline{\cal B}\mu}
 +g''_5 g_{\nu\mu}
 +g''_6 q_\nu  \gamma_\mu
 \non\\
 &&\quad
 +i g''_7 \sigma_{\mu\rho} q_\nu  q^\rho
 +g''_8 q_\nu  q_\mu
 +g''_9 q_\nu  p_{\overline{\cal B}\mu}
 \Big]
 \gamma_5
 -\Big[
 f''_1 p_{\overline{\cal B}\nu} \gamma_\mu
 +if''_2 \sigma_{\mu\rho} p_{\overline{\cal B}\nu} q^\rho
 +f''_3 p_{\overline{\cal B}\nu} q_\mu
\non\\
 &&\quad 
 +f''_4 p_{\overline{\cal B}\nu}p_{\overline{\cal B}\mu}
 +f''_5 g_{\nu\mu}
 +f''_6 q_\nu \gamma_\mu
 +if''_7 \sigma_{\mu\rho} q_\nu q^\rho
 +f''_8 q_\nu q_\mu
 +f''_9 q_\nu p_{\overline{\cal B}\mu}
 \Big]
 \Big\}
 v(p_{\overline{\cal B}},\lambda_{\overline{\cal B}}),
\label{eq: T DB}
\en
where $u^\mu,\,v^\mu$ are the Rarita-Schwinger vector spinors.
Finally the tree topological amplitude for $\overline B_q\to \DD' l\bar\nu$ decay is given by
 \be
T_{\DD}
&=&
i\frac{G_F}{\sqrt2} V_{ub} 
\bar l_L \gamma_\mu \nu_L
\non\\
&&
\times
\bar u_\nu(p_{{\cal D}},\lambda_{\cal D})
\Big\{
 \Big[
 g'''_1 \gamma_\mu
 +i g'''_2 \sigma_{\mu\rho} q^\rho
 +g'''_3 q_\mu
 +g'''_4 (p_{{\cal D}}+p_{\overline{\cal D}^\prime})_\mu
+g'''_5 (p_{{\cal D}}-p_{\overline{\cal D}^\prime})_\mu
 ]\gamma_5
\non\\
&&
-[f'''_1 \gamma_\mu 
+if'''_2 \sigma_{\mu\nu} q^\nu
+f'''_3 q_\mu 
+f'''_4 (p_{{\cal D}}+p_{\overline{\cal D}^\prime})_\mu
+f'''_5  (p_{{\cal D}}-p_{\overline{\cal D}^\prime})_\mu]\}
 \Big]
 \Big\}
 v^\nu(p_{\overline{\cal D}},\lambda_{\overline{\cal D}})
 \non\\
&& +\dots,
\label{eq: T DD}
\en
where terms such as 
$\bar u_\nu p_{\overline{\cal D}}^\nu \{\dots\} p_{\cal D\sigma} u^\sigma$,
$\bar u_\nu q^\nu \{\dots\} p_{\cal D\sigma} u^\sigma$,
$\bar u_\nu p_{\overline{\cal D}}^\nu \{\dots\} q_{\sigma} u^\sigma$,
$\bar u_\nu q^\nu \{\dots\} q_{\sigma} u^\sigma$
are not shown explicitly in the above equation.
The annihilation amplitude $A_\DD$ can be expressed similarly.
Topological amplitudes for loop induced $\overline B_q\to\bfBB' \nu\bar\nu$ decays can be obtained using the above equations and Eq. (\ref{eq: zeta}).

\begin{table}[t!]
\caption{\label{tab: GF} Values of $G^{(i)}_j$ and $F^{(i)}_j$ for Model 1 and Model 2. They are extracted from refs~\cite{Geng:2021sdl} and \cite{ Hsiao:2022uzx}, respectively, but slightly modified to match the asymptotic relations in Appendix~A and to match the $B^-\to p\bar p l\bar\nu$ data.
}
\begin{ruledtabular}
\begin{tabular}{cccccccc}
Model
          & $G^{(1)}_1$(GeV$^5)$
          & $G^{(1)}_2$(GeV$^4)$
          & $G^{(1)}_3$(GeV$^4)$
          & $G^{(1)}_4$(GeV$^4)$
          & $G^{(1)}_5$(GeV$^4)$
          \\
\hline 
Model 1 
          & 67.02
          & $-1.98$
          & $-1.98$
          & $-1.98$
          & $-1.98$
          \\
Model 2 
          & $-163.00$
          & $-11.90$
          & $-110.31$
          & $9.65$
          & $20.13$
          \\
 \hline
Model
          & $G^{(2)}_1$(GeV$^5)$
          & $G^{(2)}_2$(GeV$^4)$
          & $G^{(2)}_3$(GeV$^4)$
          & $G^{(2)}_4$(GeV$^4)$
          & $G^{(2)}_5$(GeV$^4)$
          \\
\hline 
Model 1 
          & 96.90
          & 9.89
          & 9.89
          & 9.89
          & 9.89
          \\
Model 2 
          & 94.07
          & 59.50
          & 551.56
          & $-48.27$
          & $-100.66$
          \\  
\hline
Model
          & $F^{(1)}_1$(GeV$^5)$
          & $F^{(1)}_2$(GeV$^4)$
          & $F^{(1)}_3$(GeV$^4)$
          & $F^{(1)}_4$(GeV$^4)$
          & $F^{(1)}_5$(GeV$^4)$
          \\
\hline 
Model 1 
          & $-9.06$
          & $1.98$
          & $1.98$
          & $1.98$
          & $1.98$
          \\
Model 2 
          & $168.59$
          & $11.90$
          & $110.31$
          & $-9.65$
          & $-20.13$
          \\            
\hline            
 Model
          & $F^{(2)}_1$(GeV$^5)$
          & $F^{(2)}_2$(GeV$^4)$
          & $F^{(2)}_3$(GeV$^4)$
          & $F^{(2)}_4$(GeV$^4)$
          & $F^{(2)}_5$(GeV$^4)$
          \\
\hline 
Model 1 
          & 134.94
          & $-9.89$
          & $-9.89$
          & $-9.89$
          & $-9.89$
          \\
Model 2 
          & $-71.72$
          & $-59.50$
          & $-551.56$
          & $48.27$
          & $100.66$
          \\  
\end{tabular}
\end{ruledtabular}
\end{table}

The topological amplitudes for $\overline B_q\to\BB' l\bar \nu$ decays are given in Eq. (\ref{eq: Ti}).
Fo illustration
we follow refs~\cite{Geng:2021sdl, Hsiao:2022uzx} to use
\be
g^{(i)}_j=\frac{G^{(i)}_{j}}{t^3},
\quad
f^{(i)}_j=\frac{F^{(i)}_{f_j}}{t^3},
\quad
g^{(a)}_j=f^{(a)}_j=0,
\label{eq: g f}
\en
where $G^{(i)}_j$ and $F^{(i)}_j$ are some constants to be specified later, $t\equiv m^2_{\bfBB'}$ and the last equation corresponds to the $A_\BB=PBA_\BB=0$ case. 
The values of the constants $G^{(i)}_j$ and $F^{(i)}_j$ are extracted from refs~\cite{Geng:2021sdl, Hsiao:2022uzx} 
but slightly modified to match the asymptotic relations in Appendix~A,
where it is known that there are asymptotic relations~\cite{Brodsky:1980sx} in the matrix elements of octet and decuplet baryons in the large momentum transfer region,
and to match the $B^-\to p\bar p l\bar\nu$ data.
In fact we find that the corresponding $F^{(i)}_{3, 4, 5}$ used in ref.~\cite{Geng:2021sdl} do not satisfy the correct asymptotic relations, which can however be satisfied by adding a minus sign to their $F^{(i)}_{3,4,5}$. 
Nevertheless as we shall see that the modification do not significantly affect the $B^-\to \BB' l\bar\nu$ rates.

The values of $G^{(i)}_j$ and $F^{(i)}_j$ are shown in Table~\ref{tab: GF}. 
Explicitly we use
$G^{(i)}_1=\eta_1 m_B (e^{(i)}_\parallel C_{LL}-e^{(i)}_{\overline\parallel} C_{RR})/3$, 
$F^{(i)}_1=\eta_1 m_B (e^{(i)}_\parallel C_{LL}+e^{(i)}_{\overline\parallel} C_{RR})/3$, 
$G^{(i)}_{2,3,4,5}=-F^{(i)}_{2,3,4,5}=-\eta_1\times 2e^{(i)}_F C_{LR}/3$ with $(C_{LL},C_{RR},C_{LR})=(17.78,-11.67,6.41)$ GeV$^4$ \cite{Geng:2021sdl} for Model 1,
and 
$G^{(i)}_1=\eta_2 m_B (e^{(i)}_\parallel D_{\parallel}-e^{(i)}_{\overline\parallel} D_{\overline\parallel})/3$, 
$F^{(i)}_1=\eta_2 m_B (e^{(i)}_\parallel D_{\parallel}+e^{(i)}_{\overline\parallel} D_{\overline\parallel})/3$, 
$G^{(i)}_{2,3,4,5}=-F^{(i)}_{2,3,4,5}=-\eta_2\times 2e^{(i)}_F D_{2,3,4,5}/3$ 
with $(D_{\parallel},D_{\overline\parallel})=(11.2,323.3)$ GeV$^5$ and $D_{2,3,4,5}=(47.7,442.2,-38.7,-80.7)$ GeV$^4$ \cite{Hsiao:2022uzx} for Model 2,
where the factors $\eta_1=0.93$ and $\eta_2=0.75$ are introduced to match the central value of the $B^-\to p\bar p l\bar\nu$ data and $e^{(i)}_{\parallel,\overline\parallel, F}$ are given in 
Eq.~(\ref{eq: Teee}). 
Note that the sign of $D_5$ is flipped from the one from ref. \cite{Hsiao:2022uzx}, to match the definitions of form factors $f^{(i)}_5$ and $g^{(i)}_5$ in Eq. (\ref{eq: Ti}).

\begin{table}[t!]
\caption{\label{tab: GF tilde bar} 
Values of $ G'_i$, $ F'_i$, $ G''_i$, $ F''_i$ for Model 1 and Model 2. 
}
\begin{ruledtabular}
\begin{tabular}{cccccccc}
Model
          & $ G'_1$(GeV$^5)$
          & $ G'_2$(GeV$^4)$
          & $ G'_3$(GeV$^4)$
          & $ G'_4$(GeV$^4)$
          & $ G'_5$(GeV$^4)$
          \\
\hline 
Model 1 
          & $-24.39$
          & 4.85
          & 4.85
          & 9.69
          & 0
          \\
Model 2 
          & $-209.90$
          & 29.15
          & 270.21
          & $-72.96$
          & $-12.83$
          \\  
\hline  
Model
          & $ F'_1$(GeV$^5)$
          & $ F'_2$(GeV$^4)$
          & $ F'_3$(GeV$^4)$
          & $ F'_4$(GeV$^4)$
          & $ F'_5$(GeV$^4)$
          \\
\hline 
Model 1 
          & $-117.58$
          & $-4.85$
          & $-4.85$
          & $-9.69$
          & 0
          \\
Model 2 
          & $196.21$
          & $-29.15$
          & $-270.21$
          & $72.96$
          & $12.83$
          \\
 \hline
 Model
          & $ G''_1$(GeV$^5)$
          & $ G''_2$(GeV$^4)$
          & $ G''_3$(GeV$^4)$
          & $ G''_4$(GeV$^4)$
          & $ G''_5$(GeV$^4)$
          \\
\hline 
Model 1 
          & $-24.39$
          & $4.85$
          & $4.85$
          & $0$
          & $-4.85$
          \\
Model 2 
          & $-209.90$
          & $29.15$
          & $270.21$
          & $25.66$
          & $36.48$
          \\  
\hline  
Model
          & $ F''_1$(GeV$^5)$
          & $ F''_2$(GeV$^4)$
          & $ F''_3$(GeV$^4)$
          & $ F''_4$(GeV$^4)$
          & $ F''_5$(GeV$^4)$
          \\
\hline 
Model 1 
          & $-117.58$
          & $-4.85$
          & $-4.85$
          & $0$
          & $4.85$
          \\
Model 2 
          & $196.21$
          & $-29.15$
          & $-270.21$
          & $-25.66$
          & $-36.48$
          \\                                           
\end{tabular}
\end{ruledtabular}
\end{table}

The topological amplitudes of $\overline B_q\to \BD l\bar \nu$ and $\overline B_q\to \DB l\bar \nu$ decays are given in Eqs. (\ref{eq: T BD}) and (\ref{eq: T DB}). 
For simplicity, we only concentrate on the contributions from $g'_{1,2,3,4,5}$, $f'_{1,2,3,4,5}$, $g''_{1,2,3,4,5}$ and $f''_{1,2,3,4,5}$, 
by assuming that their contributions are dominant.
This working assumption can be checked or relaxed when data of $\overline B_q\to \BD l\bar \nu$ and $\overline B_q\to \DB l\bar \nu$ decays become available.

It is known that in the asymptotic limit form factors of octet-octet and octet decuplet are related~\cite{Brodsky:1980sx}.
As shown in Appendix \ref{App: asym} in the asymptotic limit $T_{\BD}$, $T_{\BD}$ and $T_{i\BB}$ are related and have similar structure. 
These impose constrains on the form factors. 
For simplicity we assume that these form factors have similar forms as the form factors in Eq.~(\ref{eq: g f}).
Using Eqs. (\ref{eq: g f}), (\ref{eq: T BD DB asymptotic}) and (\ref{eq: Teee}), we have
\be
 g'_{1,2,3,4}&=&\frac{m_{\overline{\cal D}} G'_{1,2,3,4}}{t^4},\,\, 
 g'_{5}=\frac{m_{\overline{\cal D}} G'_{5}}{t^3},
\quad
 f'_{1,2,3,4}=\frac{m_{\overline{\cal D}} F'_{1,2,3,4}}{t^4}, \,\, f'_{5}=\frac{m_{\overline{\cal D}} F'_{5}}{t^3},
\non\\
 g''_{1,2,3,4}&=&\frac{m_{\cal D} G''_{1,2,3,4}}{t^4},\,\,  g''_{5}=\frac{m_{\cal D} G''_{5}}{t^3},
\quad
 f''_{1,2,3,4}=\frac{m_{\cal D} F_{1,2,3,4}}{t^4}, \,\, f''_{5}=\frac{m_{\cal D} F''_{5}}{t^3},
\label{eq: g f tilde bar}
\en
with
\be
 G'_{1,2,3}=-\sqrt6 G^{(i)}_{1,2,3},
\quad
 G'_{4}=-\sqrt6 (G^{(i)}_{4}+ G^{(i)}_5),
\quad
 G'_{5}=\sqrt{\frac{3}{2}} (G^{(i)}_{4}- G^{(i)}_5),
\non\\
 F'_{1,2,3}=-\sqrt6 F^{(i)}_{1,2,3},
\quad
 F'_{4}=-\sqrt6 (F^{(i)}_{4}+ F^{(i)}_5),
\quad
 F'_{5}=\sqrt{\frac{3}{2}} (F^{(i)}_{4}-F^{(i)}_5),
\en
but with $(e^{(i)}_\parallel, e^{(i)}_{\overline{\parallel}}, e^{(i)}_F)$ in $G^{(i)}_{j}, F^{(i)}_{j}$ replaced by $(e'_\parallel, e'_{\overline{\parallel}}, e'_F)$,
and
\be
 G''_{1,2,3}=-\sqrt6 G^{(i)}_{1,2,3},
\quad
 G''_{4}=\sqrt6 (G^{(i)}_{5}- G^{(i)}_4),
\quad
 G''_{5}=\sqrt{\frac{3}{2}} (G^{(i)}_{4}+ G^{(i)}_5),
\non\\
 F''_{1,2,3}=-\sqrt6 F^{(i)}_{1,2,3},
\quad
 F''_{4}=\sqrt6 (F^{(i)}_{5}- F^{(i)}_4),
\quad
 F''_{5}=\sqrt{\frac{3}{2}} (F^{(i)}_{4}+ F^{(i)}_5),
\en
but with $(e^{(i)}_\parallel, e^{(i)}_{\overline{\parallel}}, e^{(i)}_F)$  in $G^{(i)}_{j}, F^{(i)}_{j}$ replaced by $(e''_\parallel, e''_{\overline{\parallel}}, e''_F)$.
Note that the above constants are related in the asymptotic limit and, consequently, inputs from Model 1 and 2 have been used in the above relations.
The values of these constants in Model~1 and 2 are given in Table~\ref{tab: GF tilde bar}.

\begin{table}[t!]
\caption{\label{tab: GF hat} 
Values of $ G'''_i$, $ F'''_i$ for Model 1 and Model 2. 
}
\begin{ruledtabular}
\begin{tabular}{cccccccc}
Model
          & $ G'''_1$(GeV$^5)$
          & $ G'''_2$(GeV$^4)$
          & $ G'''_3$(GeV$^4)$
          & $ G'''_4$(GeV$^4)$
          & $ G'''_5$(GeV$^4)$
          \\
\hline 
Model 1 
          & $-115.47$
          & 5.94
          & 5.94
          & 5.94
          & 5.94
          \\
Model 2 
          & $115.96$
          & 35.70
          & 330.94
          & $-28.96$
          & $-60.39$
          \\  
\hline  
Model
          & $ F'''_1$(GeV$^5)$
          & $ F'''_2$(GeV$^4)$
          & $ F'''_3$(GeV$^4)$
          & $ F'''_4$(GeV$^4)$
          & $ F'''_5$(GeV$^4)$
          \\
\hline 
Model 1 
          & $-58.41$
          & $-5.94$
          & $-5.94$
          & $-5.94$
          & $-5.94$
          \\
Model 2 
          & $-132.73$
          & $-35.70$
          & $-330.94$
          & $28.96$
          & $60.39$
          \\            
\end{tabular}
\end{ruledtabular}
\end{table}

In the model calculations of $\overline B_q\to\DD' l\nu$ and $\DD' \nu\bar\nu$ decay rates,
we use Eq. (\ref{eq: T DD}) for the tree topological amplitude, 
where we neglect terms, such as 
$\bar u_\nu p_{\overline{\cal D}}^\nu \{\dots\} p_{\cal D\sigma} u^\sigma$,
$\bar u_\nu q^\nu \{\dots\} p_{\cal D\sigma} u^\sigma$,
$\bar u_\nu p_{\overline{\cal D}}^\nu \{\dots\} q_{\sigma} u^\sigma$,
$\bar u_\nu q^\nu \{\dots\} q_{\sigma} u^\sigma$,
for simplicity.
This working assumption can be checked or modified once data is available.
As in $\overline B_q\to\BB' l\bar\nu$ decays, we neglect the contribution from the annihilation topological amplitude, $A_\DD$.
Using Eqs. (\ref{eq: g f}), (\ref{eq: Teee}) and (\ref{eq: T DD asymptotic}), the form factors are given by
\be
g'''_j=m_{\cal D} m_{\overline{\cal D}}\frac{G'''_j}{t^4},
\quad
f'''_j=m_{\cal D} m_{\overline{\cal D}} \frac{F'''_j}{t^4},
\en
with
\be
G'''_j=-3G^{(i)}_j,
\quad
F'''_j=-3F^{(i)}_j,
\en
but with $(e^{(i)}_\parallel, e^{(i)}_{\overline{\parallel}}, e^{(i)}_F)$ in $G^{(i)}_{j}, F^{(i)}_{j}$ replaced by $(e'''_\parallel, e'''_{\overline{\parallel}}, e'''_F)$.
Note that in the asymptotic limit the above form factors are related to those in $\overline B_q\to\BB' l\bar\nu$ decays via Eq. (\ref{eq: Teee}), and, consequently, inputs from Model 1 and 2 have been used.
The values of these constants in Model~1 and 2 are given in Table~\ref{tab: GF hat}.

\section{Results on amplitudes}

\subsection{Decay amplitudes in terms of topological amplitudes}

Using the above Hamiltonian the decompositions of amplitudes for $\overline B_{q}\to \BB' l\bar\nu$, $\BD l \bar \nu$, $\DB l \bar \nu$, $\DD' l \bar \nu$ and 
$\overline B_{q}\to \BB' \nu \bar \nu$, $\BD \nu \bar \nu$, $\DB \nu \bar \nu$, $\DD' \nu \bar \nu$ decays are shown in Tables
\ref{tab: TPBB}, \ref{tab: TPBD}, \ref{tab: TPDB} and \ref{tab: TPDD}.
These tables are some of the main results of this work.

As shown in Table~\ref{tab: TPBB} we have three topological amplitudes, $T_{2\BB}$, $T_{1\BB}$ and $A_\BB$, in $\overline B_q\to\BB' l\bar\nu$ decays,
and three topological amplitudes, $PB_{2\BB}$, $PB_{1\BB}$ and $PBA_\BB$, in $\overline B_q\to\BB' \nu\bar\nu$ decays.
As shown in Table~\ref{tab: TPBD} we need one topological amplitude, $T_{\BD}$, in $\overline B_q\to\BD l\bar\nu$ decays,
and
one topological amplitude, $PB_{\BD}$, in $\overline B_q\to\BD \nu\bar\nu$ decays.
Similarly, as shown in Table~\ref{tab: TPDB} we have
one topological amplitude, $T_{\DB}$, in $\overline B_q\to\DB l\bar\nu$ decays,
and one topological amplitude, $PB_{\DB}$, in $\overline B_q\to\DB \nu\bar\nu$ decays.
Finally as shown in Table~\ref{tab: TPBD} we have
two topological amplitude, $T_{\DD}$ and $A_\DD$, in $\overline B_q\to\DD' l\bar\nu$ decays,
and two topological amplitudes, $PB_{\DD}$ and $PBA_\DD$, in $\overline B_q\to\DD' \nu\bar\nu$ decays.

As the numbers of independent topological amplitudes are highly limited comparing to the numbers of the decay modes,
there are plenty of relations on $\overline B_q\to\bfBB'l\bar \nu$ and $\bfBB' \nu\bar\nu$ decay amplitudes.
These relations will be given in the following discussion.


\begin{table}[t!]
\caption{\label{tab: TPBB} 
Topological amplitudes for $\overline B_{q}\to \BB' l \bar \nu$ and $\overline B_{q}\to \BB' \nu \bar \nu$ decays.
}
\begin{ruledtabular}
\begin{tabular}{cccccccc}
Mode
          & $A(\overline B_{q'}\to \BB' l \bar \nu)$
          & Mode
          & $A(\overline B_{q'}\to \BB'  l \bar \nu)$
          \\
\hline $B^-\to p \bar p l \bar \nu$
          & $T_{1\BB}+T_{2\BB}+A_\BB$
          & $B^-\to n \bar n l \bar \nu$ 
          & $T_{1\BB}+A_\BB$
          \\
$B^-\to \Sigma^+ \overline{\Sigma^+} l \bar \nu$
          & $T_{1\BB}+T_{2\BB}+A_\BB$
          & $B^-\to \Sigma^0 \overline{\Sigma^0} l \bar \nu$
          & $\frac{1}{2}(T_{1\BB}+T_{2\BB})+A_\BB$
          \\
$B^-\to  \Sigma^- \overline{\Sigma^-} l \bar \nu$ 
          & $A_\BB$
          & $B^-\to  \Xi^-\overline{\Xi^-} l \bar \nu$ 
          & $A_\BB$
          \\             
$B^-\to \Sigma^0 \overline{\Lambda} l \bar \nu$ 
          & $-\frac{1}{2\sqrt3} (T_{1\BB}-T_{2\BB})$
          & $B^-\to \Xi^0 \overline{\Xi^0} l \bar \nu$ 
          & $T_{1\BB}+A_\BB$
          \\     
$B^-\to  \Lambda \overline{\Sigma^0} l \bar \nu$ 
          & $-\frac{1}{2\sqrt3} (T_{1\BB}-T_{2\BB})$
          & $B^-\to  \Lambda \overline{\Lambda} l \bar \nu$ 
          & $\frac{1}{6} (5T_{1\BB}+T_{2\BB})+A_\BB$
          \\          
\hline $\overline B{}^0 \to p \bar n l \bar \nu$
          & $T_{2\BB}$
          & $\overline B{}^0 \to \Sigma^+ \overline{ \Sigma^0} l \bar \nu$ 
          & $-\frac{1}{\sqrt2} (T_{1\BB}+T_{2\BB})$
          \\
$\overline B{}^0 \to \Sigma^+ \overline{\Lambda} l \bar \nu$  
          & $-\frac{1}{\sqrt6} (T_{1\BB}-T_{2\BB})$
          & $\overline B{}^0 \to \Sigma^0 \overline{\Sigma^-} l \bar \nu$  
          & $\frac{1}{\sqrt2} (T_{1\BB}+T_{2\BB})$
          \\
$\overline B{}^0 \to \Lambda \overline{\Sigma^-}  l \bar \nu$  
          & $-\frac{1}{\sqrt6} (T_{1\BB}-T_{2\BB})$
          & $\overline B{}^0 \to \Xi^0 \overline{\Xi^-} l \bar \nu$  
          & $-T_{1\BB}$
          \\         
\hline $\overline B{}^0_s \to p \overline{\Sigma^0} l \bar \nu$
          & $-\frac{1}{\sqrt2} T_{1\BB}$
          & $\overline B{}^0_s \to p \overline{\Lambda} l \bar \nu$ 
          & $-\frac{1}{\sqrt 6} (T_{1\BB}+2T_{2\BB})$
          \\
$\overline B{}^0_s \to n \overline{\Sigma^-} l \bar \nu$ 
          & $-T_{1\BB}$
          & $\overline B{}^0_s \to \Sigma^+ \overline{\Xi^0} l \bar \nu$ 
          & $T_{2\BB}$
          \\
$\overline B{}^0_s \to \Sigma^0 \overline{\Xi^-} l \bar \nu$ 
          & $\frac{1}{\sqrt{2}} T_{2\BB}$ 
          & $\overline B{}^0_s \to \Lambda \overline{\Xi^-} l \bar \nu$ 
          & $\frac{1}{\sqrt{6}} (2T_{1\BB}+T_{2\BB})$ 
          \\                             
\hline \hline
Mode
          & $A(\overline B_{q'}\to \BB' \nu \bar \nu)$
          & Mode
          & $A(\overline B_{q'}\to \BB'  \nu \bar \nu)$
          \\
\hline $ B^-\to \Sigma^0 \bar p \nu \bar \nu$
          & $-\frac{1}{\sqrt2} PB_{1\BB}$
          & $ B^-\to \Sigma^- \bar n \nu \bar \nu$
          & $- PB_{1\BB}$
          \\
$B^-\to \Xi^0 \overline{\Sigma^+} \nu \bar \nu$
          & $PB_{2\BB}$
          & $ B^-\to \Xi^- \overline{\Sigma^0} \nu \bar \nu$ 
          & $\frac{1}{\sqrt2} PB_{2\BB}$ 
          \\
$ B^-\to \Xi^- \overline{\Lambda} \nu \bar \nu$ 
          & $\frac{1}{\sqrt6}(2PB_{1\BB}+PB_{2\BB})$ 
          & $ B^-\to \Lambda \bar{p} \nu \bar \nu$ 
          & $-\frac{1}{\sqrt 6} (PB_{1\BB}+2PB_{2\BB})$ 
          \\          
\hline $\overline B{}^0\to \Sigma^+ \bar p \nu \bar \nu$
          & $- PB_{1\BB}$
          & $\overline B{}^0\to \Sigma^0 \bar n \nu \bar \nu$
          & $\frac{1}{\sqrt2} PB_{1\BB}$
          \\
$\overline B{}^0\to \Xi^0 \overline{\Sigma^0} \nu \bar \nu$ 
          & $-\frac{1}{\sqrt2} PB_{2\BB}$ 
          & $\overline B{}^0\to \Xi^0 \overline{\Lambda} \nu \bar \nu$
          & $\frac{1}{\sqrt6}(2PB_{1\BB}+PB_{2\BB})$
          \\
$\overline B{}^0\to  \Xi^- \overline{\Sigma^-} \nu \bar \nu$
          & $PB_{2\BB}$
          & $\overline B{}^0\to  \Lambda \bar n \nu \bar \nu$ 
          & $-\frac{1}{\sqrt 6} (PB_{1\BB}+2PB_{2\BB})$ 
          \\         
\hline 
$\overline B{}^0_s\to p \bar p \nu \bar \nu$
          & $PBA_\BB$
          & $\overline B{}^0_s\to n \bar n \nu \bar \nu$ 
          & $PBA_\BB$
          \\
$\overline B{}^0_s \to \Sigma^+ \overline{\Sigma^+} \nu \bar \nu$
          & $PB_{1\BB}+PBA_\BB$
          & $\overline B{}^0_s \to \Sigma^0 \overline{\Sigma^0} \nu \bar \nu$
          & $PB_{1\BB}+PBA_\BB$
          \\
$\overline B{}^0_s \to \Sigma^- \overline{\Sigma^-} \nu \bar \nu$ 
          & $PB_{1\BB}+PBA_\BB$
          & $\overline B{}^0_s \to \Xi^0 \overline{\Xi^0} \nu \bar \nu$ 
          & $PB_{1\BB}+PB_{2\BB}+PBA_\BB$
          \\
$\overline B{}^0_s \to \Xi^- \overline{\Xi^-} \nu \bar \nu$ 
          & $PB_{1\BB}+PB_{2\BB}+PBA_\BB$
          & $\overline B{}^0_s \to \Lambda \overline{\Lambda} \nu \bar \nu$
          & $\frac{1}{3}(PB_{1\BB}+2PB_{2\BB})+PBA_\BB$ 
          \\                             
\end{tabular}
\end{ruledtabular}
\end{table}

\begin{table}[t!]
\caption{\label{tab: TPBD} Topological amplitudes for $\overline B_{q}\to \BD l \bar \nu$ and $\overline B_{q}\to \BD \nu \bar \nu$ decays
}
\begin{ruledtabular}
\begin{tabular}{cccccccc}
Mode
          & $A(\overline B_{q'}\to \BD l \bar \nu)$
          & Mode
          & $A(\overline B_{q'}\to \BD  l \bar \nu)$
          \\
\hline $B^-\to  p \overline{ \Delta^+} l \bar \nu$
          & $-\sqrt 2 T_\BD$
          & $B^-\to  n \overline{\Delta^0} l \bar \nu$ 
          & $-\sqrt 2 T_\BD$
          \\
$B^-\to  \Sigma^+ \overline{\Sigma^{*+}} l \bar \nu$
          & $\sqrt 2 T_\BD$
          & $B^-\to  \Sigma^0 \overline{\Sigma^{*0}} l \bar \nu$
          & $-\frac{1}{\sqrt 2} T_\BD$
          \\
$B^-\to \Xi^0 \overline{\Xi^{*0}} l \bar \nu$
          & $\sqrt 2  T_\BD$
          & $B^-\to \Lambda \overline{\Sigma^{*0}} l \bar \nu$
          & $\sqrt{\frac{3}{2}} T_\BD$
          \\      
\hline $\overline B{}^0 \to p \overline{\Delta^0} l \bar\nu$
          & $-\sqrt 2 T_\BD$
          & $\overline B{}^0 \to n \overline{\Delta^-} l \bar\nu$ 
          & $-\sqrt 6 T_\BD$
          \\
$\overline B{}^0 \to \Sigma^+ \overline{\Sigma^{*0}} l \bar\nu$
          & $T_\BD$
          & $\overline B{}^0 \to \Sigma^0 \overline{\Sigma^{*-}}  l \bar\nu$ 
          & $-T_\BD$
          \\
$\overline B{}^0 \to \Xi^0 \overline{\Xi^{*-}} l \bar\nu$ 
          & $\sqrt 2 T_\BD$
          & $\overline B{}^0 \to \Lambda \overline{\Sigma^{*-}} l \bar\nu$
          & $\sqrt 3 T_\BD$
          \\         
\hline $\overline B{}^0_s \to p \overline{\Sigma^{*0}} l \bar\nu$
          & $-T_\BD$
          & $\overline B{}^0_s \to n \overline{\Sigma^{*-}} l \bar\nu$ 
          & $-\sqrt 2 T_\BD$
          \\
$\overline B{}^0_s \to\Sigma^+ \overline{\Xi^{*0}} l \bar\nu$ 
          & $\sqrt 2 T_\BD$
          & $\overline B{}^0_s \to \Sigma^0 \overline{\Xi^{*-}} l \bar\nu$ 
          & $-T_\BD$
          \\
$\overline B{}^0_s \to \Xi^0 \overline{\Omega^-} l \bar\nu$ 
          & $\sqrt 6 T_\BD$
          & $\overline B{}^0_s \to \Lambda \overline{\Xi^{*-}} l \bar\nu$ 
          & $\sqrt 3 T_\BD$
          \\                             
\hline \hline 
Mode
          & $A(\overline B_{q'}\to \BD \nu \bar \nu)$
          & Mode
          & $A(\overline B_{q'}\to \BD  \nu \bar \nu)$
          \\
\hline
$B^-\to \Sigma^+ \overline{\Delta^{++}} \nu \bar\nu$
          & $-\sqrt 6 PB_\BD$
          & $B^-\to \Sigma^0 \overline{\Delta^+} \nu \bar\nu$ 
          & $2 PB_\BD$
          \\
$B^-\to \Sigma^- \overline{\Delta^0} \nu \bar\nu$ 
          & $\sqrt 2 PB_\BD$
          & $B^-\to \Xi^0 \overline{\Sigma^{*+}} \nu \bar\nu$ 
          & $-\sqrt 2 PB_\BD$
          \\
$B^-\to \Xi^- \overline{\Sigma^{*0}} \nu \bar\nu$ 
          & $PB_\BD$
          & 
          & 
          \\          
\hline $\overline B{}^0\to \Sigma^+ \overline{\Delta^+} \nu \bar \nu$
          & $-\sqrt 2 PB_\BD$
          & $\overline B{}^0\to \Sigma^0 \overline{\Delta^0} \nu \bar \nu$ 
          & $2 PB_\BD$
          \\
$\overline B{}^0\to\Sigma^- \overline{\Delta^-}  \nu \bar \nu$ 
          & $\sqrt 6 PB_\BD$
          & $\overline B{}^0\to \Xi^0 \overline{\Sigma^{*0}} \nu \bar \nu$ 
          & $-PB_\BD$
          \\
$\overline B{}^0\to \Xi^- \overline{\Sigma^{*-}}\nu \bar \nu$ 
          & $\sqrt 2 PB_\BD$
          & 
          & 
          \\         
\hline $\overline B{}^0_s \to \Sigma^+ \overline{\Sigma^{*+}} \nu \bar\nu$
          & $-\sqrt 2 PB_\BD$
          & $\overline B{}^0_s \to \Sigma^0 \overline{\Sigma^{*0}} \nu \bar\nu$ 
          & $\sqrt 2 PB_\BD$
          \\
$\overline B{}^0_s \to \Sigma^- \overline{\Sigma^{*-}} \nu \bar\nu$ 
          & $\sqrt 2 PB_\BD$
          & $\overline B{}^0_s \to \Xi^0 \overline{\Xi^{*0}} \nu \bar\nu$ 
          & $-\sqrt 2 PB_\BD$
          \\
$\overline B{}^0_s \to \Xi^- \overline{\Xi^{*-}}  \nu \bar\nu$ 
          & $\sqrt 2 PB_\BD$
          & 
          & 
          \\                             
\end{tabular}
\end{ruledtabular}
\end{table}

\begin{table}[t!]
\caption{\label{tab: TPDB} Topological amplitudes for $\overline B_{q}\to \DB l \bar \nu$ and $\overline B_{q}\to \DB \nu \bar \nu$ decays.}
\begin{ruledtabular}
\begin{tabular}{cccccccc}
Mode
          & $A(\overline B_{q'}\to \DB l \bar \nu)$
          & Mode
          & $A(\overline B_{q'}\to \DB  l \bar \nu)$
          \\
\hline $B^- \to \Delta^+ \bar p l \bar \nu$
          & $-\sqrt 2  T_\DB$
          & $B^- \to \Delta^0 \bar n l \bar \nu$ 
          & $-\sqrt 2  T_\DB$
          \\
$B^- \to\Sigma^{*+} \overline{\Sigma^{+}} l \bar \nu$ 
          & $\sqrt 2  T_\DB$
          & $B^- \to \Sigma^{*0} \overline{\Sigma^{0}} l \bar \nu$ 
          & $-\frac{1}{\sqrt 2}  T_\DB$
          \\
$B^- \to \Xi^{*0} \overline{\Xi^{0}} l \bar \nu$ 
          & $\sqrt 2  T_\DB$
          & $B^- \to \Sigma^{*0} \overline{\Lambda} l \bar \nu$ 
          & $\sqrt{\frac{3}{2}}  T_\DB$
          \\      
\hline $\overline B{}^0\to \Delta^{++} \bar p l \bar\nu$
          & $\sqrt 6  T_\DB$
          & $\overline B{}^0\to \Delta^+ \bar n l \bar\nu$ 
          & $\sqrt 2  T_\DB$
          \\
$\overline B{}^0\to \Sigma^{*+} \overline{\Sigma^{0}} l \bar\nu$ 
          & $-  T_\DB$
          & $\overline B{}^0\to \Sigma^{*0} \overline{\Sigma^{-}} l \bar\nu$ 
          & $-  T_\DB$
          \\
$\overline B{}^0\to \Xi^{*0} \overline{\Xi^{-}} l \bar\nu$ 
          & $-\sqrt 2  T_\DB$
          & $\overline B{}^0\to \Sigma^{*+} \overline{\Lambda} l \bar\nu$ 
          & $-\sqrt 3  T_\DB$
          \\         
\hline $\overline B{}^0_s \to \Delta^{++} \overline{\Sigma^{+}} l \bar\nu$
          & $-\sqrt 6  T_\DB$
          & $\overline B{}^0_s \to \Delta^+ \overline{\Sigma^{0}} l \bar\nu$ 
          & $2  T_\DB$
          \\
$\overline B{}^0_s \to \Delta^0 \overline{\Sigma^{-}} l \bar\nu$ 
          & $\sqrt 2  T_\DB$
          & $\overline B{}^0_s \to  \Sigma^{*+} \overline{\Xi^{0}} l \bar\nu$ 
          & $-\sqrt 2  T_\DB$
          \\
$\overline B{}^0_s \to \Sigma^{*0} \overline{\Xi^{-}} l \bar\nu$ 
          & $  T_\DB$
          & 
          & 
          \\                             
\hline \hline 
Mode
          & $A(\overline B_{q'}\to \DB \nu \bar \nu)$
          & Mode
          & $A(\overline B_{q'}\to \DB  \nu \bar \nu)$
          \\
\hline
$B^- \to \Sigma^{*0} \bar p \nu \bar\nu$
          & $-PB_\DB$
          & $B^- \to \Sigma^{*-} \bar n \nu \bar\nu$
          & $-\sqrt 2 PB_\DB$
          \\
$B^- \to \Xi^{*0} \overline{\Sigma^+} \nu \bar\nu$
          & $\sqrt 2 PB_\DB$
          & $B^- \to \Xi^{*-} \overline{\Sigma^{0}} \nu \bar\nu$
          & $- PB_\DB$
          \\
$B^- \to \Omega^- \overline{\Xi^{0}} \nu \bar\nu$ 
          & $\sqrt 6 PB_\DB$
          & $B^- \to \Xi^{*-} \overline{\Lambda} \nu \bar\nu$ 
          & $\sqrt 3 PB_\DB$
          \\          
\hline $\overline B{}^0 \to \Sigma^{*+} \bar p \nu \bar\nu$
          & $\sqrt 2 PB_\DB$
          & $\overline B{}^0 \to \Sigma^{*0} \bar n \nu \bar\nu$
          & $ PB_\DB$
          \\
$\overline B{}^0 \to  \Xi^{*0} \overline{\Sigma^{0}} \nu \bar\nu$
          & $- PB_\DB$
          & $\overline B{}^0 \to \Xi^{*-} \overline{\Sigma^{-}} \nu \bar\nu$
          & $-\sqrt 2 PB_\DB$
          \\
$\overline B{}^0 \to \Omega^- \overline{\Xi^{-}} \nu \bar\nu$
          & $-\sqrt 6 PB_\DB$
          & $\overline B{}^0 \to \Xi^{*0} \overline{\Lambda} \nu \bar\nu$
          & $-\sqrt 3 PB_\DB$
          \\         
\hline $\overline B{}^0_s \to \Sigma^{*+} \overline{\Sigma^{+}} \nu \bar\nu$
          & $-\sqrt 2 PB_\DB$
          & $\overline B{}^0_s \to \Sigma^{*0} \overline{\Sigma^{0}} \nu \bar\nu$ 
          & $\sqrt 2 PB_\DB$
          \\
$\overline B{}^0_s \to \Sigma^{*-} \overline{\Sigma^{-}} \nu \bar\nu$ 
          & $\sqrt 2 PB_\DB$
          & $\overline B{}^0_s \to \Xi^{*0} \overline{\Xi^{0}} \nu \bar\nu$ 
          & $-\sqrt 2 PB_\DB$
          \\
$\overline B{}^0_s \to \Xi^{*-} \overline{\Xi^{-}} \nu \bar\nu$ 
          & $\sqrt 2 PB_\DB$
          & 
          & 
          \\                             
\end{tabular}
\end{ruledtabular}
\end{table}

\begin{table}[t!]
\caption{\label{tab: TPDD} Topological amplitudes for $\overline B_{q}\to \DD' l \bar \nu$ and $\overline B_{q}\to \DD' \nu \bar \nu$ decays.}
\begin{ruledtabular}
\begin{tabular}{cccccccc}
Mode
          & $A(\overline B_{q'}\to \DD' l \bar \nu)$
          & Mode
          & $A(\overline B_{q'}\to \DD'  l \bar \nu)$
          \\
\hline $B^- \to \Delta^{++} \overline{\Delta^{++}} l \bar \nu$
          & $6 T_\DD+A_\DD$
          & $B^- \to \Delta^+ \overline{\Delta^+} l \bar \nu$ 
          & $4 T_\DD$
          \\
$B^- \to \Delta^{0} \overline{\Delta^0} l \bar \nu$ 
          & $2 T_\DD+A_\DD$
          & $B^- \to \Delta^- \overline{\Delta^-} l \bar \nu$ 
          & $A_\DD$
          \\
$B^- \to\Sigma^{*+} \overline{\Sigma^{*+}} l \bar \nu$ 
          & $4 T_\DD+A_\DD$
          & $B^- \to \Sigma^{*0} \overline{\Sigma^{*0}} l \bar \nu$ 
          & $2 T_\DD+A_\DD$
          \\
$B^- \to\Sigma^{*-} \overline{\Sigma^{*-}} l \bar \nu$ 
          & $A_\DD$          
          & $B^- \to \Xi^{*0} \overline{\Xi^{*0}} l \bar \nu$ 
          & $2 T_\DD+A_\DD$
          \\  
$B^- \to \Xi^{*-} \overline{\Xi^{*-}}  l \bar \nu$ 
          & $A_\DD$          
          & $B^- \to \Omega^- \overline{\Omega^-} l \bar \nu$ 
          & $A_\DD$
          \\                
\hline $\overline B{}^0 \to \Delta^{++} \overline{\Delta^+} l \bar\nu$
          & $2\sqrt 3 T_\DD$
          & $\overline B{}^0 \to  \Delta^+ \overline{\Delta^0}l \bar\nu$ 
          & $4 T_\DD$
          \\
$\overline B{}^0 \to \Delta^0 \overline{\Delta^-} l \bar\nu$ 
          & $2\sqrt 3 T_\DD$
          & $\overline B{}^0 \to \Sigma^{*+} \overline{\Sigma^{*0}} l \bar\nu$
          & $2\sqrt 2 T_\DD$
          \\
$\overline B{}^0 \to \Sigma^{*0} \overline{\Sigma^{*-}} l \bar\nu$
          & $2\sqrt 2 T_\DD$
          & $\overline B{}^0 \to \Xi^{*0} \overline{\Xi^{*-}} l \bar\nu$
          & $2 T_\DD$
          \\         
\hline $\overline B{}^0_s \to \Delta^{++} \overline{\Sigma^{*+}} l \bar \nu$
          & $2\sqrt 3 T_\DD$
          & $\overline B{}^0_s \to \Delta^+ \overline{\Sigma^{*0}} l \bar \nu$ 
          & $2\sqrt 2 T_\DD$
          \\
$\overline B{}^0_s \to \Delta^0 \overline{\Sigma^{*-}} l \bar \nu$ 
          & $2 T_\DD$
          & $\overline B{}^0_s \to \Sigma^{*+} \overline{\Xi^{*0}} l \bar \nu$ 
          & $4 T_\DD$
          \\
$\overline B{}^0_s \to \Sigma^{*0} \overline{\Xi^{*-}} l \bar \nu$ 
          & $2\sqrt 2 T_\DD$
          & $\overline B{}^0_s \to \Xi^{*0} \overline{\Omega^{-}} l \bar \nu$ 
          & $2\sqrt 3 T_\DD$
          \\                             
\hline \hline 
Mode
          & $A(\overline B_{q'}\to \DD' \nu \bar \nu)$
          & Mode
          & $A(\overline B_{q'}\to \DD'  \nu \bar \nu)$
          \\
\hline
$B^- \to  \Sigma^{*+} \overline{\Delta^{++}} \nu \bar \nu$ 
          & $2\sqrt 3 PB_\DD$
          & $B^- \to \Sigma^{*0} \overline{\Delta^{+}} \nu \bar \nu$ 
          & $2\sqrt 2 PB_\DD$
          \\
$B^- \to \Sigma^{*-} \overline{\Delta^{0}} \nu \bar \nu$
          & $2 PB_\DD$
          & $B^- \to \Xi^{*0} \overline{\Sigma^{*+}} \nu \bar \nu$
          & $4 PB_\DD$
          \\
$B^- \to \Xi^{*-} \overline{\Sigma^{*0}} \nu \bar \nu$
          & $2\sqrt 2 PB_\DD$
          & $B^- \to \Omega^- \overline{\Xi^{*0}} \nu \bar \nu$ 
          & $2\sqrt 3 PB_\DD$
          \\          
\hline $\overline B{}^0 \to \Sigma^{*+} \overline{\Delta^+} \nu \bar\nu$
          & $2 PB_\DD$
          & $\overline B{}^0 \to \Sigma^{*0} \overline{\Delta^0} \nu \bar\nu$ 
          & $2\sqrt 2 PB_\DD$
          \\
$\overline B{}^0 \to\Sigma^{*-} \overline{\Delta^-} \nu \bar\nu$ 
          & $2\sqrt 3 PB_\DD$
          & $\overline B{}^0 \to \Xi^{*0} \overline{\Sigma^{*0}} \nu \bar\nu$ 
          & $2\sqrt 2 PB_\DD$
          \\
$\overline B{}^0 \to \Xi^{*-} \overline{\Sigma^{*-}} \nu \bar\nu$ 
          & $4 PB_\DD$
          & $\overline B{}^0 \to \Omega^- \overline{\Xi^{*-}} \nu \bar\nu$ 
          & $2\sqrt 3 PB_\DD$
          \\         
\hline 
$\overline B{}^0_s\to \Delta^{++} \overline{\Delta^{++}} \nu \bar \nu$
          & $PBA_\DD$
          & $\overline B{}^0_s \to \Delta^+ \overline{\Delta^+} \nu \bar \nu$ 
          & $PBA_\DD$
          \\
$\overline B{}^0_s \to \Delta^{0} \overline{\Delta^0} \nu \bar \nu$ 
          & $PBA_\DD$
          & $\overline B{}^0_s \to \Delta^- \overline{\Delta^-} \nu \bar \nu$ 
          & $PBA_\DD$
          \\
$\overline B{}^0_s \to \Sigma^{*+} \overline{\Sigma^{*+}} \nu \bar\nu$
          & $2 PB_\DD+PBA_\DD$
          & $\overline B{}^0_s \to \Sigma^{*0} \overline{\Sigma^{*0}} \nu \bar\nu$ 
          & $2 PB_\DD+PBA_\DD$
          \\
$\overline B{}^0_s \to \Sigma^{*-} \overline{\Sigma^{*-}} \nu \bar\nu$ 
          & $2 PB_\DD+PBA_\DD$
          & $\overline B{}^0_s \to \Xi^{*0} \overline{\Xi^{*0}} \nu \bar\nu$ 
          & $4 PB_\DD+PBA_\DD$
          \\
$\overline B{}^0_s \to  \Xi^{*-} \overline{\Xi^{*-}} \nu \bar\nu$ 
          & $4 PB_\DD+PBA_\DD$
          & $\overline B{}^0_s \to \Omega^{-} \overline{\Omega^{-}} \nu \bar\nu$ 
          & $6 PB_\DD+PBA_\DD$
          \\                             
\end{tabular}
\end{ruledtabular}
\end{table}

\subsection{Relations of decay amplitudes}\label{sec: relations}

As noted previously since the number of topological amplitudes are quite limited, relations of decay amplitudes are expected.
The following relations are obtained by using the decomposition of amplitudes shown in 
Tables \ref{tab: TPBB}, \ref{tab: TPBD}, \ref{tab: TPDB} and \ref{tab: TPDD}.
 
In $\overline B_q\to\BB' l\bar\nu$ decays, we have the following relations on amplitudes,
\be
A(\overline B{}^0 \to p \bar n l \bar \nu)
&=&A(\overline B{}^0_s \to \Sigma^+ \overline{\Xi^0} l \bar \nu)
=\sqrt2 A(\overline B{}^0_s \to \Sigma^0 \overline{\Xi^-} l \bar \nu),
\en
\be
A(B^-\to n \bar n l \bar \nu)
&=& A(B^-\to \Xi^0 \overline{\Xi^0} l \bar \nu),
\en
\be
A(\overline B{}^0 \to \Xi^0 \overline{\Xi^-} l \bar \nu)
&=&\sqrt 2 A(\overline B{}^0_s \to p \overline{\Sigma^0} l \bar \nu)
=A(\overline B{}^0_s \to n \overline{\Sigma^-} l \bar \nu),
\en
\be
A(B^-\to p \bar p l \bar \nu)
&=&A(B^-\to \Sigma^+ \overline{\Sigma^+} l \bar \nu),
\en
\be
A(\overline B{}^0 \to \Sigma^+ \overline{ \Sigma^0} l \bar \nu) 
&=& -A(\overline B{}^0 \to \Sigma^0 \overline{\Sigma^-} l \bar \nu)
=\sqrt 3 A(\overline B{}^0_s \to p \overline{\Lambda} l \bar \nu),
\en
\be
\sqrt 2 A(B^-\to \Sigma^0 \overline{\Lambda} l \bar \nu)
&=&\sqrt 2 A(B^-\to  \Lambda \overline{\Sigma^0} l \bar \nu)
=A(\overline B{}^0 \to \Sigma^+ \overline{\Lambda} l \bar \nu)
=A(\overline B{}^0 \to \Lambda \overline{\Sigma^-}  l \bar \nu),
\en
\be
A(B^-\to \Sigma^- \overline{\Sigma^-} l \bar \nu)
&=&A(B^- \to \Xi^- \overline{\Xi^-} l \bar \nu),
\en
and
\be
A(B^-\to p \bar p l \bar \nu)&=&A(\overline B{}^0 \to p \bar n l \bar \nu)+A(B^-\to n \bar n l \bar \nu)
\non\\
&=&2 A(B^-\to \Sigma^0 \overline{\Sigma^0} l \bar \nu)+A(B^-\to \Sigma^- \overline{\Sigma^-} l \bar \nu),
\non\\
2\sqrt3 A(B^-\to \Sigma^0 \overline{\Lambda} l \bar \nu)&=&A(\overline B{}^0 \to p \bar n l \bar \nu)+A(\overline B{}^0 \to \Xi^0 \overline{\Xi^-} l \bar \nu),
\non\\
\sqrt6 A(B^-\to  \Lambda \overline{\Lambda} l \bar \nu)&=&A(\overline B{}^0 \to \Sigma^+ \overline{\Lambda} l \bar \nu)+\sqrt 6A(B^-\to n \bar n l \bar \nu),
\non\\
-\sqrt{6}A(\overline B{}^0_s\to p \overline{\Lambda} l \bar \nu)
&=& 2 A(\overline B{}^0_s\to \Sigma^+ \overline{\Xi^0} l \bar \nu)
-A(\overline B{}^0_s\to n\overline{\Sigma^-}  l \bar \nu),
\non\\
\sqrt 6 A(\overline B{}^0_s \to \Lambda \overline{\Xi^-} l \bar \nu)&=&A(\overline B{}^0_s \to \Sigma^+ \overline{\Xi^0} l \bar \nu)-2A(\overline B{}^0_s \to n \overline{\Sigma^-} l \bar \nu).
\en

Similarly, for $\overline B_q\to\BB' \nu\bar\nu$ decays, we have
\be
A(B^-\to \Xi^0 \overline{\Sigma^+} \nu \bar \nu)
&=&\sqrt 2 A(B^-\to \Xi^- \overline{\Sigma^0} \nu \bar \nu)
=-\sqrt2 A(\overline B{}^0\to \Xi^0 \overline{\Sigma^0} \nu \bar \nu)
\non\\
&=&A(\overline B{}^0\to  \Xi^- \overline{\Sigma^-} \nu \bar \nu),
\en
\be
-\sqrt 2 A(B^-\to \Sigma^0 \bar p \nu \bar \nu)
&=&- A(B^-\to \Sigma^- \bar n \nu \bar \nu)
=- A(\overline B{}^0\to \Sigma^+ \bar p \nu \bar \nu)
\non\\
&=& \sqrt 2 A(\overline B{}^0\to \Sigma^0 \bar n \nu \bar \nu),
\en
\be
A(\overline B{}^0_s \to \Sigma^+ \overline{\Sigma^+} \nu \bar \nu)
&=&A(\overline B{}^0_s \to \Sigma^0 \overline{\Sigma^0} \nu \bar \nu)
=A(\overline B{}^0_s \to \Sigma^- \overline{\Sigma^-} \nu \bar \nu),
\\
A(\overline B{}^0_s \to \Xi^0 \overline{\Xi^0} \nu \bar \nu)
&=&A(\overline B{}^0_s \to \Xi^- \overline{\Xi^-} \nu \bar \nu),
\en
\be
A(B^-\to \Lambda \bar{p} \nu \bar \nu)
= A(\overline B{}^0\to  \Lambda \bar n \nu \bar \nu),
\en
\be
A(\overline B{}^0_s\to p \bar p \nu \bar \nu)=A(\overline B{}^0_s\to n \bar n \nu \bar \nu),
\en
and
\be
\sqrt 6 A(\overline B{}^0\to \Xi^0 \overline{\Lambda} \nu \bar \nu)
&=&A(\overline B{}^0\to  \Xi^- \overline{\Sigma^-} \nu \bar \nu)
-A(\overline B{}^0\to \Sigma^+ \bar p \nu \bar \nu),
\non\\
\sqrt 6 A(B^-\to \Xi^- \overline{\Lambda} \nu \bar \nu)
&=&
A(B^-\to \Xi^0 \overline{\Sigma^+} \nu \bar \nu)
-2 A(B^-\to \Sigma^- \bar n \nu \bar \nu),
\non\\
-\sqrt{6}A(B^-\to \Lambda \bar{p} \nu \bar \nu)
&=& 2 A(B^-\to \Xi^0 \overline{\Sigma^+} \nu \bar \nu)
-A(B^-\to \Sigma^- \bar n \nu \bar \nu),
\non\\
\sqrt{3} A(\overline B{}^0_s \to \Lambda \overline{\Lambda} \nu \bar \nu)
&=&-\sqrt 2 A(B^-\to \Lambda \bar{p} \nu \bar \nu)
+\sqrt3 A(\overline B{}^0_s\to p \bar p \nu \bar \nu).
\en

For $\overline B_q\to\BD l\bar \nu$ decays, there is only one topological amplitude, namely $T_\BD$. Therefore, all decay amplitudes are related,
\be
-A(B^-\to  p \overline{ \Delta^+} l \bar \nu)
&=&-A(B^-\to  n \overline{\Delta^0} l \bar \nu)
=
A(B^-\to  \Sigma^+ \overline{\Sigma^{*+}} l \bar \nu)
=2 A(B^-\to  \Sigma^0 \overline{\Sigma^{*0}} l \bar \nu)
\non\\
&=&A(B^-\to \Xi^0 \overline{\Xi^{*0}} l \bar \nu)
 =\frac{2}{\sqrt 3}A(B^-\to \Lambda \overline{\Sigma^{*0}} l \bar \nu)
=-A(\overline B{}^0 \to p \overline{\Delta^0} l \bar\nu)
\non\\
&=&-\frac{1}{\sqrt 3}A(\overline B{}^0 \to n \overline{\Delta^-} l \bar\nu)
=\sqrt 2 A(\overline B{}^0 \to \Sigma^+ \overline{\Sigma^{*0}} l \bar\nu)
\non\\
&=&-\sqrt 2A(\overline B{}^0 \to \Sigma^0 \overline{\Sigma^{*-}}  l \bar\nu)
=A(\overline B{}^0 \to \Xi^0 \overline{\Xi^{*-}} l \bar\nu)
=\sqrt{\frac{2}{3}}A(\overline B{}^0 \to \Lambda \overline{\Sigma^{*-}} l \bar\nu)
\non\\
&=&-\sqrt 2 A(\overline B{}^0_s \to p \overline{\Sigma^{*0}} l \bar\nu)
=-A(\overline B{}^0_s \to n \overline{\Sigma^{*-}} l \bar\nu)
=A(\overline B{}^0_s \to\Sigma^+ \overline{\Xi^{*0}} l \bar\nu)
\non\\
&=&-\sqrt 2A(\overline B{}^0_s \to \Sigma^0 \overline{\Xi^{*-}} l \bar\nu)
=\frac{1}{\sqrt3}A(\overline B{}^0_s \to \Xi^0 \overline{\Omega^-} l \bar\nu)
\non\\  
&=&\sqrt{\frac{2}{3}}A(\overline B{}^0_s \to \Lambda \overline{\Xi^{*-}} l \bar\nu).
\en

Similarly, for $\overline B_q\to\BD \nu\bar \nu$ decays, there is only one topological amplitude, namely $PB_\BD$. Hence, all decay amplitudes are related.
Explicitly, we have the following relations,
\be
-\frac{1}{\sqrt 6}A(B^-\to \Sigma^+ \overline{\Delta^{++}} \nu \bar\nu)
&=&\frac{1}{2}A(B^-\to \Sigma^0 \overline{\Delta^+} \nu \bar\nu)
=\frac{1}{\sqrt2}A(B^-\to \Sigma^- \overline{\Delta^0} \nu \bar\nu) 
\non\\
&=&-\frac{1}{\sqrt2}A(B^-\to \Xi^0 \overline{\Sigma^{*+}} \nu \bar\nu)
=A(B^-\to \Xi^- \overline{\Sigma^{*0}} \nu \bar\nu)
\non\\
&=&-\frac{1}{\sqrt2}A(\overline B{}^0\to \Sigma^+ \overline{\Delta^+} \nu \bar \nu)
=\frac{1}{2}A(\overline B{}^0\to \Sigma^0 \overline{\Delta^0} \nu \bar \nu)
\non\\
&=&\frac{1}{\sqrt 6}A(\overline B{}^0\to\Sigma^- \overline{\Delta^-}  \nu \bar \nu) 
=-A(\overline B{}^0\to \Xi^0 \overline{\Sigma^{*0}} \nu \bar \nu)
\non\\
&=&\frac{1}{\sqrt 2}A(\overline B{}^0\to \Xi^- \overline{\Sigma^{*-}}\nu \bar \nu)
=-\frac{1}{\sqrt 2}A(\overline B{}^0_s \to \Sigma^+ \overline{\Sigma^{*+}} \nu \bar\nu)
\non\\
&=&\frac{1}{\sqrt 2}A(\overline B{}^0_s \to \Sigma^0 \overline{\Sigma^{*0}} \nu \bar\nu) 
=\frac{1}{\sqrt 2}A(\overline B{}^0_s \to \Sigma^- \overline{\Sigma^{*-}} \nu \bar\nu)
\non\\
&=&-\frac{1}{\sqrt 2}A(\overline B{}^0_s \to \Xi^0 \overline{\Xi^{*0}} \nu \bar\nu)
=\frac{1}{\sqrt 2}A(\overline B{}^0_s \to \Xi^- \overline{\Xi^{*-}}  \nu \bar\nu). 
\en

For $\overline B_q\to\DB l\bar\nu$ decays, there is only one topological amplitude $(T_\DB)$, while for $\overline B_q\to\DB\nu\bar\nu$ decays, there is also only one topological amplitude ($PB_\DB$). Hence, the decay amplitudes are highly related and we have the following relations for $\overline B_q\to \DB l\bar\nu$ decays,
\be
-\frac{1}{\sqrt2}A(B^- \to \Delta^+ \bar p l \bar \nu)
&=&-\frac{1}{\sqrt2}A(B^- \to \Delta^0 \bar n l \bar \nu) 
=\frac{1}{\sqrt2}A(B^- \to\Sigma^{*+} \overline{\Sigma^{+}} l \bar \nu) 
\non\\           
&=&-\sqrt2A(B^- \to \Sigma^{*0} \overline{\Sigma^{0}} l \bar \nu)
=\frac{1}{\sqrt2}A(B^- \to \Xi^{*0} \overline{\Xi^{0}} l \bar \nu)
\non\\
&=&\sqrt{\frac{2}{3}}A(B^- \to \Sigma^{*0} \overline{\Lambda} l \bar \nu) 
=\frac{1}{\sqrt6}A(\overline B{}^0\to \Delta^{++} \bar p l \bar\nu)
\non\\
&=&\frac{1}{\sqrt2}A(\overline B{}^0\to \Delta^+ \bar n l \bar\nu) 
=-A(\overline B{}^0\to \Sigma^{*+} \overline{\Sigma^{0}} l \bar\nu)
\non\\           
&=&-A(\overline B{}^0\to \Sigma^{*0} \overline{\Sigma^{-}} l \bar\nu) 
=-\frac{1}{\sqrt2}A(\overline B{}^0\to \Xi^{*0} \overline{\Xi^{-}} l \bar\nu) 
\non\\
&=&\frac{1}{\sqrt3}A(\overline B{}^0\to \Sigma^{*+} \overline{\Lambda} l \bar\nu)
=-\frac{1}{\sqrt6}A(\overline B{}^0_s \to \Delta^{++} \overline{\Sigma^{+}} l \bar\nu)
\non\\           
&=&\frac{1}{2}A(\overline B{}^0_s \to \Delta^+ \overline{\Sigma^{0}} l \bar\nu)
=\frac{1}{\sqrt2}A(\overline B{}^0_s \to \Delta^0 \overline{\Sigma^{-}} l \bar\nu) 
\non\\
&=&-\frac{1}{\sqrt2}A(\overline B{}^0_s \to  \Sigma^{*+} \overline{\Xi^{0}} l \bar\nu)
=A(\overline B{}^0_s \to \Sigma^{*0} \overline{\Xi^{-}} l \bar\nu),
\en  
and
\be                        
-A(B^- \to \Sigma^{*0} \bar p \nu \bar\nu)           
&=&-\frac{1}{\sqrt2}A(B^- \to \Sigma^{*-} \bar n \nu \bar\nu)    
=\frac{1}{\sqrt2}A(B^- \to \Xi^{*0} \overline{\Sigma^+} \nu \bar\nu)           
\non\\
&=&-A(B^- \to \Xi^{*-} \overline{\Sigma^{0}} \nu \bar\nu)          
=\frac{1}{\sqrt6}A(B^- \to \Omega^- \overline{\Xi^{0}} \nu \bar\nu)           
\non\\
&=&\frac{1}{\sqrt3}A(B^- \to \Xi^{*-} \overline{\Lambda} \nu \bar\nu)          
=\frac{1}{\sqrt2}A(\overline B{}^0 \to \Sigma^{*+} \bar p \nu \bar\nu)           
\non\\
&=&A(\overline B{}^0 \to \Sigma^{*0} \bar n \nu \bar\nu)           
=-A(\overline B{}^0 \to  \Xi^{*0} \overline{\Sigma^{0}} \nu \bar\nu           
\non\\
&=&-\frac{1}{\sqrt2}A(\overline B{}^0 \to \Xi^{*-} \overline{\Sigma^{-}} \nu \bar\nu)           
=-\frac{1}{\sqrt6} A(\overline B{}^0 \to \Omega^- \overline{\Xi^{-}} \nu \bar\nu)           
\non\\
&=&-\frac{1}{\sqrt3} A(\overline B{}^0 \to \Xi^{*0} \overline{\Lambda} \nu \bar\nu)           
=-\frac{1}{\sqrt2} A(\overline B{}^0_s \to \Sigma^{*+} \overline{\Sigma^{+}} \nu \bar\nu)           
\non\\
&=&\frac{1}{\sqrt2} A(\overline B{}^0_s \to \Sigma^{*0} \overline{\Sigma^{0}} \nu \bar\nu)           
=\frac{1}{\sqrt2} A(\overline B{}^0_s \to \Sigma^{*-} \overline{\Sigma^{-}} \nu \bar\nu)           
\non\\
&=&-\frac{1}{\sqrt2} A(\overline B{}^0_s \to \Xi^{*0} \overline{\Xi^{0}} \nu \bar\nu)           
=\frac{1}{\sqrt2} A(\overline B{}^0_s \to \Xi^{*-} \overline{\Xi^{-}} \nu \bar\nu),           
\en                       
for $\overline B_q\to\DB\nu\bar\nu$ decays.

For $\overline B_q\to \DD' l\bar\nu$ decays, we have two topological amplitudes, namely $T_\DD$ and $A_\DD$.
The decay amplitudes are related as following,
\be                
\sqrt3 A(\overline B{}^0 \to \Delta^{++} \overline{\Delta^+} l \bar\nu) 
&=&\frac{1}{2}A(\overline B{}^0 \to  \Delta^+ \overline{\Delta^0}l \bar\nu) 
=\sqrt3 A(\overline B{}^0 \to \Delta^0 \overline{\Delta^-} l \bar\nu) 
\non\\
&=&\frac{1}{\sqrt2} A(\overline B{}^0 \to \Sigma^{*+} \overline{\Sigma^{*0}} l \bar\nu) 
=\frac{1}{\sqrt2} A(\overline B{}^0 \to \Sigma^{*0} \overline{\Sigma^{*-}} l \bar\nu) 
\non\\
&=&A(\overline B{}^0 \to \Xi^{*0} \overline{\Xi^{*-}} l \bar\nu)     
=\frac{1}{\sqrt3} A(\overline B{}^0_s \to \Delta^{++} \overline{\Sigma^{*+}} l \bar \nu) 
\non\\
&=&\frac{1}{\sqrt2} A(\overline B{}^0_s \to \Delta^+ \overline{\Sigma^{*0}} l \bar \nu) 
=A(\overline B{}^0_s \to \Delta^0 \overline{\Sigma^{*-}} l \bar \nu) 
\non\\
&=&\frac{1}{2}A(\overline B{}^0_s \to \Sigma^{*+} \overline{\Xi^{0}} l \bar \nu) 
=\frac{1}{\sqrt2}A(\overline B{}^0_s \to \Sigma^{*0} \overline{\Xi^{*-}} l \bar \nu) 
\non\\
&=&\frac{1}{\sqrt3} A(\overline B{}^0_s \to \Xi^{*0} \overline{\Omega^{-}} l \bar \nu) 
=\frac{1}{2} A(B^- \to \Delta^+ \overline{\Delta^+} l \bar \nu),        
\en                          
\be
A(B^- \to \Delta^{0} \overline{\Delta^0} l \bar \nu) 
=
A(B^- \to \Sigma^{*0} \overline{\Sigma^{*0}} l \bar \nu) 
=
A(B^- \to \Xi^{*0} \overline{\Xi^{*0}} l \bar \nu),  
\en  
\be
A(B^- \to \Delta^- \overline{\Delta^-} l \bar \nu) 
&=&
A(B^- \to\Sigma^{*-} \overline{\Sigma^{*-}} l \bar \nu)           
=A(B^- \to \Xi^{*-} \overline{\Xi^{*-}}  l \bar \nu)           
\non\\
&=&A(B^- \to \Omega^- \overline{\Omega^-} l \bar \nu), 
\en
and          
\be
A(B^- \to \Delta^{++} \overline{\Delta^{++}} l \bar \nu) 
&=&
A(B^- \to \Delta^{0} \overline{\Delta^0} l \bar \nu) 
+A(B^- \to \Delta^- \overline{\Delta^-} l \bar \nu), 
\non\\
A(B^- \to\Sigma^{*+} \overline{\Sigma^{*+}} l \bar \nu) 
&=&
A(B^- \to \Delta^+ \overline{\Delta^+} l \bar \nu)   
+A(B^- \to\Sigma^{*-} \overline{\Sigma^{*-}} l \bar \nu). 
\en

Finally, for $\overline B_q\to \DD' \nu\bar\nu$ decays, we have two topological amplitudes, namely $PB_\DD$ and $PBA_\DD$, giving 
the following relations on the amplitudes,
\be          
\frac{1}{\sqrt3}A(B^- \to  \Sigma^{*+} \overline{\Delta^{++}} \nu \bar \nu) 
&=&\frac{1}{\sqrt2}A(B^- \to \Sigma^{*0} \overline{\Delta^{+}} \nu \bar \nu)  
=A(B^- \to \Sigma^{*-} \overline{\Delta^{0}} \nu \bar \nu) 
\non\\
&=&\frac{1}{2}A(B^- \to \Xi^{*0} \overline{\Sigma^{*+}} \nu \bar \nu) 
=\frac{1}{\sqrt2}A(B^- \to \Xi^{*-} \overline{\Sigma^{*0}} \nu \bar \nu) 
\non\\
&=&\frac{1}{\sqrt3}A(B^- \to \Omega^- \overline{\Xi^{*0}} \nu \bar \nu)          
=A(\overline B{}^0 \to \Sigma^{*+} \overline{\Delta^+} \nu \bar\nu) 
\non\\
&=&\frac{1}{\sqrt2}A(\overline B{}^0 \to \Sigma^{*0} \overline{\Delta^0} \nu \bar\nu) 
=\frac{1}{\sqrt3}A(\overline B{}^0 \to\Sigma^{*-} \overline{\Delta^-} \nu \bar\nu) 
\non\\
&=&\frac{1}{\sqrt2}A(\overline B{}^0 \to \Xi^{*0} \overline{\Sigma^{*0}} \nu \bar\nu) 
=\frac{1}{2}A(\overline B{}^0 \to \Xi^{*-} \overline{\Sigma^{*-}} \nu \bar\nu) 
\non\\
&=&\frac{1}{\sqrt3}A(\overline B{}^0 \to \Omega^- \overline{\Xi^{*-}} \nu \bar\nu), 
\en
\be
A(\overline B{}^0_s \to \Sigma^{*+} \overline{\Sigma^{*+}} \nu \bar\nu) 
&=&A(\overline B{}^0_s \to \Sigma^{*0} \overline{\Sigma^{*0}} \nu \bar\nu) 
=A(\overline B{}^0_s \to \Sigma^{*-} \overline{\Sigma^{*-}} \nu \bar\nu),  
\non\\
A(\overline B{}^0_s \to \Xi^{*0} \overline{\Xi^{*0}} \nu \bar\nu) 
&=&A(\overline B{}^0_s \to  \Xi^{*-} \overline{\Xi^{*-}} \nu \bar\nu),  
\en
\be
A(\overline B{}^0_s \to \Sigma^{*+} \overline{\Sigma^{*+}} \nu \bar\nu) 
&=& A(B^- \to \Sigma^{*-} \overline{\Delta^{0}} \nu \bar \nu) 
+A(\overline B{}^0_s \to \Delta^{0} \overline{\Delta^0} \nu \bar \nu), 
\non\\
A(\overline B{}^0_s \to \Xi^{*0} \overline{\Xi^{*0}} \nu \bar\nu) 
&=&A(\overline B{}^0 \to \Xi^{*-} \overline{\Sigma^{*-}} \nu \bar\nu) 
+A(\overline B{}^0_s \to \Delta^- \overline{\Delta^-} \nu \bar \nu), 
\non\\
A(\overline B{}^0_s \to \Omega^{-} \overline{\Omega^{-}} \nu \bar\nu)  
&=& A(\overline B{}^0 \to \Xi^{*-} \overline{\Sigma^{*-}} \nu \bar\nu)          
+A(\overline B{}^0_s \to \Sigma^{*+} \overline{\Sigma^{*+}} \nu \bar\nu), 
\en
\be
A(\overline B{}^0_s\to \Delta^{++} \overline{\Delta^{++}} \nu \bar \nu) 
&=&A(\overline B{}^0_s \to \Delta^+ \overline{\Delta^+} \nu \bar \nu) 
=A(\overline B{}^0_s \to \Delta^{0} \overline{\Delta^0} \nu \bar \nu) 
\non\\
&=&A(\overline B{}^0_s \to \Delta^- \overline{\Delta^-} \nu \bar \nu). 
\en

The above relations on amplitudes impose relations on rates.
For example, we may have three decay modes, where their rates and amplitudes are related as following
\be
\Gamma_1=\sum_i |A_1(i)|^2,
\quad
\Gamma_2=\sum_i |A_2(i)|^2,
\quad
\Gamma_3=\sum_i |A_1(i)+A_2(i)|^2,
\en
with $i$ representing the allowed momentum and helicities of final state particles, 
summing over $i$ indicating integrating over phase space and summing over final state helicities.
Note that the following discussion only applies to the SU(3) symmetric case, i.e. we are considering the relation on rates in the SU(3) symmetric limit.
Using the triangle inequality in the complex plane, we obtain
\be
&&|A_1(i)|^2+|A_2(i)|^2-2 |A_1(i)||A_2(i)|
\non\\
&&\qquad
\leq |A_1(i)+A_2(i)|^2\leq |A_1(i)|^2+|A_2(i)|^2+2 |A_1(i)||A_2(i)|.
\en 
Sum over $i$ in the above equation and make use of the following inequality,
\be
0\leq \sum_i |A_1(i)||A_2(i)|
\leq \sqrt{\sum_i |A_1(i)|^2}\sqrt{\sum_j |A_2(j)|^2},
\en
we finally obtain the triangle inequality on rates in the SU(3) symmetric limit,
\be
(\Gamma_1^{1/2}-\Gamma_2^{1/2})^2\leq \Gamma_3\leq (\Gamma_1^{1/2}+\Gamma_2^{1/2})^2.
\label{eq: triangle}
\en

\section{Results on rates}

Before we start the discussion on rates it will be useful to recall the detectability of the final state baryons.
In Table \ref{tab: sensitivity}, we identify some octet and decuplet baryons that can decay to all charged final states with unsuppressed branching ratios.
Note that modes with anti-neutron are also detectable, while 
$\Delta^+$, $\Sigma^{+,0}$, $\Xi^0$, $\Sigma^{*0}$ and $\Xi^{*-}$ can be detected by detecting a $\pi^0$ or $\gamma$.
For example, $\Delta^+$ mainly decays to $p\pi^0$ and $n\pi^+$, while $\Sigma^0$ decays to $\Lambda\gamma$. 
We should pay close attention to the modes that involve these baryons and have large decay rates in the $\overline B_q$ decays.

\begin{table}[t!]
\caption{\label{tab: sensitivity}
Octet and decuplet baryons decaying to all charged final states with unsuppressed branching ratios \cite{PDG}.
}
\begin{ruledtabular}
\begin{tabular}{ccr}
Octet/Decuplet
          & Baryons
          & All charged final states
           \\
\hline 
Octet, $\B$
          &$p$, $\Lambda$, $\Xi^-$ 
          & $\Lambda\to p\pi^-$,  $\Xi^-\to \Lambda \pi^-\to p\pi^-\pi^-$  
          \\
Decuplet, $\D$
          &$\Delta^{++,0}$, $\Sigma^{*\pm}$, $\Xi^{*0}$, $\Omega^-$
          & $\Delta^{++,0}\to p\pi^\pm$,  $\Sigma^{*\pm}\to\Lambda\pi^\pm\to p\pi^-\pi^\pm$,          
          \\
          &
          & $\Xi^{*0}\to \Xi^-\pi^+\to \Lambda \pi^-\pi^+ \to p \pi^- \pi^- \pi^+$,
          \\
          &
          & $\Omega^-\to \Lambda K^-\to p\pi^-K^-$ 
          \\
\end{tabular}
\end{ruledtabular}
\end{table}

\subsection{$\overline B_q\to\BB' l\bar\nu$ and $\overline B_q\to\BB' \nu\bar\nu$ decay rates}\label{sec: BB rates}

In this part, we will first give a generic discussion on $\overline B_q\to\BB' l\bar\nu$ and $\overline B_q\to\BB' \nu\bar\nu$ decays,
and the results will be compared to model calculations, where masses of hadrons and lifetimes are taken from ref.~\cite{PDG}.

For  $\overline B_q\to\BB' l\bar\nu$ decays, the decay amplitudes can be decomposed in terms of three independent topological amplitudes, namely $T_{2\BB}$, $T_{1\BB}$ and $A_\BB$, as shown in Table~\ref{tab: TPBB}. 
As the amplitudes of $\overline B_q\to\BB' l\bar\nu$ decays have different combinations of these topological amplitudes, the corresponding branching ratios are denoted with different parameters.
Specifically, 
we use 
$a$ for the rate with $A=T_{1\BB}+A_\BB$, 
$b$ for the rate with $A=T_{2\BB}$, 
$c$ for the rate with $A=\frac{1}{2}(T_{1\BB}+T_{2\BB})+A_\BB$, 
$d$ for the rate with $A=(T_{1\BB}-T_{2\BB})/2$, 
$e$ for the rate with $A=A_\BB$, 
$f$ for the rate with $A=\frac{1}{6}(5T_{1\BB}+T_{2\BB})+A_\BB$,
$g$ for the rate with $A=(2T_{1\BB}+ T_{2\BB})/3$,
and
$h$ for the rate with $A=\frac{1}{3}(T_{1\BB}+2T_{2\BB})+A_\BB$.
In addition, we add tildes for rates with similar amplitudes but without the $A_\BB$ terms.
For example, $\tilde a$ corresponds to the rate with $A=T_{1\BB}$.
The same set of alphabets is also used in $\overline B_q\to\BB' \nu\bar\nu$ decays as $PB_{i\,\BB}$ and $PBA_\BB$ are proportional to $T_{i\,\BB}$ and $A_\BB$ with a common proportional constant $\zeta$ as shown in Eq.~(\ref{eq: zeta}). 
Note that the above parameters correspond to the rates in the SU(3) symmetric limit.

\begin{figure}[t]
\centering
  \includegraphics[width=0.5\textwidth]{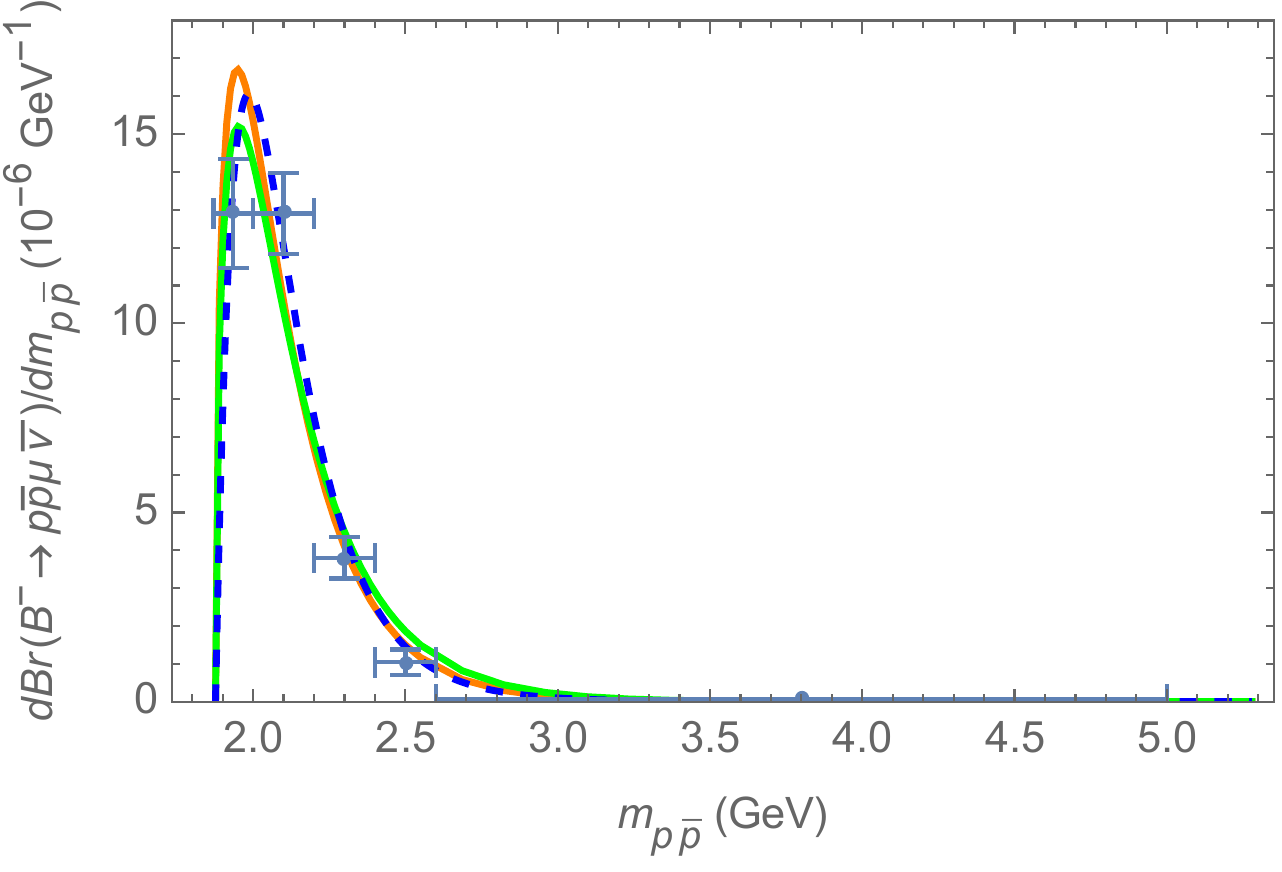}
\caption{
The experimental differential rate $dBr/dm_{p\bar p}$ of $B^-\to p\bar p \mu^-\bar\nu$ decay from LHCb~\cite{LHCb:2019cgl} can be well fitted with 
$dBr/dm_{p\bar p}=N (1/m^2_{p\bar p})^9 (m_{p\bar p}-m_p-m_{\bar p})$ with blue dashed line. 
Orange and green solid lines correspond to the differential rates from Model 1 and Model 2~with inputs basically from refs~\cite{Geng:2021sdl} and \cite{ Hsiao:2022uzx}, respectively.
See text for details.
}
 \label{fig: dBBdm}
\end{figure}

Experimentally not only data of the branching ratio of $B^-\to p\bar p l\bar\nu$ decay is obtained, 
information of differential rate is also available.
The experimental differential rate $dBr/dm_{p\bar p}$ of $B^-\to p\bar p \mu^-\bar\nu$ decay from LHCb~\cite{LHCb:2019cgl} is shown in 
Fig.~\ref{fig: dBBdm0}. 
The differential rate in Fig.~\ref{fig: dBBdm0} 
can be well fitted with
\be
\frac{dBr}{dm_{\bfBB'}}=\frac{N}{(m^2_{\bfBB'})^\gamma} (m_{\bfBB'}-m_{\bf B}-m_{\overline {\bf B}'}),
\label{eq: dBdm}
\en
where $\gamma$ and $N$ are constants. In particular, $\gamma=9$ is used in Fig.~\ref{fig: dBBdm0} for the plotted blue dashed line.
(see also Fig.~\ref{fig: dBBdm}).  
As noted in Introduction the threshold enhancement is sensitive to the position of the threshold and hence the SU(3) breaking from baryon masses are amplified producing very large SU(3) breaking effects on the integrated decay rates.

In this work we use Eq. (\ref{eq: dBdm}) to estimate the SU(3) breaking effect from threshold enhancement. 
Take $B^-\to p \bar p l \bar \nu$ and $B^-\to \Sigma^+ \overline{\Sigma^+} l \bar \nu$ decays as examples. 
As shown in Table~\ref{tab: TPBB} their amplitudes are both equal to $A=T_{1\BB}+T_{2\BB}+A_\BB$. 
Consequently, without SU(3) breaking, their rates should be identical.
However, we expect large SU(3) breaking from the threshold enhancement as the masses of $p$ and $\Sigma^+$ are different. Using Eq. (\ref{eq: dBdm}) the ratio of their branching ratios is given by 
\be
\frac{Br(B^-\to \Sigma^+ \overline{\Sigma^+} l \bar \nu)}{Br(B^-\to p \bar p l \bar \nu)}
=\frac
{\int^{m_B}_{2 m_{\Sigma^+}} dm_{\BB'} \frac{N'}{(m^2_{\BB'})^9} (m_{\BB'}-2 m_{\Sigma^+})}
{\int^{m_B}_{2m_p}dm_{\BB'}  \frac{N}{(m^2_{\BB'})^9} (m_{\BB'}-2m_p)}
=0.022 \frac{N'}{N} = 0.022\sigma,
\label{eq: sigma}
\en
where we define $N'/N\equiv\sigma$.
We see that the SU(3) breaking from the threshold enhancement is very large.
The decay rates differ by orders of magnitudes.
On the other hand, although $N'/N=\sigma$ may contain
additional SU(3) breaking from mass differences, it represents a milder SU(3) breaking effect, since the SU(3) breaking from threshold enhancement is already extracted out, the value of $\sigma$ is expected to be of order one. 
Consequently, using $Br(B^-\to p \bar p \mu \bar \nu)=(5.32\pm 0.34)\times 10^{-6}$~\cite{PDG}, we expect 
$Br(B^-\to \Sigma^+ \overline{\Sigma^+} l \bar \nu)=(5.32\times 0.022\sigma)\times 10^{-6}$ with $\sigma$ an order one parameter.
As we shall see later the above estimation agrees well with some recent theoretical calculations~\cite{Geng:2021sdl, Hsiao:2022uzx}.

With these considerations, the branching ratios of $\overline B_q\to\BB' l\bar\nu$ and $\overline B_q\to\BB' \nu\bar\nu$ decays are parametrized and are shown in Table \ref{tab: BrBB}.
SU(3) breaking effects from $B_q$ meson widths and threshold enhancement are included. 
The order one parameters 
$\alpha,\beta,\eta, \tilde \eta, \tilde{\tilde\eta}, \kappa, \tilde \kappa, \sigma, \tilde \sigma, \tilde{\tilde \sigma}, \xi, \tilde\xi, \tilde{\tilde \xi}$ 
and 
$\bar\alpha,\bar\beta,\bar\kappa, \bar{\tilde \kappa}$ denote milder SU(3) breaking,
where different parameters are used when the baryon masses are different,
tilde are used when the combinations of topological amplitudes are different, 
bar are use when the masses of baryon and anti-baryon are switched. 
From the above example, we expect these parameters to be of order one. We also expect them to be of similar size, and those with bar or tilde be close to those without bar or tilde.
We will come back to these later.

\begin{table}[t!]
\caption{\label{tab: BrBB} 
Branching ratios of $\overline B_{q}\to \BB' l \bar \nu$ and $\overline B_{q}\to \BB' \nu \bar \nu$ decays.
The $B^-\to p \bar p l \bar \nu$ rate is from experimental data \cite{Belle:2013uqr,PDG}.
Most of the parameters are expected to be of order 1. 
In particular, we expect $c\simeq \tilde c\simeq \sqrt{5.32}/2$, $a\simeq \tilde a$, $h\simeq \tilde h$ and $e\ll 5.32$,
satisfying Eqs.(\ref{eq: small e}), (\ref{eq: triangle inequality 1}) and (\ref{eq: triangle inequality 2}).
The last factors are from the SU(3) breaking from threshold enhancement,
and we expect $\alpha,\beta,\eta, \tilde \eta, \tilde{\tilde\eta}, \kappa, \tilde \kappa, \sigma, \tilde \sigma, \tilde{\tilde \sigma}, \xi, \tilde\xi, \tilde{\tilde \xi}$ 
and 
$\bar\alpha,\bar\beta,\bar\kappa, \bar{\tilde \kappa}$ being of order unity.
See text for details.
}
\begin{ruledtabular}
\begin{tabular}{cccccccc}
Mode
          & $Br(\overline B_{q}\to \BB' l \bar \nu) (10^{-6})$
          & Mode
          & $Br(\overline B_{q}\to \BB'  l \bar \nu)  (10^{-6})$
          \\
\hline $B^-\to p \bar p l \bar \nu$
          & $5.32\pm0.34$ \cite{PDG} 
          & $B^-\to n \bar n l \bar \nu$ 
          & $a\times (0.978)$ 
          \\
$B^-\to \Sigma^+ \overline{\Sigma^+} l \bar \nu$
          & $5.32\times (0.0225\sigma)$ 
          & $B^-\to \Sigma^0 \overline{\Sigma^0} l \bar \nu$
          & $c\times (0.0215 \sigma)$ 
          \\
$B^-\to  \Sigma^- \overline{\Sigma^-} l \bar \nu$ 
          & $e\times (0.0202\tilde{\tilde\sigma})$ 
          & $B^-\to  \Xi^-\overline{\Xi^-} l \bar \nu$ 
          & $e\times (0.00416\tilde{\tilde \xi})$ 
          \\           
$B^-\to \Sigma^0 \overline{\Lambda} l \bar \nu$ 
          & $\frac{d}{3}\times (0.0364\eta)$ 
          & $B^-\to \Xi^0 \overline{\Xi^0} l \bar \nu$ 
          & $a\times (0.00452\tilde\xi)$ 
          \\
$B^-\to  \Lambda \overline{\Sigma^0} l \bar \nu$ 
          & $\frac{d}{3}\times (0.0364\eta)$ 
          & $B^-\to  \Lambda \overline{\Lambda} l \bar \nu$ 
          & $f\times (0.0626\tilde \eta)$ 
          \\          
\hline $\overline B{}^0 \to p \bar n l \bar \nu$
          & $0.93 b\times (0.989)$ 
          & $\overline B{}^0 \to \Sigma^+ \overline{ \Sigma^0} l \bar \nu$ 
          & $1.85 \tilde c\times (0.0220\sigma)$ 
          \\
$\overline B{}^0 \to \Sigma^+ \overline{\Lambda} l \bar \nu$  
          & $0.62 d\times (0.0372\eta)$ 
          & $\overline B{}^0 \to \Sigma^0 \overline{\Sigma^-} l \bar \nu$  
          & $1.85 \tilde c\times (0.0208\sigma)$ 
          \\
$\overline B{}^0 \to \Lambda \overline{\Sigma^-}  l \bar \nu$  
          & $0.62 d\times (0.0352\eta)$ 
          & $\overline B{}^0 \to \Xi^0 \overline{\Xi^-} l \bar \nu$  
          & $0.9 \tilde a\times (0.00434 \tilde \xi)$ 
          \\         
\hline $\overline B{}^0_s \to p \overline{\Sigma^0} l \bar \nu$
          & $0.47 \tilde a\times (0.131 \beta)$ 
          & $\overline B{}^0_s \to p \overline{\Lambda} l \bar \nu$ 
          & $1.40 \tilde h\times (0.236\alpha)$  
          \\
$\overline B{}^0_s \to n \overline{\Sigma^-} l \bar \nu$ 
          & $0.93 \tilde a\times (0.125 \beta)$ 
          & $\overline B{}^0_s \to \Sigma^+ \overline{\Xi^0} l \bar \nu$ 
          & $0.93 b\times (0.00988\kappa)$ 
          \\
$\overline B{}^0_s \to \Sigma^0 \overline{\Xi^-} l \bar \nu$ 
          & $0.47b \times (0.00927\kappa)$ 
          & $\overline B{}^0_s \to \Lambda \overline{\Xi^-} l \bar \nu$ 
          & $1.40 g\times (0.0152\tilde \kappa)$ 
          \\                             
\hline \hline
Mode
          & $\sum_\nu Br(\overline B_{q}\to \BB' \nu \bar \nu) (10^{-8})$
          & Mode
          & $\sum_\nu Br(\overline B_{q}\to \BB'  \nu \bar \nu) (10^{-8})$
          \\
\hline $ B^-\to \Sigma^0 \bar p \nu \bar \nu$
          & $0.20 \tilde a\times (0.131\bar\beta)$ 
          & $ B^-\to \Sigma^- \bar n \nu \bar \nu$
          & $0.40 \tilde a\times (0.125\bar\beta)$ 
          \\
$ B^-\to \Xi^0 \overline{\Sigma^+} \nu \bar \nu$
          & $0.40 b\times (0.00988\bar\kappa)$ 
          & $ B^-\to \Xi^- \overline{\Sigma^0} \nu \bar \nu$ 
          & $0.20 b\times (0.00927\bar\kappa)$ 
          \\
$ B^-\to \Xi^- \overline{\Lambda} \nu \bar \nu$ 
          & $0.60 g\times (0.0152\bar{\tilde \kappa})$ 
          & $ B^-\to \Lambda \bar{p} \nu \bar \nu$ 
          & $0.60 \tilde h\times (0.236\bar\alpha)$ 
          \\          
\hline $\overline B{}^0\to \Sigma^+ \bar p \nu \bar \nu$
          & $0.37 \tilde a\times (0.134 \bar\beta)$ 
          & $\overline B{}^0\to \Sigma^0 \bar n \nu \bar \nu$
          & $0.19 \tilde a\times (0.130 \bar\beta)$ 
          \\
$\overline B{}^0\to \Xi^0 \overline{\Sigma^0} \nu \bar \nu$ 
          & $0.19 b\times (0.00968\bar\kappa)$ 
          & $\overline B{}^0\to \Xi^0 \overline{\Lambda} \nu \bar \nu$
          & $0.56 g\times (0.0159\bar{\tilde \kappa})$ 
          \\
$\overline B{}^0\to  \Xi^- \overline{\Sigma^-} \nu \bar \nu$
          & $0.37 b\times (0.00899\bar\kappa)$ 
          & $\overline B{}^0\to  \Lambda \bar n \nu \bar \nu$ 
          & $0.56 \tilde h\times (0.233\bar\alpha)$ 
          \\         
\hline $\overline B{}^0_s\to p \bar p \nu \bar \nu$
          & $0.37 e$ 
          & $\overline B{}^0_s\to n \bar n \nu \bar \nu$ 
          & $0.37 e\times (0.978)$ 
          \\
$\overline B{}^0_s \to \Sigma^+ \overline{\Sigma^+} \nu \bar \nu$
          & $0.37 a\times (0.0225\tilde \sigma)$ 
          & $\overline B{}^0_s \to \Sigma^0 \overline{\Sigma^0} \nu \bar \nu$
          & $0.37 a\times (0.0215\tilde \sigma)$ 
          \\
$\overline B{}^0_s \to \Sigma^- \overline{\Sigma^-} \nu \bar \nu$ 
          & $0.37 a\times (0.0202\tilde \sigma)$ 
          & $\overline B{}^0_s \to \Xi^0 \overline{\Xi^0} \nu \bar \nu$ 
          & $1.98\times (0.00452\xi)$ 
          \\
$\overline B{}^0_s \to \Xi^- \overline{\Xi^-} \nu \bar \nu$ 
          & $1.98\times (0.00416\xi)$ 
          & $\overline B{}^0_s \to \Lambda \overline{\Lambda} \nu \bar \nu$
          & $0.37 h\times (0.0626\tilde{\tilde \eta})$ 
          \\                             
\end{tabular}
\end{ruledtabular}
\end{table}

There are many parameters in Table \ref{tab: BrBB}. 
They are not totally independent, since we only have three independent topological amplitudes.
%
Using the triangle inequality, Eq. (\ref{eq: triangle}), the amplitude decomposition in $\overline B_q\to {\bf B}\overline {\bf B}' l\bar \nu$ decays and the decay rates as shown in Tables~\ref{tab: TPBB} and \ref{tab: BrBB}, 
we obtain the following inequalities, 
\be
\Big(\frac{\sqrt{5.32}-\sqrt e}{2}\Big)^2\lesssim &c&\lesssim \Big(\frac{\sqrt{5.32}+\sqrt e}{2}\Big)^2, 
\quad
\Big(\frac{\sqrt{5.32}-\sqrt e}{2}\Big)^2\lesssim \tilde c \lesssim\Big(\frac{\sqrt{5.32}+\sqrt e}{2}\Big)^2, 
\non\\
(\sqrt{a}-\sqrt e)^2\lesssim &\tilde a&\lesssim (\sqrt{a}+\sqrt e)^2, 
\quad
(\sqrt{h}-\sqrt e)^2\lesssim \tilde h \lesssim (\sqrt{h}+\sqrt e)^2, 
\en
\be
(\sqrt{5.32}-\sqrt a)^2\lesssim &b&\lesssim (\sqrt{5.32}+\sqrt a)^2, 
\quad
(\sqrt{\tilde c}-\sqrt {\tilde a})^2\lesssim d\lesssim (\sqrt{\tilde c}+\sqrt {\tilde a})^2, 
\non\\
\Big(\frac{\sqrt c-2\sqrt a}{3}\Big)^2\lesssim &f&\lesssim \Big(\frac{\sqrt c+2\sqrt a}{3}\Big)^2, 
\quad
\Big(\frac{2\sqrt{\tilde c}-\sqrt {\tilde a}}{3}\Big)^2\lesssim g\lesssim \Big(\frac{2\sqrt{\tilde c}+\sqrt {\tilde a}}{3}\Big)^2,
\non\\
\Big(\frac{4\sqrt c-\sqrt a}{3}\Big)^2\lesssim& h &\lesssim \Big(\frac{4\sqrt c+\sqrt a}{3}\Big)^2,
\quad
\Big(\frac{4\sqrt{\tilde c}-\sqrt {\tilde a}}{3}\Big)^2\lesssim \tilde h \lesssim \Big(\frac{4\sqrt{\tilde c}+\sqrt {\tilde a}}{3}\Big)^2,
\en
and
\be
(\sqrt{5.32}-\sqrt b)^2\lesssim &a& \lesssim (\sqrt{5.32}+\sqrt b)^2,
\quad
(\sqrt{\tilde c}-\sqrt b)^2\lesssim d\lesssim (\sqrt{\tilde c}+\sqrt b)^2,
\non\\
\Big(\frac{4\sqrt{\tilde c}-\sqrt b}{3}\Big)^2\lesssim &g&\lesssim \Big(\frac{4\sqrt{\tilde c}+\sqrt b}{3}\Big)^2,
\quad
\Big(\frac{2\sqrt{\tilde c}-\sqrt b}{3}\Big)^2\lesssim \tilde h \lesssim \Big(\frac{2\sqrt{\tilde c}+\sqrt b}{3}\Big)^2.
\en

Although the above inequalities can constrain the sizes of these parameters,
it will be useful to reduce the number of the parameters.
Note that the rates proportional to $e$ are governed by annihilation $A_\BB$ or penguin-box-annihilation $PBA_\BB$ diagrams. It is known that these contributions are usually much suppressed than tree and penguin contributions. 
For example, in two-body baryonic $B_q$ decays, $\overline B_q\to\BB'$ decays, the tree dominated mode $B^-\to p\bar p$ and penguin dominated mode $B^-\to \Lambda\bar p$ was observed with branching ratios at $10^{-8}$ and $10^{-6}$ levels, respectively \cite{LHCb:2022lff,LHCb:2017swz,LHCb:2016nbc}, while $\overline B_s\to p\bar p$ decay, which is an exchange and penguin-annihilation mode is not yet observed with the upper limit pushed down to $10^{-9}$ level \cite{LHCb:2022lff}.
It is therefore reasonable to consider the case where the annihilation $A_\BB$ and penguin-box-annihilation $PBA_\BB$ contributions are highly suppressed, i.e. $e\ll {\cal O}(1)$.
Nevertheless this assumption can be checked by searching pure annihilation (penguin-box-annuhilation) modes, $B^-\to  \Sigma^- \overline{\Sigma^-} l \bar \nu$, $B^-\to  \Xi^-\overline{\Xi^-} l \bar \nu$,  
$\overline B{}^0_s\to p \bar p \nu \bar \nu$ and $\overline B{}^0_s\to n \bar n \nu \bar \nu$ decays, as their rates are proportional to $e$. 
In particular, as the $\overline B{}^0_s\to p \bar p \nu \bar \nu$ mode has good detectability, see Table~\ref{tab: sensitivity}, it is a good place to verify the above assumption.

Applying the above assumption to the relations Eq. (\ref{eq: triangle inequality 1}), we obtain,
\be
e\ll 5.32,
\quad
c\simeq \tilde c\simeq \frac{5.32}{4},
\quad
\tilde a\simeq a,
\quad
\tilde h\simeq h,
\label{eq: small e}
\en
\be
(\sqrt{5.32}-\sqrt a)^2\lesssim &b&\lesssim (\sqrt{5.32}+\sqrt a)^2, %
\quad
(0.5\sqrt{5.32}-\sqrt {a})^2\lesssim d\lesssim (0.5\sqrt{5.32}+\sqrt { a})^2, %
\non\\
\Big(\frac{\sqrt{5.32}-4\sqrt a}{6}\Big)^2\lesssim &f&\lesssim \Big(\frac{\sqrt{5.32}+4\sqrt a}{6}\Big)^2, %
\quad
\Big(\frac{\sqrt{ 5.32}-\sqrt {a}}{3}\Big)^2\lesssim g\lesssim \Big(\frac{\sqrt{5.32}+\sqrt {a}}{3}\Big)^2,%
\non\\
\Big(\frac{2\sqrt{5.32}-\sqrt a}{3}\Big)^2\lesssim& h, \tilde h &\lesssim \Big(\frac{2\sqrt{5.32}+\sqrt a}{3}\Big)^2,
\label{eq: triangle inequality 1}
\en
and
\be
(\sqrt{5.32}-\sqrt b)^2\lesssim &a& \lesssim (\sqrt{5.32}+\sqrt b)^2,%
\quad
(0.5\sqrt{5.32}-\sqrt b)^2\lesssim d\lesssim (0.5\sqrt{5.32}+\sqrt b)^2,%
\non\\
\Big(\frac{5\sqrt{5.32}-4\sqrt b}{6}\Big)^2\lesssim &f& \lesssim \Big(\frac{5\sqrt{5.32}+4\sqrt b}{6}\Big)^2,
\,\,
\Big(\frac{2\sqrt{5.32}-\sqrt b}{3}\Big)^2\lesssim g\lesssim \Big(\frac{2\sqrt{5.32}+\sqrt b}{3}\Big)^2,
\non\\
\Big(\frac{\sqrt{5.32}-\sqrt b}{3}\Big)^2\lesssim &h, \tilde h& \lesssim \Big(\frac{\sqrt{5.32}+\sqrt b}{3}\Big)^2.
\label{eq: triangle inequality 2}
\en
These are the inequalities we shall employed in this work.

The parameters $a$, $b$, $c$ and so on in Table~\ref{tab: BrBB} need to satisfy the above triangular inequalities, 
Eqs.(\ref{eq: small e}), (\ref{eq: triangle inequality 1}) and (\ref{eq: triangle inequality 2}).
At this moment we do not have enough data to verify them.
Nevertheless, it will be useful to make use of model calculations in Sec.~\ref{sec: model calculations} for illustration.

\begin{table}[t!]
\caption{\label{tab: BrBB1} 
Branching ratios of $\overline B_{q}\to \BB' l \bar \nu$ and $\overline B_{q}\to \BB' \nu \bar \nu$ decays in Model 1 and Model 2.
$Br_1$ and $Br_2$ denote results in Model 1 and Model 2, respectively.
}
\begin{ruledtabular}
\begin{tabular}{cccccccc}
Mode
          & $Br_1 (10^{-6})$
          & $Br_2 (10^{-6})$
          & Mode
          & $Br_1 (10^{-6})$
          & $Br_2 (10^{-6})$
          \\
\hline $B^-\to p \bar p l \bar \nu$
          & $5.32$
          & $5.32$
          & $B^-\to n \bar n l \bar \nu$ 
          & $0.41$
          & $7.81$
          \\
$B^-\to \Sigma^+ \overline{\Sigma^+} l \bar \nu$
          & 0.26 
          & 0.26
          & $B^-\to \Sigma^0 \overline{\Sigma^0} l \bar \nu$
          & 0.064 
          & 0.061
          \\
$B^-\to  \Sigma^- \overline{\Sigma^-} l \bar \nu$ 
          & $\simeq 0$ 
          & $\simeq 0$ 
          & $B^-\to  \Xi^-\overline{\Xi^-} l \bar \nu$ 
          & $\simeq 0$ 
          & $\simeq 0$
          \\           
$B^-\to \Sigma^0 \overline{\Lambda} l \bar \nu$ 
          & 0.019 
          & 0.18
          & $B^-\to \Xi^0 \overline{\Xi^0} l \bar \nu$ 
          & 0.0044 
          & 0.094
          \\
$B^-\to  \Lambda \overline{\Sigma^0} l \bar \nu$ 
          & 0.019 
          & 0.18
          & $B^-\to  \Lambda \overline{\Lambda} l \bar \nu$ 
          & 0.062 
          & 0.47
          \\          
\hline $\overline B{}^0 \to p \bar n l \bar \nu$
          & 3.40 
          & 9.40
          & $\overline B{}^0 \to \Sigma^+ \overline{ \Sigma^0} l \bar \nu$ 
          & 0.12 
          & 0.12
          \\
$\overline B{}^0 \to \Sigma^+ \overline{\Lambda} l \bar \nu$  
          & 0.036 
          & 0.34
          & $\overline B{}^0 \to \Sigma^0 \overline{\Sigma^-} l \bar \nu$  
          & 0.12 
          & 0.11 
          \\
$\overline B{}^0 \to \Lambda \overline{\Sigma^-}  l \bar \nu$  
          & 0.034 
          & 0.33
          & $\overline B{}^0 \to \Xi^0 \overline{\Xi^-} l \bar \nu$  
          & 0.0040 
          & 0.084
          \\         
\hline $\overline B{}^0_s \to p \overline{\Sigma^0} l \bar \nu$
          & 0.044 
          & 0.87 
          & $\overline B{}^0_s \to p \overline{\Lambda} l \bar \nu$ 
          & 1.01 
          & 1.28 
          \\
$\overline B{}^0_s \to n \overline{\Sigma^-} l \bar \nu$ 
          & 0.086 
          & 1.68
          & $\overline B{}^0_s \to \Sigma^+ \overline{\Xi^0} l \bar \nu$ 
          & 0.10 
          & 0.24
          \\
$\overline B{}^0_s \to \Sigma^0 \overline{\Xi^-} l \bar \nu$ 
          & 0.048 
          & 0.11
          & $\overline B{}^0_s \to \Lambda \overline{\Xi^-} l \bar \nu$ 
          & 0.048 
          & 0.10
          \\                             
\hline \hline
Mode
          & $\sum_\nu Br_1 (10^{-8})$
          & $\sum_\nu Br_2 (10^{-8})$
          & Mode
          & $\sum_\nu Br_1 (10^{-8})$
          & $\sum_\nu Br_2 (10^{-8})$
          \\
\hline $ B^-\to \Sigma^0 \bar p \nu \bar \nu$
          & 0.017 
          & 0.32
          & $ B^-\to \Sigma^- \bar n \nu \bar \nu$
          & 0.031
          & 0.62
          \\
$ B^-\to \Xi^0 \overline{\Sigma^+} \nu \bar \nu$
          & 0.038 
          & 0.089
          & $ B^-\to \Xi^- \overline{\Sigma^0} \nu \bar \nu$ 
          & 0.018 
          & 0.042
          \\
$ B^-\to \Xi^- \overline{\Lambda} \nu \bar \nu$ 
          & 0.018 
          & 0.041
          & $ B^-\to \Lambda \bar{p} \nu \bar \nu$ 
          & 0.39 
          & 0.52
          \\          
\hline $\overline B{}^0\to \Sigma^+ \bar p \nu \bar \nu$
          & 0.031 
          & 0.61
          & $\overline B{}^0\to \Sigma^0 \bar n \nu \bar \nu$
          & 0.015 
          & 0.30
          \\
$\overline B{}^0\to \Xi^0 \overline{\Sigma^0} \nu \bar \nu$ 
          & 0.017 
          & 0.041
          & $\overline B{}^0\to \Xi^0 \overline{\Lambda} \nu \bar \nu$
          & 0.017
          & 0.039
          \\
$\overline B{}^0\to  \Xi^- \overline{\Sigma^-} \nu \bar \nu$
          & 0.032
          & 0.076
          & $\overline B{}^0\to  \Lambda \bar n \nu \bar \nu$
          & 0.36 
          & 0.48
          \\         
\hline $\overline B{}^0_s\to p \bar p \nu \bar \nu$
          & $\simeq 0$ 
          & $\simeq 0$
          & $\overline B{}^0_s\to n \bar n \nu \bar \nu$ 
          & $\simeq 0$ 
          & $\simeq 0$
          \\
$\overline B{}^0_s \to \Sigma^+ \overline{\Sigma^+} \nu \bar \nu$
          & 0.0081 
          & 0.17
          & $\overline B{}^0_s \to \Sigma^0 \overline{\Sigma^0} \nu \bar \nu$
          & 0.0078 
          & 0.16
          \\
$\overline B{}^0_s \to \Sigma^- \overline{\Sigma^-} \nu \bar \nu$ 
          & 0.0074 
          & 0.15
          & $\overline B{}^0_s \to \Xi^0 \overline{\Xi^0} \nu \bar \nu$ 
          & 0.028 
          & 0.028
          \\
$\overline B{}^0_s \to \Xi^- \overline{\Xi^-} \nu \bar \nu$ 
          & 0.026 
          & 0.026
          & $\overline B{}^0_s \to \Lambda \overline{\Lambda} \nu \bar \nu$
          & 0.097 
          & 0.12
          \\                             
\end{tabular}
\end{ruledtabular}
\end{table}

\begin{table}[t!]
\caption{\label{tab: alphabeta} 
Values of various parameters in Model 1 and Model 2. The bounds of the parameters $d\eta, \dots h\tilde{\tilde \eta}$ are obtained using triangular inequalities, Eqs. (\ref{eq: triangle inequality 1}) and (\ref{eq: triangle inequality 2}), while the bounds of the parameters $\alpha,\bar\alpha,\eta, \tilde \eta, \tilde{\tilde\eta}, \tilde \kappa,\bar{\tilde\kappa}$ are obtained using the values and bounds of the parameters $d\eta, \dots h\tilde{\tilde \eta}$.
}
\begin{ruledtabular}
\begin{tabular}{cccccccc}
Parameters
          & Values (Model 1)
          & Bounds (Model 1)
          & Values (Model 2)
          & Bounds (Model 2)
          \\
\hline 
$a$
          & 0.42
          & $0.15\sim 17.90$
          & 7.99
          & $0.80\sim 30.34$
          \\
$b$
          & 3.70
          & $2.75\sim 8.74$
          & 10.25
          & $0.27\sim 26.34$
          \\  
$d\eta$
          & $1.56$
          & $(0.59\sim 3.25)\eta$
          & $14.89$
          & $(4.20\sim 15.83)\eta$
          \\   
$f\tilde \eta$
          & $0.99$
          & $(0.41\sim 0.67)\tilde \eta$
          & $7.44$
          & $(2.25\sim 5.14)\tilde \eta'$
          \\
$g\tilde \kappa$
          & $2.24$
          & $(0.80\sim 0.97)\tilde \kappa$
          & $4.81$
          & $(0.22\sim 2.93)\tilde \kappa$
          \\                 
$g\bar{\tilde \kappa}$
          & $1.96$
          & $(0.80\sim 0.97)\bar{\tilde \kappa}$
          & $4.45$
          & $(0.22\sim 2.93)\bar{\tilde \kappa}$
          \\   
$\tilde h\alpha$
          & $3.08$
          & $(1.75\sim 1.99)\alpha$
          & $3.88$
          & $(0.35\sim 3.37)\alpha$
          \\  
$\tilde h\bar\alpha$
          & $2.77$
          & $(1.75\sim 1.99)\bar\alpha$
          & $3.67$
          & $(0.35\sim 3.37)\bar\alpha$
          \\  
$h\tilde{\tilde \eta}$
          & $4.17$
          & $(1.75\sim 1.99)\tilde{\tilde \eta}$
          & $5.26$
          & $(0.35\sim 3.37)\tilde{\tilde \eta}$
          \\                      
\hline  
$\alpha$
          & $-$
          & $1.55\sim 1.76$
          & $-$
          & $1.15\sim 10.93$
          \\  
$\bar\alpha$
          & $-$
          & $1.39\sim 1.59$
          & $-$
          & $1.09\sim 10.34$
          \\ 
$\beta$
          & $1.72$
          & $-$
          & $1.79$
          & $-$
          \\
$\bar\beta$
          & 1.47
          & $-$
          & $1.54$
          & $-$
          \\
$\eta$
          & $-$
          & $0.48\sim 2.62$
          & $-$
          & $0.94\sim 3.55$
          \\                  
$\tilde \eta$
          & $-$
          & $1.49\sim 2.43$
          & $-$
          & $1.45\sim 3.31$
          \\   
$\tilde{\tilde \eta}$
          & $-$
          & $2.10\sim 2.39$
          & $-$
          & $1.56\sim 14.83$
          \\               
$\sigma$
          & $2.21$
          & $-$
          & $2.13$
          & $-$
          \\
$\tilde \sigma$
          & $2.30$
          & $-$
          & $2.49$
          & $-$
          \\
$\kappa$
          & $2.96$
          & $-$
          & $2.55$
          & $-$
          \\
$\bar\kappa$
          & $2.59$
          & $-$
          & $2.21$
          & $-$
          \\  
$\tilde \kappa$
          & $-$
          & $2.31\sim 2.79$
          & $-$
          & $1.64\sim 21.73$
          \\                 
$\bar{\tilde \kappa}$
          & $-$
          & $2.02\sim 2.45$
          & $-$
          & $1.52\sim 20.11$
          \\                           
$\xi$
          & 3.19
          & $-$
          & $3.13$
          & $-$
          \\
$\tilde \xi$
          & $2.32$
          & $-$
          & $2.61$
          & $-$
          \\   
\hline
$(\frac{\bar\alpha}{\alpha} ,\frac{\bar\beta}{\beta} ,\frac{\bar{\kappa}}{\kappa},\frac{\bar{\tilde\kappa}}{\tilde\kappa})$ 
          & $(0.90,0.85, 0.87, 0.88)$ 
          & $-$
          & $(0.95, 0.86, 0.87, 0.93)$ 
          & $-$
          \\                                              
\end{tabular}
\end{ruledtabular}
\end{table}

Branching ratios of $\overline B_q\to \BB' l\bar\nu$ and $\overline B_q\to \BB' \nu\bar\nu$ decays in Model 1 and Model 2 can be obtained by using $T_{i\BB}$ and $A_\BB$, as shown in Eq. (\ref{eq: Ti}), 
with inputs as shown in Table~\ref{tab: GF}, 
and formulas of decay rates collected in Appendix \ref{App: formulas}. The results are shown in Table \ref{tab: BrBB1}.
They can be compared to the results given in refs. \cite{Geng:2021sdl,Hsiao:2022uzx},
where
$Br(B^-\to p\bar p l\bar \nu)=(5.21\pm 0.34)\times 10^{-6}$ \cite{Geng:2021sdl}, $(5.3\pm 0.2)\times 10^{-6}$ \cite{Hsiao:2022uzx},  
$Br(B^-\to n \bar n l \bar \nu)=(0.68\pm 0.10)\times 10^{-6}$,
$Br(B^-\to \Sigma^+ \overline{\Sigma^+} l \bar \nu)=(0.24\pm0.02)\times 10^{-6}$,
$Br(B^-\to \Sigma^0 \overline{\Sigma^0} l \bar \nu)=(0.06\pm0.01)\times 10^{-6}$,
$Br(B^-\to \Sigma^0 \overline{\Lambda} l \bar \nu)=(0.014\pm 0.004)\times 10^{-6}$,
$Br(B^-\to \Lambda \overline{\Sigma^0} l \bar \nu)=(0.014\pm 0.004)\times 10^{-6}$,
$Br(B^-\to \Xi^0 \overline{\Xi^0} l \bar \nu)=(0.008\pm 0.001)\times 10^{-6}$,
$Br(B^-\to  \Lambda \overline{\Lambda} l \bar \nu)=(0.08\pm 0.01)\times 10^{-6}$  \cite{Geng:2021sdl},
$Br(\overline B{}^0_s \to p \overline{\Lambda} l \bar \nu)=(2.1\pm 0.6)\times 10^{-6}$,
$\sum_\nu Br(B^- \to \Lambda\bar p \nu \bar \nu)=(3.5\pm 1.0)\times 10^{-8}$
and
$\sum_\nu Br(\overline B_s \to \Lambda\bar \Lambda \nu \bar \nu)=(0.8\pm 0.2)\times 10^{-8}$
\cite{Hsiao:2022uzx}
are reported.
We find that results in Model~1 agree with or close to those in ref. \cite{Geng:2021sdl}, while the results on $\sum_\nu Br(B^- \to \Lambda\bar p \nu \bar \nu)$ and 
$\sum_\nu Br(\overline B_s \to \Lambda\bar \Lambda \nu \bar \nu)$ in Model 2 differs to theose in ref. \cite{Hsiao:2022uzx} by factors of 7.
Results on all other modes in Table \ref{tab: BrBB1} are new.

Model 1 and Model 2 have similar results on some modes, but very different results on some other modes.
For example, their rates in the $B^-\to p\bar pl \bar \nu$ decay are identical by construction, $B^-\to \Sigma^+ \overline{\Sigma^+} l \bar \nu$ rates as well as $B^-\to \Sigma^0 \overline{\Sigma^0} l \bar \nu$ rates are similar,
but the $B^-\to \Sigma^0 \overline{\Lambda} l \bar \nu$ rate in Model 2 is larger than the one in Model 1 by one order of magnitude, 
the $\overline B{}^0 \to p \bar n l \bar \nu$ rate in Model 2 is larger than the one in Model 1 by a factor of 3,
and the $B^- \to n \bar n l \bar \nu$ rate in Model 2 is larger than the one in Model 1 by a factor of 19.
Note that the amplitudes of $B^-\to p\bar pl \bar \nu$, $B^-\to \Sigma^+ \overline{\Sigma^+} l \bar \nu$ and $B^-\to \Sigma^0 \overline{\Sigma^0} l \bar \nu$
are to $T_{1\BB}+T_{2\BB}$,
$B^-\to \Sigma^0 \overline{\Lambda} l \bar \nu$, $B^- \to n \bar n l \bar \nu$ and $\overline B{}^0 \to p \bar n l \bar \nu$ are 
proportional to $T_{1\BB}-T_{2\BB}$, $T_{1\BB}$ and $T_{2\BB}$, respectively.
These results imply that Model 1 (2) has constructive (destructive) interference of $T_{1\BB}$ and $T_{2\BB}$ in $B^-\to p\bar pl \bar \nu$ decay, 
but destructive (constructive) interference of $T_{1\BB}$ and $-T_{2\BB}$ in $B^-\to \Sigma^0 \overline{\Lambda} l \bar \nu$ decay,
and $|T_{1,2\,\BB}|$ in Model 2 are larger than those in Model 1.
These two models are complementary.
It is therefore useful to consider both of them.

In Fig.~\ref{fig: dBBdm} the differential rates $dBr/dm_{p\bar p}$ of $B^-\to p\bar p \mu^-\bar\nu$ from Model 1 and Model 2 are shown and are compared to data. 
The differential rates from Model 1 and 2 agree with data and are similar to each other.

The expectations of the orders of magnitudes of $\alpha$, $\beta$ and so on to be of order one and the triangular inequalities, 
Eqs.(\ref{eq: small e}), (\ref{eq: triangle inequality 1}) and (\ref{eq: triangle inequality 2}), on $a$, $b$ and so on 
can be checked by comparing Table \ref{tab: BrBB} with the results in Model 1 and Model 2 as shown in Table \ref{tab: BrBB1}.
The findings are shown in Table~\ref{tab: alphabeta}.
The values of the parameters $\beta, \bar\beta, \sigma, \tilde \sigma, \kappa, \bar\kappa, \xi, \tilde \xi$ are indeed of order one and are in the range of $1.47\sim 3.19$, and their values in Model 1 and Model 2 are similar with differences at most $13\%$, even though these two models have very different interference patterns. 
The ratios of $\frac{\bar\alpha}{\alpha} ,\frac{\bar\beta}{\beta} ,\frac{\bar{\kappa}}{\kappa},\frac{\bar{\tilde\kappa}}{\tilde\kappa}$ are close to one and are in the range of $0.86\sim 0.93$ and again their values in Model 1 and Model 2 are similar.
The bounds on $\alpha,\bar\alpha,\eta, \tilde \eta, \tilde{\tilde\eta}, \tilde \kappa,\bar{\tilde\kappa}$ in Model 1 are more restrictive than those in Model 2, but they all allow these parameters to be of order one.
We do not see any violation of the triangular inequalities, 
Eqs.(\ref{eq: small e}), (\ref{eq: triangle inequality 1}) and (\ref{eq: triangle inequality 2}). 
The values of $a$ and $b$ in Model 1 and 2 confirm that $T_{1\BB}$ and $T_{2\BB}$ are constructive in Model 1, but destructive in Model 2, and $|T_{1,2\,\BB}|$ in Model 2 are larger than those in Model 1.
These two models are indeed different but they give similar results on these parameters.
Furthermore, our expectations on these parameters are basically verified in these two models.

From Table \ref{tab: BrBB1} we find that the $\overline B_{q}\to \BB' l \bar \nu$ branching ratios are of the orders $10^{-8}\sim 10^{-6}$ for non-annihilation modes, while the branching ratios of $\overline B_{q}\to \BB' \nu \bar \nu$ decays are of the orders of $10^{-11}\sim 10^{-8}$ for non-penguin-box-annihilation modes.
From Tables~\ref{tab: sensitivity} and \ref{tab: BrBB1}, we see that the following modes have good detectability and relatively unsuppressed rates,
they are
$B^-\to p \bar p l \bar \nu$,
$\overline B^0\to p \bar n l \bar \nu$,
$\overline B{}^0_s \to p \overline{\Lambda} l \bar \nu$,
$B^-\to \Lambda \bar{p} \nu \bar \nu$,
$\overline B^0\to \Lambda \bar{n} \nu \bar \nu$
and
$\overline B{}^0_s \to \Lambda \overline{\Lambda} \nu \bar \nu$ decays.
It is reasonable that the $B^-\to p \bar p l \bar \nu$ decay is the first detected mode as it has a large rate with very good detectability.
In fact its rate is the largest one in Model 1, but is the third largest one in Model 2, where $\overline B^0\to p \bar n l \bar \nu$ and $B^-\to n \bar n l \bar \nu$ decays have larger rates but poorer detectability.
It will be useful to search for these modes to differentiate these two models and to understand the interference patterns of $\overline B_{q}\to \BB' l \bar \nu$ decay amplitudes.

From Tables~\ref{tab: BrBB}, we obtain 
\be
\frac{\sum_\nu Br(B^-\to \Lambda \bar{p} \nu \bar \nu)}{Br(\overline B{}^0_s \to p \overline{\Lambda} l \bar \nu)}
&=& 4.29\frac{\bar \alpha}{\alpha}\times\bigg(\frac{0.0036}{|V_{ub}|}\bigg)^2 \times10^{-3}, 
\non\\
\frac{\sum_\nu Br(\overline B^0\to \Lambda \bar{n} \nu \bar \nu)}{Br(\overline B{}^0_s \to p \overline{\Lambda} l \bar \nu)}
&=& 3.94 \frac{\bar \alpha}{\alpha}\times \bigg(\frac{0.0036}{|V_{ub}|}\bigg)^2 \times10^{-3}. 
\label{eq: ratioBB1}
\en
The ratio $\bar\alpha/\alpha$ is expected to be close to one.
In Model 1 and 2, we have $\bar\alpha/\alpha=0.90$ and 0.95, respectively, as shown in Table \ref{tab: alphabeta}, which are indeed close to one.
Hence the ratios are not sensitive to the SU(3) breakings from threshold enhancement as they are mostly cancelled out.
Furthermore, the ratios 
do not rely on the assumption of neglecting annihilation $A$ and penguin-box-annihilation $PBA$ contributions, as these modes are free from these contributions, see Table~\ref{tab: TPBB}.
As the $\overline B{}^0_s \to p \overline{\Lambda} l \bar \nu$ decay are tree level decay modes, while the 
$B^-\to \Lambda \bar{p} \nu \bar \nu$ and $\overline B^0\to \Lambda \bar{n} \nu \bar \nu$ decays are governed by penguin and box diagrams, 
the above ratios can be tests of SM.

\subsection{$\overline B_q\to\BD l\bar\nu$, $\overline B_q\to\BD \nu\bar\nu$, $\overline B_q\to\DB l\bar\nu$ and $\overline B_q\to\DB \nu\bar\nu$ decay rates}

\begin{table}[t!]
\caption{\label{tab: BrBD} Branching ratios of $\overline B_{q}\to \BD l \bar \nu$ and $\overline B_{q}\to \BD \nu \bar \nu$ decays. 
The last factors are the SU(3) breaking from threshold enhancement estimated using Eq. (\ref{eq: dBdm}) with $\gamma=7$,
and parameters $\beta', \kappa', \sigma',\xi', \omega'$ are expected to be of order 1.
}
\begin{ruledtabular}
\begin{tabular}{cccccccc}
Mode
          & $Br(\overline B_{q}\to \BD l \bar \nu) (10^{-8})$
          & Mode
          & $Br(\overline B_{q}\to \BD  l \bar \nu) (10^{-8})$
          \\
\hline $B^-\to  p \overline{ \Delta^+} l \bar \nu$
          & $2 a'$ 
          & $B^-\to  n \overline{\Delta^0} l \bar \nu$ 
          & $2 a'\times (0.993)$ 
          \\
$B^-\to  \Sigma^+ \overline{\Sigma^{*+}} l \bar \nu$
          & $2 a'\times (0.130\sigma')$ 
          & $B^-\to  \Sigma^0 \overline{\Sigma^{*0}} l \bar \nu$
          & $\frac{1}{2} a'\times (0.128\sigma')$ 
          \\
$B^-\to \Xi^0 \overline{\Xi^{*0}} l \bar \nu$
          & $2 a'\times (0.0384\xi')$ 
          & $B^-\to \Lambda \overline{\Sigma^{*0}} l \bar \nu$
          & $\frac{3}{2} a'\times (0.184\sigma')$ 
          \\      
\hline $\overline B{}^0 \to p \overline{\Delta^0} l \bar\nu$
          & $1.85 a'$ 
          & $\overline B{}^0 \to n \overline{\Delta^-} l \bar\nu$ 
          & $5.56 a'\times (0.993)$ 
          \\
$\overline B{}^0 \to \Sigma^+ \overline{\Sigma^{*0}} l \bar\nu$
          & $0.93 a'\times (0.130\sigma')$ 
          & $\overline B{}^0 \to \Sigma^0 \overline{\Sigma^{*-}}  l \bar\nu$ 
          & $0.93 a'\times (0.125\sigma')$ 
          \\
$\overline B{}^0 \to \Xi^0 \overline{\Xi^{*-}} l \bar\nu$ 
          & $1.85 a'\times (0.0379\xi')$ 
          & $\overline B{}^0 \to \Lambda \overline{\Sigma^{*-}} l \bar\nu$
          & $2.78 a'\times (0.181\sigma')$ 
          \\         
\hline $\overline B{}^0_s \to p \overline{\Sigma^{*0}} l \bar\nu$
          & $0.93 a'\times (0.444\beta')$ 
          & $\overline B{}^0_s \to n \overline{\Sigma^{*-}} l \bar\nu$ 
          & $1.86 a'\times (0.434\beta')$ 
          \\
$\overline B{}^0_s \to\Sigma^+ \overline{\Xi^{*0}} l \bar\nu$ 
          & $1.86 a'\times (0.0662\kappa')$ 
          & $\overline B{}^0_s \to \Sigma^0 \overline{\Xi^{*-}} l \bar\nu$ 
          & $1.86 a'\times (0.0643\kappa')$ 
          \\
$\overline B{}^0_s \to \Xi^0 \overline{\Omega^-} l \bar\nu$ 
          & $5.59 a'\times (0.0215\omega')$ 
          & $\overline B{}^0_s \to \Lambda \overline{\Xi^{*-}} l \bar\nu$ 
          & $2.79 a'\times (0.0906\kappa')$ 
          \\                             
\hline \hline 
Mode
          & $\sum_\nu Br(\overline B_{q}\to \BD \nu \bar \nu) (10^{-10})$
          & Mode
          & $\sum_\nu Br(\overline B_{q}\to \BD  \nu \bar \nu) (10^{-10})$
          \\
\hline
$B^-\to \Sigma^+ \overline{\Delta^{++}} \nu \bar\nu$
          & $2.40 a'\times (0.269\beta')$ 
          & $B^-\to \Sigma^0 \overline{\Delta^+} \nu \bar\nu$ 
          & $1.60 a'\times (0.264\beta')$ 
          \\
$B^-\to \Sigma^- \overline{\Delta^0} \nu \bar\nu$ 
          & $0.80 a'\times (0.258\beta')$ 
          & $B^-\to \Xi^0 \overline{\Sigma^{*+}} \nu \bar\nu$ 
          & $0.80 a'\times (0.0734\kappa')$ 
          \\
$B^-\to \Xi^- \overline{\Sigma^{*0}} \nu \bar\nu$ 
          & $0.40 a'\times (0.0709\kappa')$ 
          & 
          & 
          \\          
\hline $\overline B{}^0\to \Sigma^+ \overline{\Delta^+} \nu \bar \nu$
          & $0.74 a'\times (0.269\beta')$ 
          & $\overline B{}^0\to \Sigma^0 \overline{\Delta^0} \nu \bar \nu$ 
          & $1.48 a'\times (0.264\beta')$ 
          \\
$\overline B{}^0\to\Sigma^- \overline{\Delta^-}  \nu \bar \nu$ 
          & $2.22 a'\times (0.258\beta')$ 
          & $\overline B{}^0\to \Xi^0 \overline{\Sigma^{*0}} \nu \bar \nu$ 
          & $0.37 a'\times (0.0731\kappa')$ 
          \\
$\overline B{}^0\to \Xi^- \overline{\Sigma^{*-}}\nu \bar \nu$ 
          & $0.74 a'\times (0.0698\kappa')$ 
          & 
          & 
          \\         
\hline $\overline B{}^0_s \to \Sigma^+ \overline{\Sigma^{*+}} \nu \bar\nu$
          & $0.75 a'\times (0.130\sigma')$ 
          & $\overline B{}^0_s \to \Sigma^0 \overline{\Sigma^{*0}} \nu \bar\nu$ 
          & $0.75 a'\times (0.128\sigma')$ 
          \\
$\overline B{}^0_s \to \Sigma^- \overline{\Sigma^{*-}} \nu \bar\nu$ 
          & $0.75 a'\times (0.123\sigma')$ 
          & $\overline B{}^0_s \to \Xi^0 \overline{\Xi^{*0}} \nu \bar\nu$ 
          & $0.75 a'\times (0.0384\xi')$ 
          \\
$\overline B{}^0_s \to \Xi^- \overline{\Xi^{*-}}  \nu \bar\nu$ 
          & $0.75 a'\times (0.0368\xi')$ 
          & 
          & 
          \\                             
\end{tabular}
\end{ruledtabular}
\end{table}

\begin{table}[t!]
\caption{\label{tab: BrDB} Branching ratios of $\overline B_{q}\to \DB l \bar \nu$ and $\overline B_{q}\to \DB \nu \bar \nu$ decays. 
The last factors are from the SU(3) breaking from threshold enhancement estimated using Eq. (\ref{eq: dBdm}) with $\gamma=7$,
and parameters $\beta'', \kappa'', \sigma'', \xi'', \omega''$ are expected to be of order one.}
\begin{ruledtabular}
\begin{tabular}{cccccccc}
Mode
          & $Br(\overline B_{q}\to \DB l \bar \nu) (10^{-8})$
          & Mode
          & $Br(\overline B_{q}\to \DB  l \bar \nu) (10^{-8})$
          \\
\hline $B^- \to \Delta^+ \bar p l \bar \nu$
          & $2 a''$ 
          & $B^- \to \Delta^0 \bar n l \bar \nu$ 
          & $2 a''\times (0.991)$ 
          \\
$B^- \to\Sigma^{*+} \overline{\Sigma^{+}} l \bar \nu$ 
          & $2 a''\times (0.0660\sigma'')$ 
          & $B^- \to \Sigma^{*0} \overline{\Sigma^{0}} l \bar \nu$ 
          & $\frac{1}{2} a''\times (0.0643\sigma'')$ 
          \\
$B^- \to \Xi^{*0} \overline{\Xi^{0}} l \bar \nu$ 
          & $2 a''\times (0.0130\xi'')$ 
          & $B^- \to \Sigma^{*0} \overline{\Lambda} l \bar \nu$ 
          & $\frac{3}{2} a''\times (0.104\sigma'')$ 
          \\      
\hline $\overline B{}^0\to \Delta^{++} \bar p l \bar\nu$
          & $5.56 a''$ 
          & $\overline B{}^0\to \Delta^+ \bar n l \bar\nu$ 
          & $1.85 a''\times (0.991)$ 
          \\
$\overline B{}^0\to \Sigma^{*+} \overline{\Sigma^{0}} l \bar\nu$ 
          & $0.93 a''\times (0.0646\sigma'')$ 
          & $\overline B{}^0\to \Sigma^{*0} \overline{\Sigma^{-}} l \bar\nu$ 
          & $0.93 a''\times (0.0624\sigma'')$ 
          \\
$\overline B{}^0\to \Xi^{*0} \overline{\Xi^{-}} l \bar\nu$ 
          & $1.85 a''\times (0.0125\xi'')$ 
          & $\overline B{}^0\to \Sigma^{*+} \overline{\Lambda} l \bar\nu$ 
          & $2.78 a''\times (0.105\sigma'')$ 
          \\         
\hline $\overline B{}^0_s \to \Delta^{++} \overline{\Sigma^{+}} l \bar\nu$
          & $5.59 a''\times (0.173\beta'')$ 
          & $\overline B{}^0_s \to \Delta^+ \overline{\Sigma^{0}} l \bar\nu$ 
          & $3.73 a''\times (0.170\beta'')$ 
          \\
$\overline B{}^0_s \to \Delta^0 \overline{\Sigma^{-}} l \bar\nu$ 
          & $1.86 a''\times (0.164\beta'')$ 
          & $\overline B{}^0_s \to  \Sigma^{*+} \overline{\Xi^{0}} l \bar\nu$ 
          & $1.86 a''\times (0.0308\kappa'')$
          \\
$\overline B{}^0_s \to \Sigma^{*0} \overline{\Xi^{-}} l \bar\nu$ 
          & $0.93 a''\times (0.0294\kappa'')$ 
          & 
          & 
          \\                             
\hline \hline 
Mode
          & $\sum_\nu Br(\overline B_{q}\to \DB \nu \bar \nu) (10^{-10})$
          & Mode
          & $\sum_\nu Br(\overline B_{q}\to \DB  \nu \bar \nu) (10^{-10})$
          \\
\hline
$B^- \to \Sigma^{*0} \bar p \nu \bar\nu$
          & $0.40 a''\times (0.339\beta'')$ 
          & $B^- \to \Sigma^{*-} \bar n \nu \bar\nu$
          & $0.80 a''\times (0.328\beta'')$ 
          \\
$B^- \to \Xi^{*0} \overline{\Sigma^+} \nu \bar\nu$
          & $0.80 a''\times (0.0270\kappa'')$ 
          & $B^- \to \Xi^{*-} \overline{\Sigma^{0}} \nu \bar\nu$
          & $0.40 a''\times (0.0260\kappa'')$ 
          \\
$B^- \to \Omega^- \overline{\Xi^{0}} \nu \bar\nu$ 
          & $2.40 a''\times (0.00602 \omega'')$ 
          & $B^- \to \Xi^{*-} \overline{\Lambda} \nu \bar\nu$ 
          & $1.20 a''\times (0.0408\kappa'')$ 
          \\          
\hline $\overline B{}^0 \to \Sigma^{*+} \bar p \nu \bar\nu$
          & $0.74 a''\times (0.341\beta'')$ 
          & $\overline B{}^0 \to \Sigma^{*0} \bar n \nu \bar\nu$
          & $0.37 a''\times (0.336\beta'')$ 
          \\
$\overline B{}^0 \to  \Xi^{*0} \overline{\Sigma^{0}} \nu \bar\nu$
          & $0.37 a''\times (0.0263\kappa'')$ 
          & $\overline B{}^0 \to \Xi^{*-} \overline{\Sigma^{-}} \nu \bar\nu$
          & $0.74 a''\times (0.0251\kappa'')$ 
          \\
$\overline B{}^0 \to \Omega^- \overline{\Xi^{-}} \nu \bar\nu$
          & $2.24 a''\times (0.00580\omega'')$ 
          & $\overline B{}^0 \to \Xi^{*0} \overline{\Lambda} \nu \bar\nu$
          & $1.11 a''\times (0.0416\kappa'')$ 
          \\         
\hline $\overline B{}^0_s \to \Sigma^{*+} \overline{\Sigma^{+}} \nu \bar\nu$
          & $0.75 a''\times (0.0660\sigma'')$ 
          & $\overline B{}^0_s \to \Sigma^{*0} \overline{\Sigma^{0}} \nu \bar\nu$ 
          & $0.75 a''\times (0.0643\sigma'')$ 
          \\
$\overline B{}^0_s \to \Sigma^{*-} \overline{\Sigma^{-}} \nu \bar\nu$ 
          & $0.75 a''\times (0.0611\sigma'')$ 
          & $\overline B{}^0_s \to \Xi^{*0} \overline{\Xi^{0}} \nu \bar\nu$ 
          & $0.75 a''\times (0.0130\xi'')$ 
          \\
$\overline B{}^0_s \to \Xi^{*-} \overline{\Xi^{-}} \nu \bar\nu$ 
          & $0.75 a''\times (0.0123\xi'')$ 
          & 
          & 
          \\                             
\end{tabular}
\end{ruledtabular}
\end{table}

\begin{table}[t!]
\caption{\label{tab: BrBD1} Branching ratios of $\overline B_{q}\to \BD l \bar \nu$ and $\overline B_{q}\to \BD \nu \bar \nu$ decays in Model 1 and Model 2.
$Br_1$ and $Br_2$ denote results in Model 1 and Model 2, respectively.
}
\begin{ruledtabular}
\begin{tabular}{cccccccc}
Mode
          & $Br_1(10^{-8})$
          & $Br_2(10^{-8})$
          & Mode
          & $Br_1(10^{-8})$
          & $Br_2(10^{-8})$
          \\
\hline $B^-\to  p \overline{ \Delta^+} l \bar \nu$
          & 2.30 
          & 14.48
          & $B^-\to  n \overline{\Delta^0} l \bar \nu$ 
          & 2.28 
          & 14.38
          \\
$B^-\to  \Sigma^+ \overline{\Sigma^{*+}} l \bar \nu$
          & 0.22 
          & 1.48
          & $B^-\to  \Sigma^0 \overline{\Sigma^{*0}} l \bar \nu$
          & 0.054 
          & 0.36
          \\
$B^-\to \Xi^0 \overline{\Xi^{*0}} l \bar \nu$
          & 0.046 
          & 0.32
          & $B^-\to \Lambda \overline{\Sigma^{*0}} l \bar \nu$
          & 0.25 
          & 1.65
          \\      
\hline $\overline B{}^0 \to p \overline{\Delta^0} l \bar\nu$
          & 2.13 
          & 13.44
          & $\overline B{}^0 \to n \overline{\Delta^-} l \bar\nu$ 
          & 6.35 
          & 40.01
          \\
$\overline B{}^0 \to \Sigma^+ \overline{\Sigma^{*0}} l \bar\nu$
          & 0.10 
          & 0.68 
          & $\overline B{}^0 \to \Sigma^0 \overline{\Sigma^{*-}}  l \bar\nu$ 
          & 0.098 
          & 0.66
          \\
$\overline B{}^0 \to \Xi^0 \overline{\Xi^{*-}} l \bar\nu$ 
          & 0.042 
          & 0.29
          & $\overline B{}^0 \to \Lambda \overline{\Sigma^{*-}} l \bar\nu$
          & 0.45 
          & 2.30
          \\         
\hline $\overline B{}^0_s \to p \overline{\Sigma^{*0}} l \bar\nu$
          & 0.48 
          & 3.08
          & $\overline B{}^0_s \to n \overline{\Sigma^{*-}} l \bar\nu$ 
          & 0.94 
          & 5.99
          \\
$\overline B{}^0_s \to\Sigma^+ \overline{\Xi^{*0}} l \bar\nu$ 
          & 0.10 
          & 0.70
          & $\overline B{}^0_s \to \Sigma^0 \overline{\Xi^{*-}} l \bar\nu$ 
          & 0.049 
          & 0.34
          \\
$\overline B{}^0_s \to \Xi^0 \overline{\Omega^-} l \bar\nu$ 
          & 0.069 
          & 0.49
          & $\overline B{}^0_s \to \Lambda \overline{\Xi^{*-}} l \bar\nu$ 
          & 0.22 
          & 1.51
          \\                             
\hline \hline 
Mode
          & $\sum_\nu Br_1 (10^{-10})$
          & $\sum_\nu Br_2 (10^{-10})$
          & Mode
          & $\sum_\nu Br_1 (10^{-10})$
          & $\sum_\nu Br_2 (10^{-10})$
          \\
\hline
$B^-\to \Sigma^+ \overline{\Delta^{++}} \nu \bar\nu$
          & 0.64 
          & 4.27
          & $B^-\to \Sigma^0 \overline{\Delta^+} \nu \bar\nu$ 
          & 0.42 
          & 2.79
          \\
$B^-\to \Sigma^- \overline{\Delta^0} \nu \bar\nu$ 
          & 0.20 
          & 1.36
          & $B^-\to \Xi^0 \overline{\Sigma^{*+}} \nu \bar\nu$ 
          & 0.043 
          & 0.30 
          \\
$B^-\to \Xi^- \overline{\Sigma^{*0}} \nu \bar\nu$ 
          & 0.021 
          & 0.14
          & 
          \\          
\hline $\overline B{}^0\to \Sigma^+ \overline{\Delta^+} \nu \bar \nu$
          & 0.20 
          & 1.32
          & $\overline B{}^0\to \Sigma^0 \overline{\Delta^0} \nu \bar \nu$ 
          & 0.39 
          & 2.59
          \\
$\overline B{}^0\to\Sigma^- \overline{\Delta^-}  \nu \bar \nu$ 
          & 0.57 
          & 3.78
          & $\overline B{}^0\to \Xi^0 \overline{\Sigma^{*0}} \nu \bar \nu$ 
          & 0.020 
          & 0.14
          \\
$\overline B{}^0\to \Xi^- \overline{\Sigma^{*-}}\nu \bar \nu$ 
          & 0.038 
          & 0.26
          & 
          \\         
\hline $\overline B{}^0_s \to \Sigma^+ \overline{\Sigma^{*+}} \nu \bar\nu$
          & 0.097 
          & 0.65
          & $\overline B{}^0_s \to \Sigma^0 \overline{\Sigma^{*0}} \nu \bar\nu$ 
          & 0.095 
          & 0.64
          \\
$\overline B{}^0_s \to \Sigma^- \overline{\Sigma^{*-}} \nu \bar\nu$ 
          & 0.090 
          & 0.61
          & $\overline B{}^0_s \to \Xi^0 \overline{\Xi^{*0}} \nu \bar\nu$ 
          & 0.021 
          & 0.14
          \\
$\overline B{}^0_s \to \Xi^- \overline{\Xi^{*-}}  \nu \bar\nu$ 
          & 0.019 
          & 0.14
          & 
          \\                             
\end{tabular}
\end{ruledtabular}
\end{table}

\begin{table}[t!]
\caption{\label{tab: BrDB1} 
Same as Table~\ref{tab: BrBD1} but for branching ratios of $\overline B_{q}\to \DB l \bar \nu$ and $\overline B_{q}\to \DB \nu \bar \nu$ decays.
}
\begin{ruledtabular}
\begin{tabular}{cccccccc}
Mode
          Mode
          & $Br_1(10^{-8})$
          & $Br_2(10^{-8})$
          & Mode
          & $Br_1(10^{-8})$
          & $Br_2(10^{-8})$
          \\
\hline $B^- \to \Delta^+ \bar p l \bar \nu$
          & 2.70 
          & 12.38
          & $B^- \to \Delta^0 \bar n l \bar \nu$ 
          & 2.68 
          & 12.28
          \\
$B^- \to\Sigma^{*+} \overline{\Sigma^{+}} l \bar \nu$ 
          & 0.26 
          & 1.45
          & $B^- \to \Sigma^{*0} \overline{\Sigma^{0}} l \bar \nu$ 
          & 0.065 
          & 0.36
          \\
$B^- \to \Xi^{*0} \overline{\Xi^{0}} l \bar \nu$ 
          & 0.056 
          & 0.37
          & $B^- \to \Sigma^{*0} \overline{\Lambda} l \bar \nu$ 
          & 0.30 
          & 1.63
          \\      
\hline $\overline B{}^0\to \Delta^{++} \bar p l \bar\nu$
          & 7.51 
          & 34.45
          & $\overline B{}^0\to \Delta^+ \bar n l \bar\nu$ 
          & 2.48 
          & 11.40
          \\
$\overline B{}^0\to \Sigma^{*+} \overline{\Sigma^{0}} l \bar\nu$ 
          & 0.12 
          & 0.66
          & $\overline B{}^0\to \Sigma^{*0} \overline{\Sigma^{-}} l \bar\nu$ 
          & 0.12 
          & 0.64
          \\
$\overline B{}^0\to \Xi^{*0} \overline{\Xi^{-}} l \bar\nu$ 
          & 0.050 
          & 0.33
          & $\overline B{}^0\to \Sigma^{*+} \overline{\Lambda} l \bar\nu$ 
          & 0.56 
          & 3.04
          \\         
\hline $\overline B{}^0_s \to \Delta^{++} \overline{\Sigma^{+}} l \bar\nu$
          & 2.05 
          & 9.47
          & $\overline B{}^0_s \to \Delta^+ \overline{\Sigma^{0}} l \bar\nu$ 
          & 1.34 
          & 6.20
          \\
$\overline B{}^0_s \to \Delta^0 \overline{\Sigma^{-}} l \bar\nu$ 
          & 0.65 
          & 3.02
          & $\overline B{}^0_s \to  \Sigma^{*+} \overline{\Xi^{0}} l \bar\nu$ 
          & 0.14 
          & 0.80
          \\
$\overline B{}^0_s \to \Sigma^{*0} \overline{\Xi^{-}} l \bar\nu$ 
          & 0.069 
          & 0.38
          & 
          \\                             
\hline \hline 
Mode
          & $\sum_\nu Br_1 (10^{-10})$
          & $\sum_\nu Br_2 (10^{-10})$
          & Mode
          & $\sum_\nu Br_1 (10^{-10})$
          & $\sum_\nu Br_2 (10^{-10})$
          \\
\hline
$B^- \to \Sigma^{*0} \bar p \nu \bar\nu$
          & 0.21 
          & 1.16
          & $B^- \to \Sigma^{*-} \bar n \nu \bar\nu$
          & 0.42 
          & 2.26
          \\
$B^- \to \Xi^{*0} \overline{\Sigma^+} \nu \bar\nu$
          & 0.045 
          & 0.29
          & $B^- \to \Xi^{*-} \overline{\Sigma^{0}} \nu \bar\nu$
          & 0.022 
          & 0.14
          \\
$B^- \to \Omega^- \overline{\Xi^{0}} \nu \bar\nu$ 
          & 0.030 
          & 0.23
          & $B^- \to \Xi^{*-} \overline{\Lambda} \nu \bar\nu$ 
          & 0.099 
          & 0.64
          \\          
\hline $\overline B{}^0 \to \Sigma^{*+} \bar p \nu \bar\nu$
          & 0.40 
          & 2.16
          & $\overline B{}^0 \to \Sigma^{*0} \bar n \nu \bar\nu$
          & 0.20 
          & 1.07
          \\
$\overline B{}^0 \to  \Xi^{*0} \overline{\Sigma^{0}} \nu \bar\nu$
          & 0.020 
          & 0.13
          & $\overline B{}^0 \to \Xi^{*-} \overline{\Sigma^{-}} \nu \bar\nu$
          & 0.039 
          & 0.25
          \\
$\overline B{}^0 \to \Omega^- \overline{\Xi^{-}} \nu \bar\nu$
          & 0.027 
          & 0.21
          & $\overline B{}^0 \to \Xi^{*0} \overline{\Lambda} \nu \bar\nu$
          & 0.094 
          & 0.60
          \\         
\hline $\overline B{}^0_s \to \Sigma^{*+} \overline{\Sigma^{+}} \nu \bar\nu$
          & 0.12 
          & 0.63
          & $\overline B{}^0_s \to \Sigma^{*0} \overline{\Sigma^{0}} \nu \bar\nu$ 
          & 0.11 
          & 0.62
          \\
$\overline B{}^0_s \to \Sigma^{*-} \overline{\Sigma^{-}} \nu \bar\nu$ 
          & 0.11 
          & 0.59
          & $\overline B{}^0_s \to \Xi^{*0} \overline{\Xi^{0}} \nu \bar\nu$ 
          & 0.025 
          & 0.16
          \\
$\overline B{}^0_s \to \Xi^{*-} \overline{\Xi^{-}} \nu \bar\nu$ 
          & 0.024 
          & 0.16
          & 
          \\                             
\end{tabular}
\end{ruledtabular}
\end{table}

\begin{table}[t!]
\caption{\label{tab: alphabeta BD DB} 
Values of various parameters in Model 1 and Model 2. 
}
\begin{ruledtabular}
\begin{tabular}{cccccccc}
Parameters
          & Values (Model 1)
          & Values (Model 2)
          \\
\hline 
$a'$
          & 1.15
          & 7.24
          \\
\hline          
$\beta'$
          & 1.01
          & 1.03
          \\  
$\kappa'$
          & 0.72
          & 0.78
          \\                 
$\sigma'$
          & 0.74
          & 0.79
          \\                 
$\xi'$
          & 0.52
          & 0.57
          \\
$\omega'$
          & 0.50
          & 0.56
          \\   
     
\hline
\hline
Parameters
          & Values (Model 1)
          & Values (Model 2)
          \\
\hline
$a''$
          & 1.35
          & 6.19
          \\
\hline          
$\beta''$
          & 1.57
          & 1.58
          \\  
$\kappa''$
          & 1.86
          & 2.24
          \\                 
$\sigma''$
          & 1.49
          & 1.78
          \\                 
$\xi''$
          & 1.58
          & 2.29
          \\ 
$\omega''$
          & 1.53
          & 2.59
          \\                                                      
\end{tabular}
\end{ruledtabular}
\end{table}

\begin{figure}[t]
\centering
 \subfigure[]{
  \includegraphics[width=0.5\textwidth]{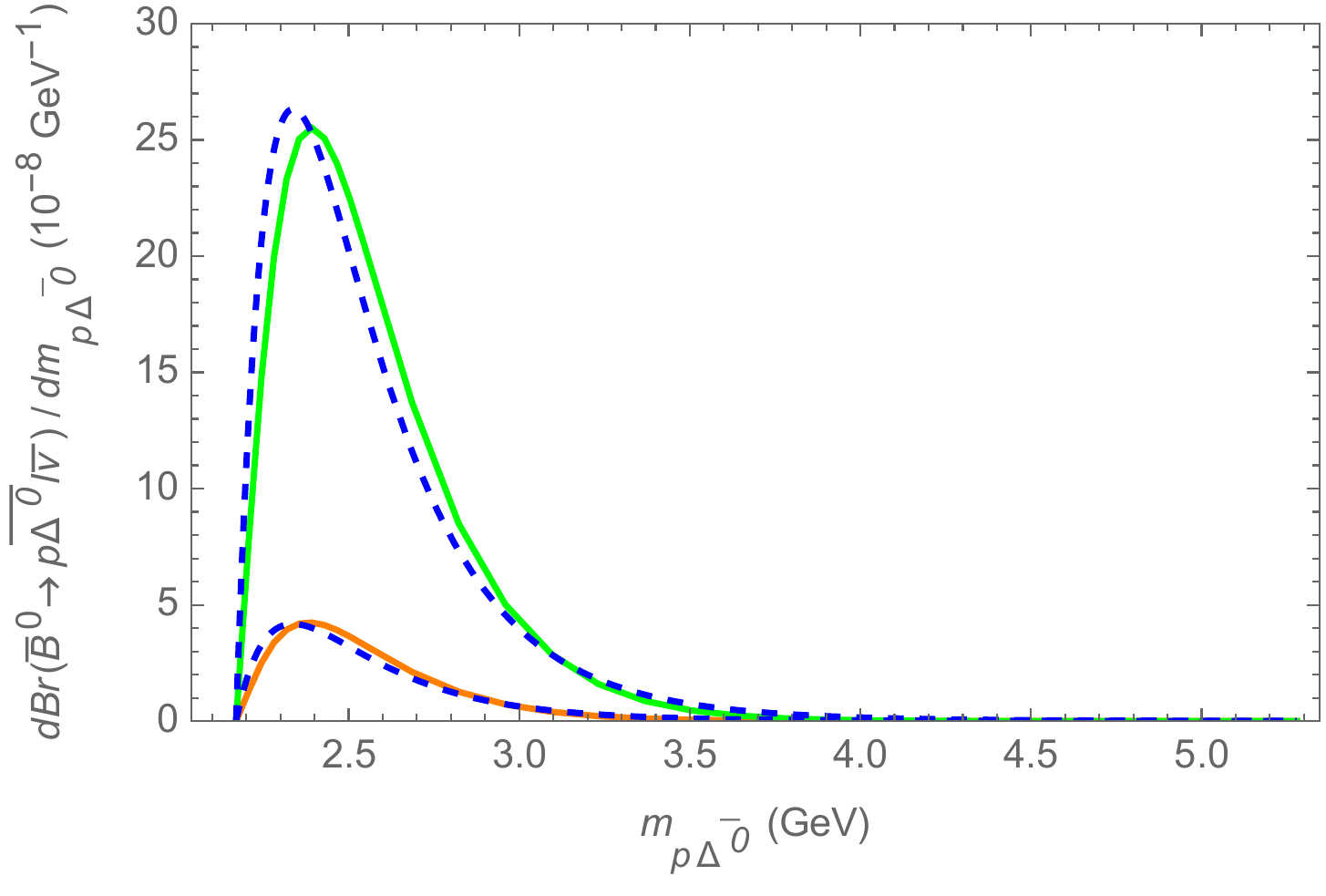}
}
\subfigure[]{
  \includegraphics[width=0.5\textwidth]{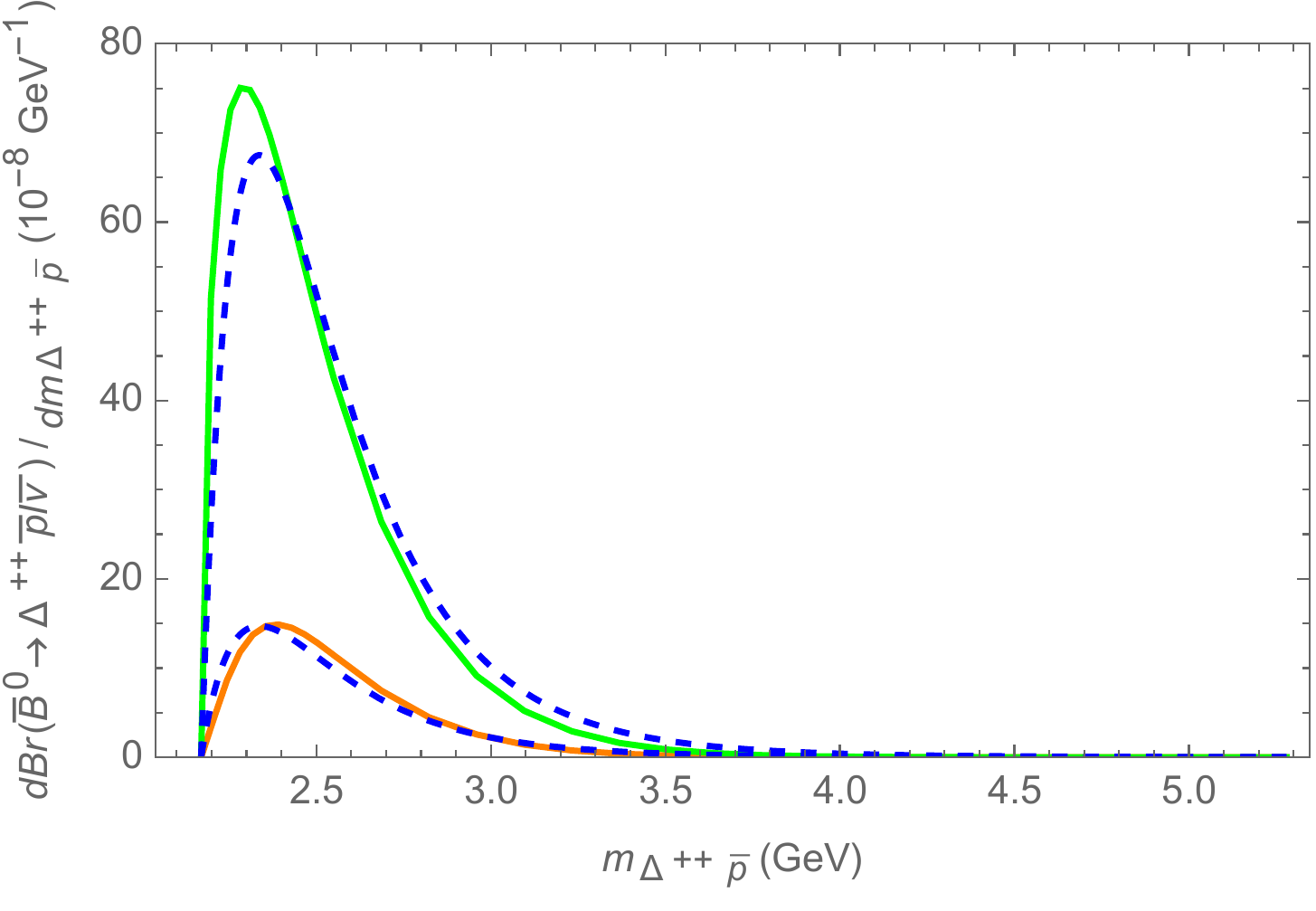}
}
\caption{
(a) Differential rate $dBr/dm_{p\overline{\Delta^0}}$ of $\overline B^0\to p\overline{\Delta^0} l^-\bar\nu$ decay from Model 1 (orange solid line) and Model 2 (green solid line). 
The dashed lines are $dBr/dm_{p\overline{\Delta^0}}$ using Eq. (\ref{eq: dBdm}) with $\gamma=7$. 
(b)~Same as (a) but for the differential rate $dBr/dm_{\Delta^{++} \bar p}$ of $\overline B^0\to \Delta^{++} \bar p l^-\bar\nu$ decay.
}
 \label{fig: dBDdm}
\end{figure}

We now consider the rates of $\overline B_q\to\BD l\bar\nu$ and $\overline B_q\to\BD \nu\bar\nu$ decays
and the rates of $\overline B_q\to\DB l\bar\nu$ and $\overline B_q\to\DB \nu\bar\nu$ decays.
As shown in Table~\ref{tab: TPBD}, in $\overline B_q\to\BD l\nu$ decays, there is only one topological amplitude, namely $T_\BD$, 
while in $\overline B_q\to\BD \nu\bar\nu$ decays, there is also only one topological amplitude, namely $PB_\BD$, 
but $PB_\BD$ and $T_\BD$ are related by $\zeta$ as shown in Eq. (\ref{eq: zeta}).
Similar features also hold in $\overline B_q\to\DB l\bar\nu$ and $\overline B_q\to\DB \nu\bar\nu$ decays by using Table~\ref{tab: TPDB} and Eq. (\ref{eq: zeta}), but with $T_\DB$ and $PB_\DB$.
The decay rates of $\overline B_q\to\BD l\nu, \BD \nu\bar\nu$ decays and 
$\overline B_q\to\DB' l\bar\nu, \DB \nu\bar\nu$ decays are parametrized in term of $a'$ and $a''$, respectively,
where the rates correspond to $A=T_\BD$ and $T_\DB$ are denoted as $a'$ and $a''$, respectively.
Note that the above parameters correspond to the rates in the SU(3) symmetric limit.

As in $\overline B_q\to\BB' l\bar\nu$ and $\overline B_q\to\BB' \nu\bar\nu$ decays, we expect to see threshold enhancement in the differential rates of these modes.
Likewise the SU(3) breaking on $\overline B_q\to\BD l\bar\nu$ and $\overline B_q\to\DB l\bar\nu$ rates from the threshold enhancement can be estimated as in the $\overline B_q\to \BB' l\bar\nu$ case, once the corresponding differential rates of some modes are known.
However, at the moment no such information is available yet. 
We should make use of some model calculations to obtain informations of the differential rates of these modes. 
As we shall see  
the SU(3) breaking from the threshold enhancement can be estimated using Eq. (\ref{eq: dBdm}) and similar procedure in the discussion around Eq.~(\ref{eq: sigma}) but with $\gamma=7$.
In Tables~\ref{tab: BrBD} and \ref{tab: BrDB} 
the decay rates of $\overline B_q\to\BD l\nu, \BD \nu\bar\nu$ decays and 
$\overline B_q\to\DB l\bar\nu, \DB \nu\bar\nu$ decays are shown.
The parameters $\beta^{\prime,\prime\prime}$, $\kappa^{\prime,\prime\prime}$, $\sigma^{\prime,\prime\prime}$ and so on 
are used to denote milder SU(3) breaking effects and are expected to be of order one.
The relative sizes of $\overline B_{q}\to \BD l \bar \nu (\nu \bar \nu)$ and $\overline B_{q}\to \DB l \bar \nu (\nu \bar \nu)$ decay rates can be readily read from Tables~\ref{tab: BrBD} and \ref{tab: BrDB}.

At this moment we do not have enough data to verify Tables~\ref{tab: BrBD} and \ref{tab: BrDB}.
As in the case of $\overline B_q\to\BB' l\bar\nu$ and $\overline B_q\to\BB' \nu\bar\nu$ decays we will make use of Model 1 and Model 2 for illustration.
Using inputs from Sec.~\ref{sec: model calculations} and the formulas given Appendix \ref{App: formulas},
the branching ratios of $\overline B_{q}\to \BD l \bar \nu$ and $\overline B_{q}\to \BD \nu \bar \nu$ in Model 1 and 2 are obtained and are shown in Table~\ref{tab: BrBD1},
while the branching ratios of $\overline B_{q}\to \DB l \bar \nu$ and $\overline B_{q}\to \DB \nu \bar \nu$ in Model 1 and 2 are obtained and are shown in Table~\ref{tab: BrDB1}. These results are new.

From Tables~\ref{tab: BrBD1} and \ref{tab: BrDB1}, we see that the branching ratios of $\overline B_{q}\to \BD l \bar \nu$ and $\overline B_{q}\to \DB l \bar \nu$ decays are in the ranges of $10^{-9}\sim 10^{-8}$ and $10^{-9}\sim 10^{-7}$ in Model 1 and 2, respectively,
while $\sum_\nu Br(\overline B_{q}\to \BD l \nu \nu)$ and $\sum_\nu Br(\overline B_{q}\to \DB l \nu \nu)$ are in the ranges of 
$10^{-12}\sim 10^{-10}$ and $10^{-11}\sim 10^{-10}$ in Model 1 and 2, respectively.
The rates in Model 2 are greater than those in Model 1 by a factor of $4\sim 7$. 
This mostly corresponds to the fact that $|T_\BD|$ and $|T_\DB|$ in Model 2 are greater than those in Model 1, 
as reflected through the sizes of $a'$ and $a''$ as shown in Table~\ref{tab: alphabeta BD DB}.
This is not surprising as $|T_{1\BB}|$ and $|T_{2\BB}|$ in $\overline B_{q}\to \BB' l \bar \nu$ decays
in Model 2 are greater than those in Model 1, 
as reflected in the sizes of $a$ and $b$ as shown in Table~\ref{tab: alphabeta}.
From Table~\ref{tab: alphabeta BD DB}, we see that the parameters $\beta^{\prime,\prime\prime}, \kappa^{\prime,\prime\prime}, \sigma^{\prime,\prime\prime}, \xi^{\prime,\prime\prime}$ and $\omega'$ denoting milder SU(3) breaking are indeed of order one and are similar in Model 1 and 2 in most cases.

From Tables~\ref{tab: BrBD}, \ref{tab: BrBD1} and \ref{tab: sensitivity}, we find that
$\overline B{}^0 \to p \overline{\Delta^0} l \bar\nu$ and
$\overline B{}^0\to \Sigma^0 \overline{\Delta^0} \nu \bar \nu$ have relatively unsuppressed rates and good detectability. 
In particular, we have the follow rate ratio of the loop induced mode and tree induced modes, 
\be
\frac{\sum_\nu Br(\overline B{}^0\to \Sigma^0 \overline{\Delta^0} \nu \bar \nu)}{Br(\overline B{}^0 \to p \overline{\Delta^0} l \bar\nu)}
&=& 2.11 \beta' \times\bigg(\frac{0.0036}{|V_{ub}|}\bigg)^2 \times10^{-3},
\label{eq: ratio BD}
\en
where $\beta'$ is of order one. In fact as shown in Table~\ref{tab: alphabeta BD DB}, we have $\beta'=1.01$ and $1.03$ in Model 1 and 2, respectively.
The ratio in Eq. (\ref{eq: ratio BD}) can be a test of SM.

From Tables~\ref{tab: BrDB}, \ref{tab: BrDB1} and \ref{tab: sensitivity}, we find that
$\overline B{}^0\to \Delta^{++} \bar p l \bar\nu$,
$\overline B{}^0\to \Sigma^{*+} \overline{\Lambda} l \bar\nu$, 
$B^-\to \Delta^+\bar p l\bar \nu$, 
$B^-\to \Delta^0\bar n l\bar \nu$, 
$\overline B{}^0 \to \Sigma^{*+} \bar p \nu \bar\nu$
and
$B^- \to \Sigma^{*-} \bar n \nu \bar\nu$ decays
have good detectability and relatively unsuppressed rates. 
The rate ratios of these loop induced modes and tree induced modes can be sensible tests of SM.
For example, we have the following rate ratio, 
\be
\frac{\sum_\nu Br(\overline B{}^0 \to \Sigma^{*+} \bar p \nu \bar\nu)}{Br(\overline B{}^0\to \Delta^{++} \bar p l \bar\nu)}
&=& 4.55 \kappa'' \times\bigg(\frac{0.0036}{|V_{ub}|}\bigg)^2 \times10^{-4},
\label{eq: ratio DB}
\en
where $\kappa''$ is of order one.
In fact, as shown in Table~\ref{tab: alphabeta BD DB}, 
we have $\kappa''=1.86$ and $2.24$ in Model 1 and 2, respectively.

The differential rates $dBr/dm_{p\overline{\Delta^0}}$ of $\overline B^0\to p\overline{\Delta^0} l^-\bar\nu$ decay and $dBr/dm_{\Delta^{++} \bar p}$ of $\overline B^0\to \Delta^{++} \bar p l^-\bar\nu$ decay from Model 1 and Model 2 are plotted in Fig.~\ref{fig: dBDdm}.
They can be compared to the dashed lines plotted using Eq. (\ref{eq: dBdm}) with $\gamma=7$. 
They clearly exhibit threshold enhancement as expected.

\begin{table}[t!]
\caption{\label{tab: BrDD} Branching ratios of $\overline B_{q}\to \DD' l \bar \nu$ and $\overline B_{q}\to \DD' \nu \bar \nu$ decays. 
Parameters $a'''$, $b'''$, $c'''$, $d'''$ are expected to be of similar sizes, while $e'''$ is expected to be much suppressed.
The last factors are SU(3) breaking from threshold enhancement, estimated uisng Eq. (\ref{eq: dBdm}) with $\gamma=10$,
and parameters $\beta''',\sigma''',\kappa''',\xi''',\upsilon''',\omega'''$ are expected to be of order 1.}
\begin{ruledtabular}
\begin{tabular}{cccccccc}
Mode
          & $Br(\overline B_{q}\to \DD' l \bar \nu) (10^{-8})$
          & Mode
          & $Br(\overline B_{q}\to \DD'  l \bar \nu) (10^{-8})$
          \\
\hline         
$B^- \to \Delta^{++} \overline{\Delta^{++}} l \bar \nu$
          & $36 b'''$ 
          & $B^- \to \Delta^+ \overline{\Delta^+} l \bar \nu$ 
          & $16 a'''$ 
          \\
$B^- \to \Delta^{0} \overline{\Delta^0} l \bar \nu$ 
          & $4 d'''$ 
          & $B^- \to \Delta^- \overline{\Delta^-} l \bar \nu$ 
          & $e'''$ 
          \\
$B^- \to\Sigma^{*+} \overline{\Sigma^{*+}} l \bar \nu$ 
          & $16 c'''\times (0.125\sigma''')$ 
          & $B^- \to \Sigma^{*0} \overline{\Sigma^{*0}} l \bar \nu$ 
          & $4 d'''\times (0.124\sigma''')$ 
          \\
$B^- \to\Sigma^{*-} \overline{\Sigma^{*-}} l \bar \nu$ 
          & $e'''\times (0.118\sigma''')$ 
          & $B^- \to \Xi^{*0} \overline{\Xi^{*0}} l \bar \nu$ 
          & $4 d'''\times (0.0198\xi''')$ 
          \\  
$B^- \to \Xi^{*-} \overline{\Xi^{*-}}  l \bar \nu$ 
          & $e'''\times (0.0191\xi''')$
          & $B^- \to \Omega^- \overline{\Omega^-} l \bar \nu$ 
          & $e'''\times (0.00407\upsilon''')$ 
          \\                                     
\hline $\overline B{}^0 \to \Delta^{++} \overline{\Delta^+} l \bar\nu$
          & $11.1 a'''$ 
          & $\overline B{}^0 \to  \Delta^+ \overline{\Delta^0}l \bar\nu$ 
          & $14.8 a'''$ 
          \\
$\overline B{}^0 \to \Delta^0 \overline{\Delta^-} l \bar\nu$ 
          & $11.1 a'''$ 
          & $\overline B{}^0 \to \Sigma^{*+} \overline{\Sigma^{*0}} l \bar\nu$
          & $7.4 a'''\times (0.124\sigma''')$ 
          \\
$\overline B{}^0 \to \Sigma^{*0} \overline{\Sigma^{*-}} l \bar\nu$
          & $7.4 a'''\times (0.121\sigma''')$ 
          & $\overline B{}^0 \to \Xi^{*0} \overline{\Xi^{*-}} l \bar\nu$
          & $3.7 a'''\times (0.0195\xi''')$ 
          \\         
\hline $\overline B{}^0_s \to \Delta^{++} \overline{\Sigma^{*+}} l \bar \nu$
          & $11.2 a'''\times (0.343\beta''')$ 
          & $\overline B{}^0_s \to \Delta^+ \overline{\Sigma^{*0}} l \bar \nu$ 
          & $7.5 a'''\times (0.341\beta''')$ 
          \\
$\overline B{}^0_s \to \Delta^0 \overline{\Sigma^{*-}} l \bar \nu$ 
          & $3.7 a'''\times (0.333\beta''')$ 
          & $\overline B{}^0_s \to \Sigma^{*+} \overline{\Xi^{*0}} l \bar \nu$ 
          & $14.9 a'''\times (0.0486\kappa''')$ 
          \\
$\overline B{}^0_s \to \Sigma^{*0} \overline{\Xi^{*-}} l \bar \nu$ 
          & $7.5 a'''\times (0.0474\kappa''')$ 
          & $\overline B{}^0_s \to \Xi^{*0} \overline{\Omega^{-}} l \bar \nu$ 
          & $11.2 a'''\times (0.00883\omega''')$ 
          \\                             
\hline \hline 
Mode
          & $\sum_\nu Br(\overline B_{q}\to \DD' \nu \bar \nu) (10^{-10})$
          & Mode
          & $\sum_\nu Br(\overline B_{q}\to \DD'  \nu \bar \nu) (10^{-10})$
          \\
\hline
$B^- \to  \Sigma^{*+} \overline{\Delta^{++}} \nu \bar \nu$ 
          & $4.80 a'''\times (0.343\bar\beta''')$ 
          & $B^- \to \Sigma^{*0} \overline{\Delta^{+}} \nu \bar \nu$ 
          & $3.20 a'''\times (0.341\bar\beta''')$ 
          \\
$B^- \to \Sigma^{*-} \overline{\Delta^{0}} \nu \bar \nu$
          & $1.60 a'''\times (0.333\bar\beta''')$ 
          & $B^- \to \Xi^{*0} \overline{\Sigma^{*+}} \nu \bar \nu$
          & $6.40 a'''\times (0.0486\bar\kappa''')$ 
          \\
$B^- \to \Xi^{*-} \overline{\Sigma^{*0}} \nu \bar \nu$
          & $3.20 a'''\times (0.0474\bar\kappa''')$ 
          & $B^- \to \Omega^- \overline{\Xi^{*0}} \nu \bar \nu$ 
          & $4.80 a'''\times (0.00883\bar\omega''')$ 
          \\          
\hline $\overline B{}^0 \to \Sigma^{*+} \overline{\Delta^+} \nu \bar\nu$
          & $1.48 a'''\times (0.343\bar\beta''')$ 
          & $\overline B{}^0 \to \Sigma^{*0} \overline{\Delta^0} \nu \bar\nu$ 
          & $2.97 a'''\times (0.341\bar\beta''')$ 
          \\
$\overline B{}^0 \to\Sigma^{*-} \overline{\Delta^-} \nu \bar\nu$ 
          & $4.45 a'''\times (0.333\bar\beta''')$ 
          & $\overline B{}^0 \to \Xi^{*0} \overline{\Sigma^{*0}} \nu \bar\nu$ 
          & $2.97 a'''\times (0.0484\bar\kappa''')$ 
          \\
$\overline B{}^0 \to \Xi^{*-} \overline{\Sigma^{*-}} \nu \bar\nu$ 
          & $5.93 a'''\times (0.0464\bar\kappa''')$ 
          & $\overline B{}^0 \to \Omega^- \overline{\Xi^{*-}} \nu \bar\nu$ 
          & $4.45 a'''\times (0.00867\bar\omega''')$ 
          \\         
\hline $\overline B{}^0_s\to \Delta^{++} \overline{\Delta^{++}} \nu \bar \nu$
          & $0.37e'''$ 
          & $\overline B{}^0_s \to \Delta^+ \overline{\Delta^+} \nu \bar \nu$ 
          & $0.37 e'''$ 
          \\
$\overline B{}^0_s \to \Delta^{0} \overline{\Delta^0} \nu \bar \nu$ 
          & $0.37 e'''$ 
          & $\overline B{}^0_s \to \Delta^- \overline{\Delta^-} \nu \bar \nu$ 
          & $0.37 e'''$ 
          \\
$\overline B{}^0_s \to \Sigma^{*+} \overline{\Sigma^{*+}} \nu \bar\nu$
          & $1.49 d'''\times (0.125\sigma''')$ 
          & $\overline B{}^0_s \to \Sigma^{*0} \overline{\Sigma^{*0}} \nu \bar\nu$ 
          & $1.49 d'''\times (0.124\sigma''')$ 
          \\
$\overline B{}^0_s \to \Sigma^{*-} \overline{\Sigma^{*-}} \nu \bar\nu$ 
          & $1.49 d'''\times (0.118\sigma''')$ 
          & $\overline B{}^0_s \to \Xi^{*0} \overline{\Xi^{*0}} \nu \bar\nu$ 
          & $5.96 c'''\times (0.0198\xi''')$ 
          \\
$\overline B{}^0_s \to  \Xi^{*-} \overline{\Xi^{*-}} \nu \bar\nu$ 
          & $5.96 c''''\times (0.0191\xi''')$ 
          & $\overline B{}^0_s \to \Omega^{-} \overline{\Omega^{-}} \nu \bar\nu$ 
          & $13.42 b''''\times (0.00407\upsilon''')$ 
          \\                             
\end{tabular}
\end{ruledtabular}
\end{table}

\subsection{$\overline B_q\to\DD' l\bar\nu$ and $\overline B_q\to\DD' \nu\bar\nu$ decay rates}

As shown in Table~\ref{tab: TPDD}, the $\overline B_q\to\DD' l\nu$ decay amplitudes are governed by tree $T_\DD$ and annihilation $A_\DD$ amplitudes,
while the $\overline B_q\to\DD' \nu\bar\nu$ decay amplitudes are governed by penguin-box $PB$ and penguin-box-annihilation $PBA$ amplitudes.
The penguin-box and tree amplitudes are related by a proportional constant $\zeta$, while the penguin-box-annihilation and annihilation amplitudes are related by the same constant, see Eq. (\ref{eq: zeta}).
The $\overline B_q\to\DD' l\nu$ 
decay rates can be parametrized by 5 parameters, namely, $a'''$, $b'''$, $c'''$, $d'''$ and $e'''$, where the first four are contributed from tree and annihilation amplitudes, 
with the following amplitudes $T_\DD$, $T_\DD+A_\DD/6$, $T_\DD+A_\DD/4$, $T_\DD+A_\DD/2$, respectively,
while the last one is only from the annihilation amplitude, $A_\DD$.
The same set of parameters can be used in $\overline B_q\to\DD' \nu\bar\nu$ 
decay rates as the topological amplitudes are proportional to those in $\overline B_q\to\DD' l\nu$  
decays by the common factor, $\zeta$. 
Note that the above parameters correspond to the rates in the SU(3) symmetric limit.

Using the triangle inequality Eq. (\ref{eq: triangle}), we obtain the following relations, 
\be
(\sqrt{a'''}-\sqrt{e'''}/6)^2\lesssim &b'''&\lesssim (\sqrt{a'''}+\sqrt{e'''}/6)^2, 
\non\\
(\sqrt{a'''}-\sqrt{e'''}/4)^2\lesssim &c'''&\lesssim (\sqrt{a'''}+\sqrt{e'''}/4)^2, 
\non\\
(\sqrt{a'''}-\sqrt{e'''}/2)^2\lesssim &d'''&\lesssim (\sqrt{a'''}+\sqrt{e'''}/2)^2.
\en
As in the case of $\overline B_q\to\BB' l\bar\nu$ decays, it is expected that the contributions from annihilation amplitudes to be much suppressed than others. 
Consequently, we should have $e'''\ll a'''$ and the above inequalities lead to the following relations,
\be
a'''\simeq b'''\simeq c'''\simeq d'''\gg e'''.
\label{eq: a b c d e}
\en

As in other 
$\overline B_q\to\bfBB' l\bar\nu (\nu\bar\nu)$ decays, 
threshold enhancements in the differential rates of 
$\overline B_q\to\DD' l\bar\nu$ and $\overline B_q\to\DD' \nu\bar\nu$ decays are anticipated.
They will lead to large SU(3) breaking effect on $\overline B_q\to\DD' l\bar\nu$ and $\overline B_q\to\DD' \nu\bar\nu$ decay rates.
The SU(3) breaking on rates from the threshold enhancement can be estimated as in the $\overline B_q\to \BB' l\bar\nu$ case,
once the differential rate of a $\overline B_q\to\DD' l\bar\nu$ decay mode is measured.
Apparently, no such information is available at this moment. 
We should make use of some model calculations to obtain informations of the differential rates of these modes for illustration. 
As we shall see, in the model calculations 
the SU(3) breaking can be estimated using Eq. (\ref{eq: dBdm}) with $\gamma=10$ employing a similar procedure in the discussion around Eq.~(\ref{eq: sigma}),
and the related parameters $\beta''',\sigma''',\kappa''',\xi''',\upsilon''',\omega'''$ denoting milder SU(3) breaking are expected to be of order 1.

With these considerations the rates of $\overline B_q\to\DD' l\nu$ and $\DD' \nu\bar\nu$ decays are shown 
in Table \ref{tab: BrDD}. 
With parameters $a'''\simeq b'''\simeq c'''\simeq d'''\gg e'''$ and SU(3) breaking parameters $\beta''',\sigma''',\kappa''',\xi''',\upsilon''',\omega'''$ of order 1,
the relative sizes of $\overline B_q\to\DD' l\nu$ and $\DD' \nu\bar\nu$ decay rates can be readily read from the table. 
From Tables \ref{tab: BrDD} and \ref{tab: sensitivity} we note that
$B^- \to \Delta^{++} \overline{\Delta^{++}} l \bar \nu$ decay mode has the largest rate and good detectability. It is also among the least suppressed modes by SU(3) breaking effect from the threshold enhancement even if $\gamma=10$ in Eq. (\ref{eq: dBdm}) is not borne out. 
For $\overline B_q\to\DD' \nu\bar\nu$ modes, $B^- \to  \Sigma^{*+} \overline{\Delta^{++}} \nu \bar \nu$ decay has relatively unsuppressed rate and good detectability.
The above assumption of neglecting annihilation contributions can be checked by searching the $B^- \to\Sigma^{*-} \overline{\Sigma^{*-}} l \bar \nu$ decay mode, which is a pure annihilation mode but with final states of good detectability.

\begin{table}[t!]
\caption{\label{tab: BrDD1} Branching ratios of $\overline B_{q}\to \DD' l \bar \nu$ and $\overline B_{q}\to \DD' \nu \bar \nu$ decays in Model 1 and 2.
$Br_1$ and $Br_2$ denote results in Model 1 and Model 2, respectively. Those with vanishing rates are pure annihilation or penguin-box annihilation modes.
}
\begin{ruledtabular}
\begin{tabular}{cccccccc}
Mode
          & $Br_1(10^{-8})$
          & $Br_2(10^{-8})$
          & Mode
          & $Br_1(10^{-8})$
          & $Br_2(10^{-8})$
          \\
\hline         
$B^- \to \Delta^{++} \overline{\Delta^{++}} l \bar \nu$
          & 16.92 
          & 48.26
          & $B^- \to \Delta^+ \overline{\Delta^+} l \bar \nu$ 
          & 7.52 
          & 21.45
          \\
$B^- \to \Delta^{0} \overline{\Delta^0} l \bar \nu$ 
          & 1.88 
          & 5.36
          & $B^- \to \Delta^- \overline{\Delta^-} l \bar \nu$ 
          & 0 
          & 0
          \\
$B^- \to\Sigma^{*+} \overline{\Sigma^{*+}} l \bar \nu$ 
          & 1.32 
          & 3.83
          & $B^- \to \Sigma^{*0} \overline{\Sigma^{*0}} l \bar \nu$ 
          & 0.33 
          & 0.95
          \\
$B^- \to\Sigma^{*-} \overline{\Sigma^{*-}} l \bar \nu$ 
          & 0 
          & 0        
          & $B^- \to \Xi^{*0} \overline{\Xi^{*0}} l \bar \nu$ 
          & 0.060 
          & 0.18
          \\  
$B^- \to \Xi^{*-} \overline{\Xi^{*-}}  l \bar \nu$ 
          & 0 
          & 0         
          & $B^- \to \Omega^- \overline{\Omega^-} l \bar \nu$ 
          & 0 
          & 0
          \\                                     
\hline $\overline B{}^0 \to \Delta^{++} \overline{\Delta^+} l \bar\nu$
          & 5.23 
          & 14.93
          & $\overline B{}^0 \to  \Delta^+ \overline{\Delta^0}l \bar\nu$ 
          & 6.98 
          & 19.90
          \\
$\overline B{}^0 \to \Delta^0 \overline{\Delta^-} l \bar\nu$ 
          & 5.23 
          & 14.93
          & $\overline B{}^0 \to \Sigma^{*+} \overline{\Sigma^{*0}} l \bar\nu$
          & 0.61 
          & 1.77
          \\
$\overline B{}^0 \to \Sigma^{*0} \overline{\Sigma^{*-}} l \bar\nu$
          &  0.60 
          & 1.72
          & $\overline B{}^0 \to \Xi^{*0} \overline{\Xi^{*-}} l \bar\nu$
          & 0.055 
          & 0.16
          \\         
\hline $\overline B{}^0_s \to \Delta^{++} \overline{\Sigma^{*+}} l \bar \nu$
          & 2.59 
          & 7.45
          & $\overline B{}^0_s \to \Delta^+ \overline{\Sigma^{*0}} l \bar \nu$ 
          & 1.72 
          & 4.94
          \\
$\overline B{}^0_s \to \Delta^0 \overline{\Sigma^{*-}} l \bar \nu$ 
          & 0.84 
          & 2.42
          & $\overline B{}^0_s \to \Sigma^{*+} \overline{\Xi^{*0}} l \bar \nu$ 
          & 0.64 
          & 1.85
          \\
$\overline B{}^0_s \to \Sigma^{*0} \overline{\Xi^{*-}} l \bar \nu$ 
          & 0.31 
          & 0.91
          & $\overline B{}^0_s \to \Xi^{*0} \overline{\Omega^{-}} l \bar \nu$ 
          & 0.093 
          & 0.27
          \\                             
\hline \hline 
Mode
          & $\sum_\nu Br_1 (10^{-10})$
          & $\sum_\nu Br_2 (10^{-10})$
          & Mode
          & $\sum_\nu Br_1 (10^{-10})$
          & $\sum_\nu Br_2 (10^{-10})$
          \\
\hline
$B^- \to  \Sigma^{*+} \overline{\Delta^{++}} \nu \bar \nu$ 
          & 0.92 
          & 2.70
          & $B^- \to \Sigma^{*0} \overline{\Delta^{+}} \nu \bar \nu$ 
          & 0.61 
          & 1.79
          \\
$B^- \to \Sigma^{*-} \overline{\Delta^{0}} \nu \bar \nu$
          & 0.30 
          & 0.88
          & $B^- \to \Xi^{*0} \overline{\Sigma^{*+}} \nu \bar \nu$
          & 0.22 
          & 0.66
          \\
$B^- \to \Xi^{*-} \overline{\Sigma^{*0}} \nu \bar \nu$
          & 0.11 
          & 0.32
          & $B^- \to \Omega^- \overline{\Xi^{*0}} \nu \bar \nu$ 
          & 0.031 
          & 0.095
          \\          
\hline $\overline B{}^0 \to \Sigma^{*+} \overline{\Delta^+} \nu \bar\nu$
          & 0.28 
          & 0.83
          & $\overline B{}^0 \to \Sigma^{*0} \overline{\Delta^0} \nu \bar\nu$ 
          & 0.57 
          & 1.66
          \\
$\overline B{}^0 \to\Sigma^{*-} \overline{\Delta^-} \nu \bar\nu$ 
          & 0.83 
          & 2.44
          & $\overline B{}^0 \to \Xi^{*0} \overline{\Sigma^{*0}} \nu \bar\nu$ 
          & 0.10 
          & 0.30
          \\
$\overline B{}^0 \to \Xi^{*-} \overline{\Sigma^{*-}} \nu \bar\nu$ 
          & 0.20 
          & 0.59
          & $\overline B{}^0 \to \Omega^- \overline{\Xi^{*-}} \nu \bar\nu$ 
          & 0.029 
          & 0.086
          \\         
\hline $\overline B{}^0_s\to \Delta^{++} \overline{\Delta^{++}} \nu \bar \nu$
          & 0 
          & 0
          & $\overline B{}^0_s \to \Delta^+ \overline{\Delta^+} \nu \bar \nu$ 
          & 0 
          & 0
          \\
$\overline B{}^0_s \to \Delta^{0} \overline{\Delta^0} \nu \bar \nu$ 
          & 0 
          & 0
          & $\overline B{}^0_s \to \Delta^- \overline{\Delta^-} \nu \bar \nu$ 
          & 0 
          & 0
          \\
$\overline B{}^0_s \to \Sigma^{*+} \overline{\Sigma^{*+}} \nu \bar\nu$
          & 0.15 
          & 0.43
          & $\overline B{}^0_s \to \Sigma^{*0} \overline{\Sigma^{*0}} \nu \bar\nu$ 
          &  0.14 
          & 0.43
          \\
$\overline B{}^0_s \to \Sigma^{*-} \overline{\Sigma^{*-}} \nu \bar\nu$ 
          & 0.14 
          & 0.41
          & $\overline B{}^0_s \to \Xi^{*0} \overline{\Xi^{*0}} \nu \bar\nu$ 
          & 0.11 
          & 0.33
          \\
$\overline B{}^0_s \to  \Xi^{*-} \overline{\Xi^{*-}} \nu \bar\nu$ 
          & 0.11 
          & 0.32
          & $\overline B{}^0_s \to \Omega^{-} \overline{\Omega^{-}} \nu \bar\nu$ 
          & 0.049 
          & 0.15
          \\                             
\end{tabular}
\end{ruledtabular}
\end{table}

\begin{figure}[t]
\centering
  \includegraphics[width=0.5\textwidth]{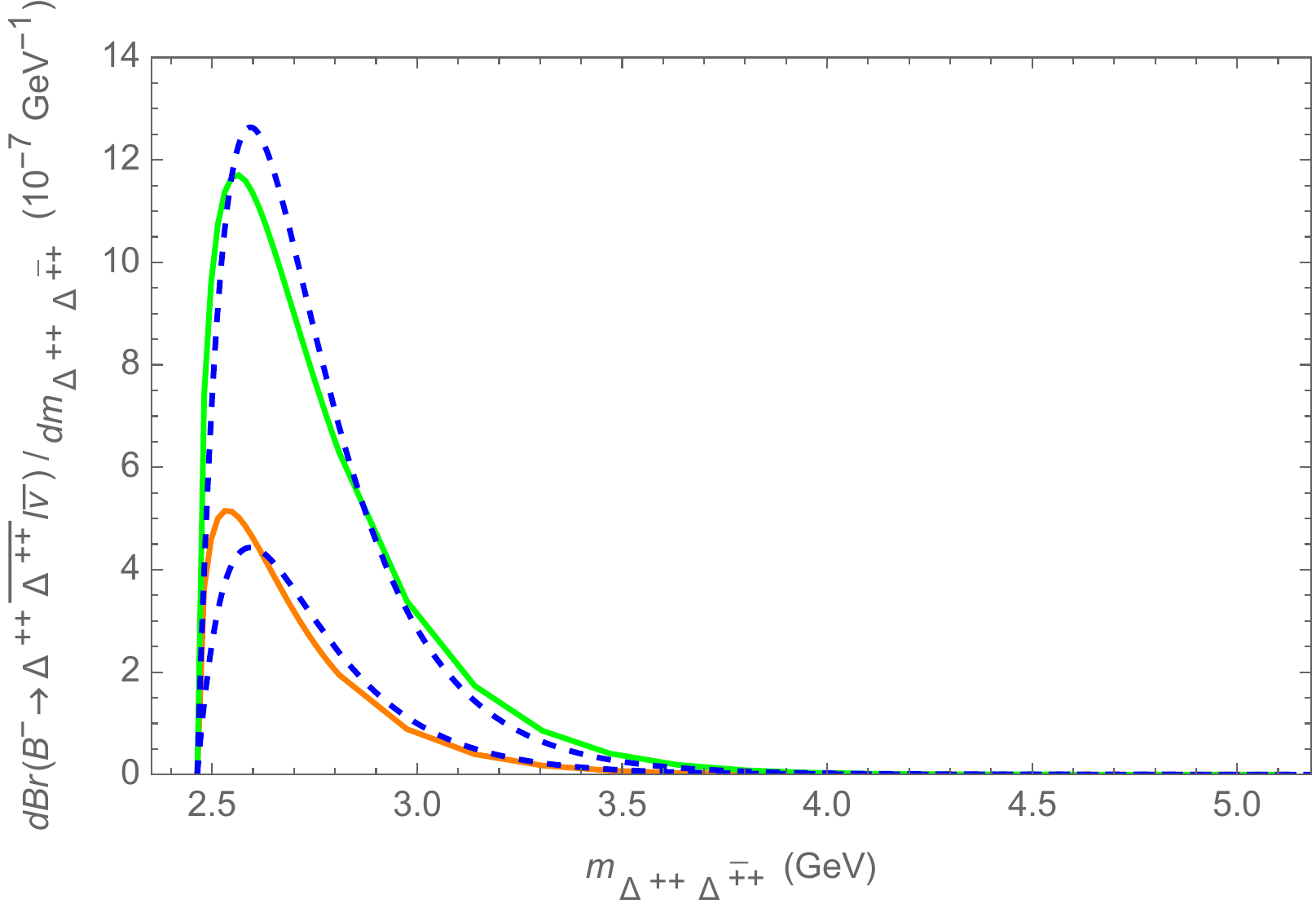}
\caption{
Differential rate $dBr/dm_{\Delta^{++}\overline{\Delta^{++}}}$ of $B^-\to \Delta^{++}\overline{\Delta^{++}} l^-\bar\nu$ decay from Model 1 (orange solid line) and Model 2 (green solid line). 
The dashed lines are $dBr/dm_{\Delta^{++}\overline{\Delta^{++}}}$ using Eq. (\ref{eq: dBdm}) with $\gamma=10$. 
}
 \label{fig: dDDdm}
\end{figure}

\begin{table}[t!]
\caption{\label{tab: alphabeta DD} 
Values of various parameters in Model 1 and Model 2. 
Other parameters can be obtained by using this table and Eq. (\ref{eq: a b c d e}).
}
\begin{ruledtabular}
\begin{tabular}{cccccccc}
Parameters
          & Values (Model 1)
          & Values (Model 2)
          \\
\hline 
$a'''$
          & 0.47
          & 1.34
          \\
$e'''$
          & 0
          & 0
          \\
\hline                    
$\beta'''$
          & 1.43
          & 1.45
          \\  
$\kappa'''$
          & 1.87
          & 1.91
          \\                 
$\sigma'''$
          & 1.40
          & 1.43
          \\                 
$\xi'''$
          & 1.62
          & 1.65
          \\
$\upsilon'''$
          & 1.93
          & 2.03
          \\          
$\omega'''$
          & 2.01
          & 2.06
          \\   
$\bar\beta'''$
          & 1.19
          & 1.22
          \\  
$\bar\kappa'''$
          & 1.51
          & 1.58
          \\                 
$\bar\omega'''$
          & 1.58
          & 1.66
          \\  
\hline
$\bar\beta'''/\beta'''$
          & 0.83
          & 0.85
          \\     
$\bar\kappa'''/\kappa'''$
          & 0.81
          & 0.83
          \\    
$\bar\omega'''/\omega'''$
          & 0.79
          & 0.81
          \\                                                                          
\end{tabular}
\end{ruledtabular}
\end{table}

Using inputs from Sec.~\ref{sec: model calculations} and the formulas given Appendix \ref{App: formulas},
the branching ratios of $\overline B_{q}\to \DD' l \bar \nu$ and $\overline B_{q}\to \DD' \nu \bar \nu$ in Model 1 and 2 are obtained and are shown in Table~\ref{tab: BrDD1}.
These results are new.

From Table~\ref{tab: BrDD1},
we see that the branching ratios of $\overline B_{q}\to \DD' l \bar \nu$ decays of non-annihilation modes are in the ranges of 
$10^{-9}\sim 10^{-7}$ in Model 1 and 2,
while $\sum_\nu Br(\overline B_{q}\to \DD' l \nu \nu)$ are in the ranges of 
$10^{-12}\sim 10^{-10}$ and $10^{-11}\sim 10^{-10}$ in Model 1 and 2, respectively.
The rates in Model 2 are greater than those in Model 1 by a factor of $3$. 
This corresponds to the fact that $|T_\DD|$ in Model 2 is greater than one the in Model 1
as reflected through the sizes of $a'''$ in these two models as shown in Table~\ref{tab: alphabeta DD}.
This is not surprising as we noted previously, the sizes of topological amplitudes $|T_{1\BB}|$, $|T_{2\BB}|$
$|T_{\BD}|$ and $|T_\DB|$ in Model 2
are greater than those in Model 1. 

The differential rates $dBr/dm_{\Delta^{++}\overline{\Delta^{++}}}$ of $B^-\to \Delta^{++}\overline{\Delta^{++}} l^-\bar\nu$ decay from Model 1 and Model 2 are shown in Fig.~\ref{fig: dDDdm}, they can be compared to  
the one plotted using Eq. (\ref{eq: dBdm}) with $\gamma=10$.
They clearly exhibit threshold enhancement as in other $\overline B_{q}\to \bfBB' l \bar \nu$ decays.

From Table~\ref{tab: alphabeta DD}, we see that the parameters $\beta^{\prime\prime\prime}, \kappa^{\prime\prime\prime}, \sigma^{\prime\prime\prime}, \xi^{\prime\prime\prime}$, $\upsilon'''$, $\omega'$, $\bar\beta^{\prime\prime\prime}, \bar\kappa^{\prime\prime\prime}$ and $\bar\omega'''$ denoting milder SU(3) breaking are indeed of order one and are similar in Model 1 and 2.
Furthermore,
$\bar\beta^{\prime\prime\prime}/\beta^{\prime\prime\prime}$, $\bar\kappa'''/\kappa'''$
and $\bar\omega'''/\omega'''$ are close to one as expected.

By taking into account the sensitivity of detection, see Table \ref{tab: sensitivity}, and decay rates, see Tables \ref{tab: BrDD} and \ref{tab: BrDD1},
we find that the following decay modes have relatively unsuppressed rates and good detectability, they are
$B^- \to \Delta^{++} \overline{\Delta^{++}} l \bar \nu$,
$B^- \to \Delta^{0} \overline{\Delta^0} l \bar \nu$,
$B^- \to\Sigma^{*+} \overline{\Sigma^{*+}} l \bar \nu$,
$\overline B{}^0_s \to \Delta^{++} \overline{\Sigma^{*+}} l \bar \nu$,
$\overline B{}^0_s \to \Delta^0 \overline{\Sigma^{*-}} l \bar \nu$,
$\overline B{}^0_s \to \Sigma^{*+} \overline{\Xi^{*0}} l \bar \nu$, 
$B^- \to  \Sigma^{*+} \overline{\Delta^{++}} \nu \bar \nu$,
$B^- \to \Sigma^{*-} \overline{\Delta^{0}} \nu \bar \nu$,
$B^- \to \Xi^{*0} \overline{\Sigma^{*+}} \nu \bar \nu$,
$\overline B{}^0_s \to \Sigma^{*+} \overline{\Sigma^{*+}} \nu \bar\nu$,
$\overline B{}^0_s \to \Sigma^{*-} \overline{\Sigma^{*-}} \nu \bar\nu$,
$\overline B{}^0_s \to \Xi^{*0} \overline{\Xi^{*0}} \nu \bar\nu$
decays.

Ratios of rates from loop induced modes and tree induced modes are sensible test of the SM.
From Table \ref{tab: BrDD}, we obtain
\be
\frac{\sum_\nu Br(\overline B{}^0_s \to \Sigma^{*+} \overline{\Sigma^{*+}} \nu \bar\nu)}{Br(B^- \to \Delta^{0} \overline{\Delta^0} l \bar \nu)}
&=& 4.66 \sigma''' \times\bigg(\frac{0.0036}{|V_{ub}|}\bigg)^2 \times10^{-4},
\non\\
\frac{\sum_\nu Br(\overline B{}^0_s \to \Sigma^{*-} \overline{\Sigma^{*-}} \nu \bar\nu)}{Br(B^- \to \Delta^{0} \overline{\Delta^0} l \bar \nu)}
&=& 4.40 \sigma''' \times\bigg(\frac{0.0036}{|V_{ub}|}\bigg)^2 \times10^{-4},
\label{eq: ratio DD 1}
\en
and
\be
\frac{\sum_\nu Br(B^- \to  \Sigma^{*+} \overline{\Delta^{++}} \nu \bar \nu)}{Br(\overline B{}^0_s \to \Delta^{++} \overline{\Sigma^{*+}} l \bar \nu)}
&=& 4.29 \frac{\bar\beta'''}{\beta'''} \times\bigg(\frac{0.0036}{|V_{ub}|}\bigg)^2 \times10^{-3},
\non\\
\frac{\sum_\nu Br(B^- \to \Sigma^{*-} \overline{\Delta^{0}} \nu \bar \nu)}{Br(\overline B{}^0_s \to \Delta^0 \overline{\Sigma^{*-}} l \bar \nu)}
&=& 4.29 \frac{\bar\beta'''}{\beta'''} \times\bigg(\frac{0.0036}{|V_{ub}|}\bigg)^2 \times10^{-3},
\non\\
\frac{\sum_\nu Br(B^- \to \Xi^{*0} \overline{\Sigma^{*+}} \nu \bar \nu)}{Br(\overline B{}^0_s \to \Sigma^{*+} \overline{\Xi^{*0}} l \bar \nu)}
&=& 4.29 \frac{\bar\kappa'''}{\kappa'''} \times\bigg(\frac{0.0036}{|V_{ub}|}\bigg)^2 \times10^{-3},
\label{eq: ratio DD 2}
\en
where $\sigma'''$ is expected to be of order one, while $\bar\beta'''/\beta'''$ and $\bar\kappa'''/\kappa'''$ are expected to be close to one. In fact as shown in Table \ref{tab: alphabeta DD}, we have $\sigma'''=1.40 (1.43)$, $\bar\beta'''/\beta'''=0.83 (0.85)$ and $\bar\kappa'''/\kappa'''=0.81 (0.83)$ in Model 1 (2), which are indeed agree with the above expectations.
Note that the ratios in Eqs. (\ref{eq: ratio DD 1}) and (\ref{eq: ratio DD 2}) do not involve the small $e'''$ assumption, and the ratios in Eq. (\ref{eq: ratio DD 2}) are less sensitive to the SU(3) breaking from threshold enhancement.
These ratios can be checked experimentally.

\section{Discussions and Conclusion}

We study the decay amplitudes and rates of $\overline B_q\to {{\rm\bf B}\overline{\rm\bf B}}' l \bar\nu$ and $\overline B_q\to {{\rm\bf B}\overline{\rm\bf B}}' \nu \bar\nu$ decays with all low lying octet $({\cal B})$ and decuplet $({\cal D})$ baryons using a topological amplitude approach. 
The decay amplitudes are decomposed into combinations of topological amplitudes. 
In $\overline B_q\to\BB' l\bar\nu$ decays we need three topological amplitudes, 
namely two tree amplitudes, $T_{2\BB}$, $T_{1\BB}$ and one annihilation amplitude, $A_\BB$.
In $\overline B_q\to\BD l\bar\nu$ decays only one tree amplitude, $T_{\BD}$, is needed.
Likewise in $\overline B_q\to\DB l\bar\nu$ decays, we only need one tree amplitude, $T_{\DB}$.
Lastly in $\overline B_q\to\DD' l\bar\nu$ decays, two topological amplitudes, namely a tree amplitude, $T_{\DD}$, and an annihilation amplitude, $A_\DD$, are needed.
In loop induced decay modes,
we have three topological amplitudes, namely two penguin-box amplitudes $PB_{2\BB}$, $PB_{1\BB}$ and one penguin-box-annihilation amplitude, $PBA_\BB$, in $\overline B_q\to\BB' \nu\bar\nu$ decays,
one topological amplitude, namely a penguin-box amplitude, $PB_{\BD}$, in $\overline B_q\to\BD \nu\bar\nu$ decays,
one topological amplitude, namely a penguin-box amplitude, $PB_{\DB}$, in $\overline B_q\to\DB \nu\bar\nu$ decays,
two topological amplitudes, namely a penguin-box amplitude, $PB_{\DD}$, and a penguin-box-annihilation amplitude, $PBA_\DD$, in $\overline B_q\to\DD' \nu\bar\nu$ decays.
As the numbers of independent topological amplitudes are highly limited,
there are plenty of relations on these $\overline B_q\to\bfBB'l\bar \nu$ and $\bfBB' \nu\bar\nu$ decay amplitudes.
Furthermore, the loop topological amplitudes and tree topological amplitudes have simple relations, as their ratios are determined by the CKM factors and loop functions.
 
 It is known that the $B^-\to p\bar p \mu^-\bar\nu$ differential rate exhibits threshold enhancement, which is expected to hold in all other $\overline B_{q}\to \bfBB' l \bar \nu (\nu\bar\nu)$ decay modes.
These $B_q$ decays have large phase space and one normally does expect the SU(3) breaking in baryon masses to have large SU(3) breaking effects on the $B_q$ decay rates. 
However, the threshold enhancement effectively squeezes the phase space to the threshold region and thus mimics the decay just above threshold situation.
It amplifies the effects of SU(3) breaking in final state baryon masses, 
consequently, 
the decay rates may differ by orders of magnitudes even if their amplitudes are of similar sizes.
In this work, the $B^-\to p\bar p \mu^-\bar\nu$ differential rate and model calculations with available theoretical inputs from ref. \cite{Geng:2021sdl, Hsiao:2022uzx}, which can reproduce the observed differential rate, are used to estimate the SU(3) breaking from threshold enhancement. 
We find that the differential rates $dBr/dm_{\bfBB'}$ of $\overline B_{q}\to \bfBB' l \bar \nu (\nu\bar\nu)$ decays can be parametrized as 
\be
\frac{dBr}{dm_{\bfBB'}}=\frac{N}{(m^2_{\bfBB'})^\gamma} (m_{\bfBB'}-m_{\bf B}-m_{\overline {\bf B}'}),
\en
with $\gamma$ and $N$ some constants, and we obtain
$\gamma=9, 7, 7, 10$ for $\overline B_{q}\to \BB' l \bar \nu (\nu\bar\nu)$, $\overline B_{q}\to \BD l \bar \nu (\nu\bar\nu)$, $\overline B_{q}\to \DB l \bar \nu (\nu\bar\nu)$ and $\overline B_{q}\to \DD' l \bar \nu (\nu\bar\nu)$ decays, respectively.
SU(3) breaking from threshold enhancement can be estimated using the above equation. 
The estimations on SU(3) breaking from threshold enhancement are supported by model calculations and can be improved once differential rates of other modes are measured.

Note that as shown in Figs. \ref{fig: dBBdm0} and \ref{fig: dBBdm}, although $\gamma=9$ agrees with $dBr/dm_{p\bar p}$ of $B^-\to p\bar p \mu^-\bar\nu$ decay from LHCb~\cite{LHCb:2019cgl} and the theoretical calculations using inputs from refs~\cite{Geng:2021sdl} and \cite{ Hsiao:2022uzx}, which made use of QCD counting rules,
there are only four data points with non-negligible uncertainties in the plots. Therefore the reliability of the value of $\gamma$ remains to be checked when more data is available. 
In $\overline B_{q}\to \BD l \bar \nu (\nu\bar\nu)$, $\overline B_{q}\to \DB l \bar \nu (\nu\bar\nu)$ and $\overline B_{q}\to \DD' l \bar \nu (\nu\bar\nu)$ decays, the values of $\gamma$ are determined by comparing to numerical results of the theoretical calculations. It should be noted that there are some assumptions employed in the theoretical calculations. Therefore these $\gamma$ should be taken as illustrations for the moment.
The estimations on SU(3) breaking from threshold enhancement can be improved once differential rates of these modes are measured.
In particular, in $\overline B_q\to \BD l\bar\nu\, (\DB l\bar\nu)$ decays, as there is only one topological amplitude, namely $T_\BD$ ($T_\DB$), the rates of all other modes [including $\overline B_q\to \BD \nu\bar\nu\, (\DB \nu\bar\nu)$ modes] can be estimated without resorting to model calculations, once the total rate and the differential rate of a single mode is measured. 
The same situation also applies to $\overline B_{q}\to \DD' l \bar \nu (\nu\bar\nu)$ decays as long as the decuplet and anti-decuplet baryons are not related by charge conjugation, as there is only one topological amplitude, namely $T_\DD$ ($PB_\DD$) in these decays. 
It is therefore interesting to see the experimental results in these modes.

In the model calculations,
we find that the $\overline B_{q}\to \BB' l \bar \nu$ branching ratios are of the orders $10^{-8}\sim 10^{-6}$ for non-annihilation modes, while the branching ratios of $\overline B_{q}\to \BB' \nu \bar \nu$ decays are of the orders of $10^{-11}\sim 10^{-8}$ for non-penguin-box-annihilation modes.
The branching ratios of $\overline B_{q}\to \BD l \bar \nu$ and $\overline B_{q}\to \DB l \bar \nu$ decays are in the ranges of $10^{-9}\sim 10^{-7}$
while $\sum_\nu Br(\overline B_{q}\to \BD l \nu \nu)$ and $\sum_\nu Br(\overline B_{q}\to \DB l \nu \nu)$ are in the ranges of 
$10^{-12}\sim 10^{-10}$.
The branching ratios of $\overline B_{q}\to \DD' l \bar \nu$ decays of non-annihilation modes are in the ranges of 
$10^{-9}\sim 10^{-7}$,
while $\sum_\nu Br(\overline B_{q}\to \DD' l \nu \nu)$ are in the ranges of 
$10^{-12}\sim 10^{-10}$.

Modes with relatively unsuppressed rates and good detectability are identified as following. 
In $\overline B_q\to\BB' l\bar\nu$ and $\overline B_q\to\BB' \nu\bar\nu$ decays, we have
$B^-\to p \bar p l \bar \nu$,
$\overline B^0\to p \bar n l \bar \nu$,
$\overline B{}^0_s \to p \overline{\Lambda} l \bar \nu$,
$B^-\to \Lambda \bar{p} \nu \bar \nu$,
$\overline B^0\to \Lambda \bar{n} \nu \bar \nu$
and
$\overline B{}^0_s \to \Lambda \overline{\Lambda} \nu \bar \nu$ decays.
In $\overline B_q\to\BD l\bar\nu$ and $\overline B_q\to\BD \nu\bar\nu$ decays,
$\overline B{}^0 \to p \overline{\Delta^0} l \bar\nu$ and
$\overline B{}^0\to \Sigma^0 \overline{\Delta^0} \nu \bar \nu$ 
have unsuppressed rates and good detectability. 
While in $\overline B_q\to\DB l\bar\nu$ and $\overline B_q\to\DB \nu\bar\nu$ decays,
$\overline B{}^0\to \Delta^{++} \bar p l \bar\nu$,
$\overline B{}^0\to \Sigma^{*+} \overline{\Lambda} l \bar\nu$, 
$B^-\to \Delta^+\bar p l\bar \nu$, 
$B^-\to \Delta^0\bar n l\bar \nu$, 
$\overline B{}^0 \to \Sigma^{*+} \bar p \nu \bar\nu$
and
$B^- \to \Sigma^{*-} \bar n \nu \bar\nu$ decay modes are identified. 
Finally in $\overline B_q\to\DD' l\bar\nu$ and $\overline B_q\to\DD' \nu\bar\nu$ decays,
we find that the following decay modes have unsuppressed rates and good detectability,
they are $B^- \to \Delta^{++} \overline{\Delta^{++}} l \bar \nu$,
$B^- \to \Delta^{0} \overline{\Delta^0} l \bar \nu$,
$B^- \to\Sigma^{*+} \overline{\Sigma^{*+}} l \bar \nu$,
$\overline B{}^0_s \to \Delta^{++} \overline{\Sigma^{*+}} l \bar \nu$,
$\overline B{}^0_s \to \Delta^0 \overline{\Sigma^{*-}} l \bar \nu$,
$\overline B{}^0_s \to \Sigma^{*+} \overline{\Xi^{*0}} l \bar \nu$, 
$B^- \to  \Sigma^{*+} \overline{\Delta^{++}} \nu \bar \nu$,
$B^- \to \Sigma^{*-} \overline{\Delta^{0}} \nu \bar \nu$,
$B^- \to \Xi^{*0} \overline{\Sigma^{*+}} \nu \bar \nu$,
$\overline B{}^0_s \to \Sigma^{*+} \overline{\Sigma^{*+}} \nu \bar\nu$,
$\overline B{}^0_s \to \Sigma^{*-} \overline{\Sigma^{*-}} \nu \bar\nu$,
$\overline B{}^0_s \to \Xi^{*0} \overline{\Xi^{*0}} \nu \bar\nu$ decays.
These modes can be searched experimentally in near future.

Ratios of rates of some loop induced $\overline B_q\to {{\rm\bf B}\overline{\rm\bf B}}' \nu \bar\nu$ decays and tree induced $\overline B_q\to {{\rm\bf B}\overline{\rm\bf B}}' l \bar\nu$ decays are predicted and can be checked experimentally. They can be tests of the SM.  
In particular, we predict
\be
\frac{\sum_\nu Br(B^-\to \Lambda \bar{p} \nu \bar \nu)}{Br(\overline B{}^0_s \to p \overline{\Lambda} l \bar \nu)}
&=& 4.29\frac{\bar \alpha}{\alpha}\times\bigg(\frac{0.0036}{|V_{ub}|}\bigg)^2 \times10^{-3}, 
\non\\
\frac{\sum_\nu Br(\overline B^0\to \Lambda \bar{n} \nu \bar \nu)}{Br(\overline B{}^0_s \to p \overline{\Lambda} l \bar \nu)}
&=& 3.94 \frac{\bar \alpha}{\alpha}\times \bigg(\frac{0.0036}{|V_{ub}|}\bigg)^2 \times10^{-3},
\label{eq: ratios BB1}
\en
for $\overline B_q\to \BB' \nu\bar\nu, \BB' l\bar\nu$ decays,
\be
\frac{\sum_\nu Br(\overline B{}^0\to \Sigma^0 \overline{\Delta^0} \nu \bar \nu)}{Br(\overline B{}^0 \to p \overline{\Delta^0} l \bar\nu)}
&=& 2.11 \beta' \times\bigg(\frac{0.0036}{|V_{ub}|}\bigg)^2 \times10^{-3},
\label{eq: ratios BD1}
\en
for $\overline B_q\to \BD \nu\bar\nu, \BD l\bar\nu$ decays,
\be
\frac{\sum_\nu Br(\overline B{}^0 \to \Sigma^{*+} \bar p \nu \bar\nu)}{Br(\overline B{}^0\to \Delta^{++} \bar p l \bar\nu)}
&=& 4.55 \kappa'' \times\bigg(\frac{0.0036}{|V_{ub}|}\bigg)^2 \times10^{-4},
\label{eq: ratios DB1}
\en
for $\overline B_q\to \DB \nu\bar\nu, \DB l\bar\nu$ decays, and
\be
\frac{\sum_\nu Br(\overline B{}^0_s \to \Sigma^{*+} \overline{\Sigma^{*+}} \nu \bar\nu)}{Br(B^- \to \Delta^{0} \overline{\Delta^0} l \bar \nu)}
&=& 4.66 \sigma''' \times\bigg(\frac{0.0036}{|V_{ub}|}\bigg)^2 \times10^{-4},
\non\\
\frac{\sum_\nu Br(\overline B{}^0_s \to \Sigma^{*-} \overline{\Sigma^{*-}} \nu \bar\nu)}{Br(B^- \to \Delta^{0} \overline{\Delta^0} l \bar \nu)}
&=& 4.40 \sigma''' \times\bigg(\frac{0.0036}{|V_{ub}|}\bigg)^2 \times10^{-4},
\label{eq: ratios DD1}
\en
and
\be
\frac{\sum_\nu Br(B^- \to  \Sigma^{*+} \overline{\Delta^{++}} \nu \bar \nu)}{Br(\overline B{}^0_s \to \Delta^{++} \overline{\Sigma^{*+}} l \bar \nu)}
&=& 4.29 \frac{\bar\beta'''}{\beta'''} \times\bigg(\frac{0.0036}{|V_{ub}|}\bigg)^2 \times10^{-3},
\non\\
\frac{\sum_\nu Br(B^- \to \Sigma^{*-} \overline{\Delta^{0}} \nu \bar \nu)}{Br(\overline B{}^0_s \to \Delta^0 \overline{\Sigma^{*-}} l \bar \nu)}
&=& 4.29 \frac{\bar\beta'''}{\beta'''} \times\bigg(\frac{0.0036}{|V_{ub}|}\bigg)^2 \times10^{-3},
\non\\
\frac{\sum_\nu Br(B^- \to \Xi^{*0} \overline{\Sigma^{*+}} \nu \bar \nu)}{Br(\overline B{}^0_s \to \Sigma^{*+} \overline{\Xi^{*0}} l \bar \nu)}
&=& 4.29 \frac{\bar\kappa'''}{\kappa'''} \times\bigg(\frac{0.0036}{|V_{ub}|}\bigg)^2 \times10^{-3},
\label{eq: ratios DD2}
\en
for $\overline B_q\to \DD' \nu\bar\nu, \DD' l\bar\nu$ decays.
The parameters 
$\beta'$, $\kappa''$ and $\sigma'''$ are expected to be of order one,
while the ratios $\bar\alpha/\alpha$, $\bar\beta'''/\beta'''$ and $\bar\kappa'''/\kappa'''$ are expected to be close to one.
These expectations are supported by model calculations.
Note that the ratios in Eqs. (\ref{eq: ratios BB1}) and (\ref{eq: ratios DD2}) are insensitive to the SU(3) breaking from threshold enhancement,
while those in Eqs. (\ref{eq: ratios BD1}), (\ref{eq: ratios DB1}) and (\ref{eq: ratios DD1}) do depend on the estimations of SU(3) breaking from threshold enhancement, which, however, can be checked and improved when more modes are discovered.
The ratios which are insensitive to the modeling of SU(3) breaking from threshold enhancement can be tests of the SM.

The approach developed in this work can be applied to some other related modes. In particular, $\overline B_{q}\to \bfBB' l^+ l^-$ decays can be studied using a
similar method. Given the fact that the final states have good detectability, 
these are interesting modes to be studied~\cite{referee}. 
Further investigation is needed as the governing operators in $H_{\rm eff}$, see for example \cite{Altmannshofer:2008dz}, are more complicated, 
where they have structures beyond the simple $V-A$ form considered in this work [see Eq. (\ref{eq: AA})], and the (differential) decay rates can be rather different.
Nevertheless as long as the SU(3) flavor structure of the amplitudes is concerned, 
the decompositions of $\overline B_{q}\to \bfBB' l^+ l^-$ decay amplitudes are identical to those in $\overline B_{q}\to \bfBB' \nu \bar \nu$ amplitudes presented in 
Tables~\ref{tab: TPBB}, \ref{tab: TPBD}, \ref{tab: TPDB} and \ref{tab: TPDD}.
Relations of $\overline B_{q}\to \bfBB' \nu \bar \nu$ amplitudes as shown in Sec.~\ref{sec: relations} are applicable to $\overline B_{q}\to \bfBB' l^+ l^-$ amplitudes.
Consequently, although the absolute sizes of decay rates and the shapes of differential rates need further investigation, 
the relative decay rates of $\overline B_{q}\to \bfBB' l^+ l^-$ decays can be estimated using the results given in this work, especially for modes with simple topological structures, such as $\overline B_{q}\to \BD' \nu \bar \nu$, $\overline B_{q}\to \DB' \nu \bar \nu$ decays and some $\overline B_{q}\to \DD' \nu \bar \nu$ decay modes.
By na\"ively  using the results on $\overline B_{q}\to \bfBB' \nu \bar \nu$ decay rates in Tables~\ref{tab: BrBD}, \ref{tab: BrDB} and \ref{tab: BrDD}, it is expected that the following 
$\overline B_{q}\to \bfBB' l^+ l^-$ modes should have relatively unsuppressed rates and good detectability.
These modes are
$\overline B{}^0\to \Sigma^0 \overline{\Delta^0} l^+ l^-$, 
$\overline B{}^0 \to \Sigma^{*+} \bar p l^+ l^-$,
$B^- \to  \Sigma^{*+} \overline{\Delta^{++}} l^+ l^-$,
$B^- \to \Sigma^{*-} \overline{\Delta^{0}} l^+ l^-$,
and $B^- \to \Xi^{*0} \overline{\Sigma^{*+}} l^+ l^-$
decays.
In addition to the above modes, although having much complicated topological structure, the $\overline B_{q}\to \BB' l^+ l^-$ decay modes should also be searched, especially the $B^-\to \Lambda \bar{p} l^+ l^-$ decay, which has good detectability.  
It will be interesting to search for them.

\begin{acknowledgments}
The author would like to thank Yu-Kuo Hsiao for discussion.
This work is supported in part by
the National Science and Technology Council of R.O.C.
under Grant Nos. NSTC-111-2112-M-033-007 and NSTC-112-2112-M-033 -006.
\end{acknowledgments}

\appendix

\section{$\overline B_q\to {\mathbf B}\bar {\mathbf B}^\prime $ matrix elements in the asymptotic limit}\label{App: asym}

We discuss $\overline B_q\to {\mathbf B}\bar {\mathbf B}^\prime $ transition matrix elements in the asymptotic limit in this appendix.
We follow ref.~\cite{Brodsky:1980sx} to obtain the asymptotic limit of these matrix elements.
The wave function of a octet or decuplet baryon with helicity $\lambda=-1/2$ can be expressed as 
\begin{equation}
|{\mathbf B}\,;\downarrow\rangle\sim
\frac{1}{\sqrt3}(|{\mathbf B}\,;\downarrow\uparrow\downarrow\rangle
                +|{\mathbf B}\,;\downarrow\downarrow\uparrow\rangle
                +|{\mathbf B}\,;\uparrow\downarrow\downarrow\rangle), 
\end{equation}
which are composed of 13-, 12- and 23-symmetric terms,
respectively.
For octet baryons, 
we have
\begin{eqnarray}
|p\,;\downarrow\uparrow\downarrow\rangle&=&
\left[\frac{d(1)u(3)+u(1)d(3)}{\sqrt6} u(2)
 -\sqrt{\frac{2}{3}} u(1)d(2)u(3)\right]
|\downarrow\uparrow\downarrow\rangle,
\nonumber\\
|n\,;\downarrow\uparrow\downarrow\rangle&=&
(-|p\,;\downarrow\uparrow\downarrow\rangle
\,\,{\rm with}\,\,\,u \leftrightarrow d),
\nonumber\\
|\Sigma^+\,;\downarrow\uparrow\downarrow\rangle&=&
(-|p\,;\downarrow\uparrow\downarrow\rangle
\,\,{\rm with}\,\,d \rightarrow s),
\nonumber\\
|\Sigma^0\,;\downarrow\uparrow\downarrow\rangle&=&
\bigg[-\frac{u(1)d(3)+d(1)u(3)}{\sqrt3}\,s(2)
      +\frac{u(2)d(3)+d(2)u(3)}{2\sqrt3}\,s(1)
\nonumber\\
      &&\,\,+\frac{u(1)d(2)+d(1)u(2)}{2\sqrt3}\,s(3)\bigg]
|\downarrow\uparrow\downarrow\rangle,
\nonumber\\
|\Sigma^-\,;\downarrow\uparrow\downarrow\rangle&=&
(|p\,;\downarrow\uparrow\downarrow\rangle
\,\,{\rm with}\,\,u, d\rightarrow d,s),
\nonumber\\
|\Lambda\,;\downarrow\uparrow\downarrow\rangle&=&
\bigg[\frac{d(2)u(3)-u(2)d(3)}{2}\,s(1)
      +\frac{u(1)d(2)-d(1)u(2)}{2}\,s(3)\bigg]
|\downarrow\uparrow\downarrow\rangle,
\non\\
|\Xi^0\,;\downarrow\uparrow\downarrow\rangle&=&
(|p\,;\downarrow\uparrow\downarrow\rangle
\,\,{\rm with}\,\,u, d\rightarrow s, u),
\nonumber\\
|\Xi^-\,;\downarrow\uparrow\downarrow\rangle&=&
(-|p\,;\downarrow\uparrow\downarrow\rangle
\,\,{\rm with}\,\,u\rightarrow s),
\en
and for decuplet baryons, we have
\be
|\Delta^{++};\downarrow\uparrow\downarrow\rangle&=&u(1)u(2)u(3)|\downarrow\uparrow\downarrow\rangle,\qquad\qquad
|\Delta^{-};\downarrow\uparrow\downarrow\rangle=d(1)d(2)d(3)|\downarrow\uparrow\downarrow\rangle,
\nonumber\\
|\Delta^{+};\downarrow\uparrow\downarrow\rangle&=&
\frac{1}{\sqrt3}[u(1)u(2)d(3)+u(1)d(2)u(3)+d(1)u(2)u(3)]|\downarrow\uparrow\downarrow\rangle,
\nonumber\\
|\Delta^{0};\downarrow\uparrow\downarrow\rangle&=&(|\Delta^{+};\downarrow\uparrow\downarrow\rangle\,\,{\rm
with}\,\,u \leftrightarrow d),\qquad
|\Sigma^{*+};\downarrow\uparrow\downarrow\rangle=(|\Delta^{+};\downarrow\uparrow\downarrow\rangle\,\,{\rm
with}\,\,d \leftrightarrow s),
\nonumber\\
|\Sigma^{*0};\downarrow\uparrow\downarrow\rangle&=&\frac{1}{\sqrt6}[u(1)d(2)s(3)+{\rm
permutation}]|\downarrow\uparrow\downarrow\rangle,
\non\\
|\Omega^-;\downarrow\uparrow\downarrow\rangle&=&(|\Delta^{++};\downarrow\uparrow\downarrow\rangle
\,\,{\rm with}\,\,u\rightarrow s),
\end{eqnarray}
for the $|{\mathbf B}\,;\downarrow\uparrow\downarrow\rangle$ parts.
while the 12- and 23-symmetric parts can be easily obtained by suitable permutation.

\begin{table}[t!]
\caption{\label{tab: eBB} The coefficients $(e_{\parallel}, e_{\overline\parallel}, e_F)$ for various $\la {\mathbf B} \bar{\mathbf B}'|\bar q_L \gamma_\mu b_L|\bar B_{q'}\ra$ matrix elements.
}
\begin{ruledtabular}
\begin{tabular}{cccccccc}
$\la {\mathbf B} \bar{\mathbf B}'|\bar q_L \gamma_\mu b_L|\bar B_{q'}\ra$
          & $(e_{\parallel}, e_{\overline\parallel}, e_F)$
          & $\la {\mathbf B} \bar{\mathbf B}'|\bar q_L \gamma_\mu b_L|\bar B_{q'}\ra$
          & $(e_{\parallel}, e_{\overline\parallel}, e_F)$
          \\
\hline $\la p \bar p|\bar u_L \gamma_\mu b_L|B^-\ra$
          & $(5,1,-2)$
          & $\la n \bar n|\bar u_L \gamma_\mu b_L|B^-\ra$
          & $(1,2,\frac{1}{2})$
          \\
$\la \Sigma^+ \overline{\Sigma^+} |\bar u_L \gamma_\mu b_L|B^-\ra$
          & $(5,1,-2)$
          & $\la \Sigma^0 \overline{\Sigma^0} |\bar u_L \gamma_\mu b_L|B^-\ra$
          & $(\frac{5}{2},\frac{1}{2},-1)$
          \\
$\la \Sigma^0 \overline{\Lambda} |\bar u_L \gamma_\mu b_L|B^-\ra$
          & $(\frac{\sqrt{3}}{2},-\frac{\sqrt{3}}{2},-\frac{\sqrt{3}}{2})$
          & $\la \Xi^0 \overline{\Xi^0} |\bar u_L \gamma_\mu b_L|B^-\ra$
          & $(1,2,\frac{1}{2})$
          \\
$\la \Lambda \overline{\Sigma^0} |\bar u_L \gamma_\mu b_L|B^-\ra$
          & $(\frac{\sqrt{3}}{2},-\frac{\sqrt{3}}{2},-\frac{\sqrt{3}}{2})$
          & $\la \Lambda \overline{\Lambda} |\bar u_L \gamma_\mu b_L|B^-\ra$
          & $(\frac{3}{2},\frac{3}{2},0)$
          \\          
\hline $\la p \bar n|\bar u_L \gamma_\mu b_L|\overline B{}^0\ra$
          & $(4,-1,-\frac{5}{2})$
          & $\la \Sigma^+ \overline{ \Sigma^0}|\bar u_L \gamma_\mu b_L|\overline B{}^0\ra$
          & $(-\frac{5}{\sqrt 2}, -\frac{1}{\sqrt 2},\sqrt2)$
          \\
$\la \Sigma^+ \overline{\Lambda} |\bar u_L \gamma_\mu b_L|\overline B{}^0\ra$
          & $(\sqrt{\frac{3}{2}}, -\sqrt{\frac{3}{2}},-\sqrt{\frac{3}{2}})$
          & $\la \Sigma^0 \overline{\Sigma^-} |\bar u_L \gamma_\mu b_L|\overline B{}^0\ra$
          & $(\frac{5}{\sqrt 2}, \frac{1}{\sqrt 2},-\sqrt2)$
          \\
$\la \Lambda \overline{\Sigma^-} |\bar u_L \gamma_\mu b_L|\overline B{}^0\ra$
          & $(\sqrt{\frac{3}{2}}, -\sqrt{\frac{3}{2}},-\sqrt{\frac{3}{2}})$
          & $\la \Xi^0 \overline{\Xi^-} |\bar u_L \gamma_\mu b_L|\overline B{}^0\ra$
          & $(-1,-2,-\frac{1}{2})$
          \\         
\hline $\la p \overline{\Sigma^0}|\bar u_L \gamma_\mu b_L|\overline B{}^0_s\ra$
          & $(-\frac{1}{\sqrt2},-\sqrt2,-\frac{1}{2\sqrt2})$
          & $\la p \overline{\Lambda}|\bar u_L \gamma_\mu b_L|\overline B{}^0_s\ra$
          & $(-3\sqrt{\frac{3}{2}}, 0, \frac{3}{2}\sqrt{\frac{3}{2}})$
          \\
$\la n \overline{\Sigma^-} |\bar u_L \gamma_\mu b_L|\overline B{}^0_s\ra$
          & $(-1, -2,-\frac{1}{2})$
          & $\la \Sigma^+ \overline{\Xi^0} |\bar u_L \gamma_\mu b_L|\overline B{}^0_s\ra$
          & $(4, -1,-\frac{5}{2})$
          \\
$\la \Sigma^0 \overline{\Xi^-} |\bar u_L \gamma_\mu b_L|\overline B{}^0_s\ra$
          & $(2\sqrt{2}, -\frac{1}{\sqrt 2},-\frac{5}{2\sqrt2})$
          & $\la \Lambda \overline{\Xi^-} |\bar u_L \gamma_\mu b_L|\overline B{}^0_s\ra$
          & $(\sqrt 6, \sqrt{\frac{3}{2}},-\frac{1}{2}\sqrt{\frac{3}{2}})$
          \\                             
\hline \hline $\la \Sigma^0 \bar p|\bar s_L \gamma_\mu b_L|B^-\ra$
          & $(-\frac{1}{\sqrt2},-\sqrt2,-\frac{1}{2\sqrt2})$
          & $\la \Sigma^- \bar n|\bar s_L \gamma_\mu b_L|B^-\ra$
          & $(-1,-2,-\frac{1}{2})$
          \\
$\la \Xi^0 \overline{\Sigma^+} |\bar s_L \gamma_\mu b_L|B^-\ra$
          & $(4,-1,-\frac{5}{2})$
          & $\la \Xi^- \overline{\Sigma^0} |\bar s_L \gamma_\mu b_L|B^-\ra$
          & $(2\sqrt2,-\frac{1}{\sqrt2},-\frac{5}{2\sqrt2})$
          \\
$\la \Xi^- \overline{\Lambda} |\bar s_L \gamma_\mu b_L|B^-\ra$
          & $(\sqrt6, \sqrt{\frac{3}{2}},-\frac{1}{2}\sqrt{\frac{3}{2}})$
          & $\la \Lambda \bar{p} |\bar s_L \gamma_\mu b_L|B^-\ra$
          & $(-3\sqrt{\frac{3}{2}},0, \frac{3}{2}\sqrt{\frac{3}{2}})$
          \\          
\hline $\la \Sigma^+ \bar p|\bar s_L \gamma_\mu b_L|\overline B{}^0\ra$
          & $(-1,-2,-\frac{1}{2})$
          & $\la \Sigma^0 \bar n|\bar s_L \gamma_\mu b_L|\overline B{}^0\ra$
          & $(\frac{1}{\sqrt 2},\sqrt2, \frac{1}{2\sqrt 2})$
          \\
$\la \Xi^0 \overline{\Sigma^0} |\bar s_L \gamma_\mu b_L|\overline B{}^0\ra$
          & $(-2\sqrt2, \frac{1}{\sqrt2},\frac{5}{2\sqrt2})$
          & $\la \Xi^0 \overline{\Lambda} |\bar s_L \gamma_\mu b_L|\overline B{}^0\ra$
          & $(\sqrt6, \sqrt{\frac{3}{2}},-\frac{1}{2}\sqrt{\frac{3}{2}})$
          \\
$\la \Xi^- \overline{\Sigma^-} |\bar s_L \gamma_\mu b_L|\overline B{}^0\ra$
          & $(4, -1,-\frac{5}{2})$
          & $\la \Lambda \bar n |\bar s_L \gamma_\mu b_L|\overline B{}^0\ra$
          & $(-3\sqrt{\frac{3}{2}},0, \frac{3}{2}\sqrt{\frac{3}{2}})$
          \\         
\hline $\la \Sigma^+ \overline{\Sigma^+}|\bar s_L \gamma_\mu b_L|\overline B{}^0_s\ra$
          & $(1,2,\frac{1}{2})$
          & $\la \Sigma^0 \overline{\Sigma^0}|\bar s_L \gamma_\mu b_L|\overline B{}^0_s\ra$
          & $(1,2,\frac{1}{2})$
          \\
$\la \Sigma^- \overline{\Sigma^-}|\bar s_L \gamma_\mu b_L|\overline B{}^0_s\ra$
          & $(1,2,\frac{1}{2})$
          & $\la \Xi^0 \overline{\Xi^0} |\bar s_L \gamma_\mu b_L|\overline B{}^0_s\ra$
          & $(5, 1,-2)$
          \\
$\la \Xi^- \overline{\Xi^-} |\bar s_L \gamma_\mu b_L|\overline B{}^0_s\ra$
          & $(5, 1,-2)$
          & $\la \Lambda \overline{\Lambda} |\bar s_L \gamma_\mu b_L|\overline B{}^0_s\ra$
          & $(3, 0,-\frac{3}{2})$
          \\                             
\end{tabular}
\end{ruledtabular}
\end{table}

\begin{table}[t!]
\caption{\label{tab: eBD} The coefficients $(e_{\parallel}, e_{\overline\parallel}, e_F)$ for various $\la {\mathbf B} \bar{\mathbf B}'|\bar q_L \gamma_\mu b_L|\bar B_{q'}\ra$ matrix elements.
}
\begin{ruledtabular}
\begin{tabular}{cccccccc}
$\la {\mathbf B} \bar{\mathbf B}'|\bar q_L \gamma_\mu b_L|\bar B_{q'}\ra$
          & $(e_{\parallel}, e_{\overline\parallel}, e_F)$
          & $\la {\mathbf B} \bar{\mathbf B}'|\bar q_L \gamma_\mu b_L|\bar B_{q'}\ra$
          & $(e_{\parallel}, e_{\overline\parallel}, e_F)$
          \\
\hline $\la p \overline{ \Delta^+} |\bar u_L \gamma_\mu b_L|B^-\ra$
          & $(-\sqrt 2,\sqrt2,-\frac{1}{\sqrt2})$
          & $\la n \overline{\Delta^0}|\bar u_L \gamma_\mu b_L|B^-\ra$
          & $(-\sqrt 2,\sqrt2,-\frac{1}{\sqrt2})$
          \\
$\la \Sigma^+ \overline{\Sigma^{*+}} |\bar u_L \gamma_\mu b_L|B^-\ra$
          & $(\sqrt 2, -\sqrt2, \frac{1}{\sqrt2})$
          & $\la \Sigma^0 \overline{\Sigma^{*0}} |\bar u_L \gamma_\mu b_L|B^-\ra$
          & $(-\frac{1}{\sqrt 2}, \frac{1}{\sqrt2}, -\frac{1}{2\sqrt2})$
          \\
$\la \Xi^0 \overline{\Xi^{*0}} |\bar u_L \gamma_\mu b_L|B^-\ra$
          & $(\sqrt 2, -\sqrt 2, \frac{1}{\sqrt 2})$
          & $\la \Lambda \overline{\Sigma^{*0}} |\bar u_L \gamma_\mu b_L|B^-\ra$
          & $(\sqrt{\frac{3}{2}}, -\sqrt{\frac{3}{2}}, \frac{1}{2}\sqrt{\frac{3}{2}})$
          \\      
\hline $\la p \overline{\Delta^0}|\bar u_L \gamma_\mu b_L|\overline B{}^0\ra$
          & $(-\sqrt 2, \sqrt2, -\frac{1}{\sqrt2})$
          & $\la n \overline{\Delta^-}|\bar u_L \gamma_\mu b_L|\overline B{}^0\ra$
          & $(-\sqrt 6, \sqrt 6, -\sqrt{\frac{3}{2}})$
          \\
$\la \Sigma^+ \overline{\Sigma^{*0}} |\bar u_L \gamma_\mu b_L|\overline B{}^0\ra$
          & $(1, -1, \frac{1}{2})$
          & $\la \Sigma^0 \overline{\Sigma^{*-}} |\bar u_L \gamma_\mu b_L|\overline B{}^0\ra$
          & $(-1, 1, -\frac{1}{2})$
          \\
$\la \Xi^0 \overline{\Xi^{*-}} |\bar u_L \gamma_\mu b_L|\overline B{}^0\ra$
          & $(\sqrt 2,  -\sqrt2, \frac{1}{\sqrt2})$
          & $\la \Lambda \overline{\Sigma^{*-}} |\bar u_L \gamma_\mu b_L|\overline B{}^0\ra$
          & $(\sqrt 3, -\sqrt 3, \frac{\sqrt 3}{2})$
          \\         
\hline $\la p \overline{\Sigma^{*0}}|\bar u_L \gamma_\mu b_L|\overline B{}^0_s\ra$
          & $(-1, 1, -\frac{1}{2})$
          & $\la n \overline{\Sigma^{*-}} |\bar u_L \gamma_\mu b_L|\overline B{}^0_s\ra$
          & $(-\sqrt 2,  \sqrt2, -\frac{1}{\sqrt2})$
          \\
$\la \Sigma^+ \overline{\Xi^{*0}} |\bar u_L \gamma_\mu b_L|\overline B{}^0_s\ra$
          & $(\sqrt 2,  -\sqrt2, \frac{1}{\sqrt2})$
          & $\la \Sigma^0 \overline{\Xi^{*-}} |\bar u_L \gamma_\mu b_L|\overline B{}^0_s\ra$
          & $(-1, 1, -\frac{1}{2})$
          \\
$\la \Xi^0 \overline{\Omega^-} |\bar u_L \gamma_\mu b_L|\overline B{}^0_s\ra$
          & $(\sqrt 6,  -\sqrt6, \sqrt{\frac{3}{2}})$
          & $\la \Lambda \overline{\Xi^{*-}} |\bar u_L \gamma_\mu b_L|\overline B{}^0_s\ra$
          & $(\sqrt 3,  -\sqrt3, \frac{\sqrt3}{2})$
          \\                             
\hline \hline $\la \Sigma^+ \overline{\Delta^{++}}|\bar s_L \gamma_\mu b_L|B^-\ra$
          & $(-\sqrt 6,  \sqrt6, -\sqrt{\frac{3}{2}})$
          & $\la \Sigma^0 \overline{\Delta^+}|\bar s_L \gamma_\mu b_L|B^-\ra$
          & $(2,-2,1)$
          \\
$\la \Sigma^- \overline{\Delta^0} |\bar s_L \gamma_\mu b_L|B^-\ra$
          & $(\sqrt 2,  -\sqrt2, \frac{1}{\sqrt2})$
          & $\la \Xi^0 \overline{\Sigma^{*+}} |\bar s_L \gamma_\mu b_L|B^-\ra$
          & $(-\sqrt 2,  \sqrt2, -\frac{1}{\sqrt2})$
          \\
$\la \Xi^- \overline{\Sigma^{*0}} |\bar s_L \gamma_\mu b_L|B^-\ra$
          & $(1, -1, \frac{1}{2})$
          & 
          & 
          \\          
\hline $\la \Sigma^+ \overline{\Delta^+}|\bar s_L \gamma_\mu b_L|\overline B{}^0\ra$
          & $(-\sqrt 2,  \sqrt2, -\frac{1}{\sqrt2})$
          & $\la \Sigma^0 \overline{\Delta^0}|\bar s_L \gamma_\mu b_L|\overline B{}^0\ra$
          & $(2,-2,1)$
          \\
$\la \Sigma^- \overline{\Delta^-} |\bar s_L \gamma_\mu b_L|\overline B{}^0\ra$
          & $(\sqrt 6,  -\sqrt6, \sqrt{\frac{3}{2}})$
          & $\la \Xi^0 \overline{\Sigma^{*0}} |\bar s_L \gamma_\mu b_L|\overline B{}^0\ra$
          & $(-1, 1, -\frac{1}{2})$
          \\
$\la \Xi^- \overline{\Sigma^{*-}} |\bar s_L \gamma_\mu b_L|\overline B{}^0\ra$
          & $(\sqrt 2,  -\sqrt2, \frac{1}{\sqrt2})$
          & 
          & 
          \\         
\hline $\la \Sigma^+ \overline{\Sigma^{*+}}|\bar s_L \gamma_\mu b_L|\overline B{}^0_s\ra$
          & $(-\sqrt 2,  \sqrt2, -\frac{1}{\sqrt2})$
          & $\la \Sigma^0 \overline{\Sigma^{*0}}|\bar s_L \gamma_\mu b_L|\overline B{}^0_s\ra$
          & $(\sqrt 2,  -\sqrt2, \frac{1}{\sqrt2})$
          \\
$\la \Sigma^- \overline{\Sigma^{*-}} |\bar s_L \gamma_\mu b_L|\overline B{}^0_s\ra$
          & $(\sqrt 2,  -\sqrt2, \frac{1}{\sqrt2})$
          & $\la \Xi^0 \overline{\Xi^{*0}} |\bar s_L \gamma_\mu b_L|\overline B{}^0_s\ra$
          & $(-\sqrt 2,  \sqrt2, -\frac{1}{\sqrt2})$
          \\
$\la \Xi^- \overline{\Xi^{*-}} |\bar s_L \gamma_\mu b_L|\overline B{}^0_s\ra$
          & $(\sqrt 2,  -\sqrt2, \frac{1}{\sqrt2})$
          & 
          & 
          \\                             
\end{tabular}
\end{ruledtabular}
\end{table}

\begin{table}[t!]
\caption{\label{tab: eDB} The coefficients $(e_{\parallel}, e_{\overline\parallel}, e_F)$ for various $\la {\mathbf B} \bar{\mathbf B}'|\bar q_L \gamma_\mu b_L|\bar B_{q'}\ra$ matrix elements.
}
\begin{ruledtabular}
\begin{tabular}{cccccccc}
$\la {\mathbf B} \bar{\mathbf B}'|\bar q_L \gamma_\mu b_L|\bar B_{q'}\ra$
          & $(e_{\parallel}, e_{\overline\parallel}, e_F)$
          & $\la {\mathbf B} \bar{\mathbf B}'|\bar q_L \gamma_\mu b_L|\bar B_{q'}\ra$
          & $(e_{\parallel}, e_{\overline\parallel}, e_F)$
          \\
\hline $\la \Delta^+ \bar p |\bar u_L \gamma_\mu b_L|B^-\ra$
          & $(-\sqrt 2,\sqrt2,-\frac{1}{\sqrt2})$
          & $\la \Delta^0 \bar n|\bar u_L \gamma_\mu b_L|B^-\ra$
          & $(-\sqrt 2,\sqrt2,-\frac{1}{\sqrt2})$
          \\
$\la \Sigma^{*+} \overline{\Sigma^{+}} |\bar u_L \gamma_\mu b_L|B^-\ra$
          & $(\sqrt 2, -\sqrt2, \frac{1}{\sqrt2})$
          & $\la \Sigma^{*0} \overline{\Sigma^{0}} |\bar u_L \gamma_\mu b_L|B^-\ra$
          & $(-\frac{1}{\sqrt 2}, \frac{1}{\sqrt2}, -\frac{1}{2\sqrt2})$
          \\
$\la \Xi^{*0} \overline{\Xi^{0}} |\bar u_L \gamma_\mu b_L|B^-\ra$
          & $(\sqrt 2, -\sqrt 2, \frac{1}{\sqrt 2})$
          & $\la \Sigma^{*0} \overline{\Lambda} |\bar u_L \gamma_\mu b_L|B^-\ra$
          & $(\sqrt{\frac{3}{2}}, -\sqrt{\frac{3}{2}}, \frac{1}{2}\sqrt{\frac{3}{2}})$
          \\      
\hline $\la \Delta^{++} \bar p|\bar u_L \gamma_\mu b_L|\overline B{}^0\ra$
          & $(\sqrt 6, -\sqrt 6, \sqrt{\frac{3}{2}})$
          & $\la \Delta^+ \bar n |\bar u_L \gamma_\mu b_L|\overline B{}^0\ra$
          & $(\sqrt 2, -\sqrt2, \frac{1}{\sqrt2})$
          \\
$\la \Sigma^{*+} \overline{\Sigma^{0}} |\bar u_L \gamma_\mu b_L|\overline B{}^0\ra$
          & $(-1, 1, -\frac{1}{2})$
          & $\la \Sigma^{*0} \overline{\Sigma^{-}} |\bar u_L \gamma_\mu b_L|\overline B{}^0\ra$
          & $(-1, 1, -\frac{1}{2})$
          \\
$\la \Xi^{*0} \overline{\Xi^{-}} |\bar u_L \gamma_\mu b_L|\overline B{}^0\ra$
          & $(-\sqrt 2,  \sqrt2, -\frac{1}{\sqrt2})$
          & $\la \Sigma^{*+} \overline{\Lambda} |\bar u_L \gamma_\mu b_L|\overline B{}^0\ra$
          & $(-\sqrt 3, \sqrt 3, -\frac{\sqrt 3}{2})$
          \\         
\hline $\la \Delta^{++} \overline{\Sigma^{+}}|\bar u_L \gamma_\mu b_L|\overline B{}^0_s\ra$
          & $(-\sqrt 6,  \sqrt6, -\sqrt{\frac{3}{2}})$
          & $\la \Delta^+ \overline{\Sigma^{0}} |\bar u_L \gamma_\mu b_L|\overline B{}^0_s\ra$
          & $(2,-2,1)$
          \\
$\la \Delta^0 \overline{\Sigma^{-}} |\bar u_L \gamma_\mu b_L|\overline B{}^0_s\ra$
          & $(\sqrt 2,  -\sqrt2, \frac{1}{\sqrt2})$
          & $\la \Sigma^{*+} \overline{\Xi^{0}} |\bar u_L \gamma_\mu b_L|\overline B{}^0_s\ra$
          & $(-\sqrt 2,  \sqrt2, -\frac{1}{\sqrt2})$
          \\
$\la \Sigma^{*0} \overline{\Xi^{-}} |\bar u_L \gamma_\mu b_L|\overline B{}^0_s\ra$
          & $(1, -1, \frac{1}{2})$
          & 
          & 
          \\                             
\hline \hline $\la \Sigma^{*0} \bar p|\bar s_L \gamma_\mu b_L|B^-\ra$
          & $(-1, 1, -\frac{1}{2})$
          & $\la \Sigma^{*-} \bar n|\bar s_L \gamma_\mu b_L|B^-\ra$
          & $(-\sqrt 2,  \sqrt2, -\frac{1}{\sqrt2})$
          \\
$\la \Xi^{*0} \overline{\Sigma^+} |\bar s_L \gamma_\mu b_L|B^-\ra$
          & $(\sqrt 2,  -\sqrt2, \frac{1}{\sqrt2})$
          & $\la \Xi^{*-} \overline{\Sigma^{0}} |\bar s_L \gamma_\mu b_L|B^-\ra$
          & $(-1, 1, -\frac{1}{2})$
          \\
$\la \Omega^- \overline{\Xi^{0}} |\bar s_L \gamma_\mu b_L|B^-\ra$
          & $(\sqrt 6, -\sqrt 6, \sqrt{\frac{3}{2}})$
          & $\la \Xi^{*-} \overline{\Lambda} |\bar s_L \gamma_\mu b_L|B^-\ra$
          & $(\sqrt 3, -\sqrt 3, \frac{\sqrt 3}{2})$
          \\          
\hline $\la \Sigma^{*+} \bar p|\bar s_L \gamma_\mu b_L|\overline B{}^0\ra$
          & $(\sqrt 2,  -\sqrt2, \frac{1}{\sqrt2})$
          & $\la \Sigma^{*0} \bar n|\bar s_L \gamma_\mu b_L|\overline B{}^0\ra$
          & $(1, -1, \frac{1}{2})$
          \\
$\la \Xi^{*0} \overline{\Sigma^{0}} |\bar s_L \gamma_\mu b_L|\overline B{}^0\ra$
          & $(-1, 1, -\frac{1}{2})$
          & $\la \Xi^{*-} \overline{\Sigma^{-}} |\bar s_L \gamma_\mu b_L|\overline B{}^0\ra$
          & $(-\sqrt 2,  \sqrt2, -\frac{1}{\sqrt2})$
          \\
$\la \Omega^- \overline{\Xi^{-}} |\bar s_L \gamma_\mu b_L|\overline B{}^0\ra$
          & $(-\sqrt 6,  \sqrt6, -\sqrt{\frac{3}{2}})$
          & $\la \Xi^{*0} \overline{\Lambda} |\bar s_L \gamma_\mu b_L|\overline B{}^0\ra$
          & $(-\sqrt 3, \sqrt 3, -\frac{\sqrt 3}{2})$
          \\         
\hline $\la \Sigma^{*+} \overline{\Sigma^{+}}|\bar s_L \gamma_\mu b_L|\overline B{}^0_s\ra$
          & $(-\sqrt 2,  \sqrt2, -\frac{1}{\sqrt2})$
          & $\la \Sigma^{*0} \overline{\Sigma^{0}}|\bar s_L \gamma_\mu b_L|\overline B{}^0_s\ra$
          & $(\sqrt 2,  -\sqrt2, \frac{1}{\sqrt2})$
          \\
$\la \Sigma^{*-} \overline{\Sigma^{-}} |\bar s_L \gamma_\mu b_L|\overline B{}^0_s\ra$
          & $(\sqrt 2,  -\sqrt2, \frac{1}{\sqrt2})$
          & $\la \Xi^{*0} \overline{\Xi^{0}} |\bar s_L \gamma_\mu b_L|\overline B{}^0_s\ra$
          & $(-\sqrt 2,  \sqrt2, -\frac{1}{\sqrt2})$
          \\
$\la \Xi^{*-} \overline{\Xi^{-}} |\bar s_L \gamma_\mu b_L|\overline B{}^0_s\ra$
          & $(\sqrt 2,  -\sqrt2, \frac{1}{\sqrt2})$
          & 
          & 
          \\                             
\end{tabular}
\end{ruledtabular}
\end{table}

\begin{table}[t!]
\caption{\label{tab: eDD} The coefficients $(e_{\parallel}, e_{\overline\parallel}, e_F)$ for various $\la {\mathbf B} \bar{\mathbf B}'|\bar q_L \gamma_\mu b_L|\bar B_{q'}\ra$ matrix elements.
}
\begin{ruledtabular}
\begin{tabular}{cccccccc}
$\la {\mathbf B} \bar{\mathbf B}'|\bar q_L \gamma_\mu b_L|\bar B_{q'}\ra$
          & $(e_{\parallel}, e_{\overline\parallel}, e_F)$
          & $\la {\mathbf B} \bar{\mathbf B}'|\bar q_L \gamma_\mu b_L|\bar B_{q'}\ra$
          & $(e_{\parallel}, e_{\overline\parallel}, e_F)$
          \\
\hline $\la \Delta^{++} \overline{\Delta^{++}} |\bar u_L \gamma_\mu b_L|B^-\ra$
          & $(6,3,3)$
          & $\la \Delta^+ \overline{\Delta^+}|\bar u_L \gamma_\mu b_L|B^-\ra$
          & $(4,2,2)$
          \\
$\la \Delta^{*0} \overline{\Delta^0} |\bar u_L \gamma_\mu b_L|B^-\ra$
          & $(2,1,1)$
          & $\la \Sigma^{*+} \overline{\Sigma^{*+}} |\bar u_L \gamma_\mu b_L|B^-\ra$
          & $(4,2,2)$
          \\
$\la \Sigma^{*0} \overline{\Sigma^{*0}} |\bar u_L \gamma_\mu b_L|B^-\ra$
          & $(2,1,1)$
          & $\la \Xi^{*0} \overline{\Xi^{*0}} |\bar u_L \gamma_\mu b_L|B^-\ra$
          & $(2,1,1)$
          \\      
\hline $\la \Delta^{++} \overline{\Delta^+}|\bar u_L \gamma_\mu b_L|\overline B{}^0\ra$
          & $(2\sqrt 3, \sqrt 3, \sqrt 3)$
          & $\la \Delta^+ \overline{\Delta^0} |\bar u_L \gamma_\mu b_L|\overline B{}^0\ra$
          & $(4,2,2)$
          \\
$\la \Delta^0 \overline{\Delta^-} |\bar u_L \gamma_\mu b_L|\overline B{}^0\ra$
          & $(2\sqrt 3, \sqrt 3, \sqrt 3)$
          & $\la \Sigma^{*+} \overline{\Sigma^{*0}} |\bar u_L \gamma_\mu b_L|\overline B{}^0\ra$
          & $(2\sqrt 2, \sqrt 2, \sqrt 2)$
          \\
$\la \Sigma^{*0} \overline{\Sigma^{*-}} |\bar u_L \gamma_\mu b_L|\overline B{}^0\ra$
          & $(2\sqrt 2, \sqrt 2, \sqrt 2)$
          & $\la \Xi^{*0} \overline{\Xi^{*-}} |\bar u_L \gamma_\mu b_L|\overline B{}^0\ra$
          & $(2,1,1)$
          \\         
\hline $\la \Delta^{++} \overline{\Sigma^{*+}}|\bar u_L \gamma_\mu b_L|\overline B{}^0_s\ra$
          & $(2\sqrt 3, \sqrt 3, \sqrt 3)$
          & $\la \Delta^+ \overline{\Sigma^{*0}} |\bar u_L \gamma_\mu b_L|\overline B{}^0_s\ra$
          & $(2\sqrt 2, \sqrt 2, \sqrt 2)$
          \\
$\la \Delta^0 \overline{\Sigma^{*-}} |\bar u_L \gamma_\mu b_L|\overline B{}^0_s\ra$
          & $(2,1,1)$
          & $\la \Sigma^{*+} \overline{\Xi^{0}} |\bar u_L \gamma_\mu b_L|\overline B{}^0_s\ra$
          & $(4,2,2)$
          \\
$\la \Sigma^{*0} \overline{\Xi^{*-}} |\bar u_L \gamma_\mu b_L|\overline B{}^0_s\ra$
          & $(2\sqrt 2, \sqrt 2, \sqrt 2)$
          & $\la \Xi^{*0} \overline{\Omega^{-}} |\bar u_L \gamma_\mu b_L|\overline B{}^0_s\ra$
          & $(2\sqrt 3, \sqrt 3, \sqrt 3)$
          \\                             
\hline \hline $\la \Sigma^{*+} \overline{\Delta^{++}}|\bar s_L \gamma_\mu b_L|B^-\ra$
          & $(2\sqrt 3, \sqrt 3, \sqrt 3)$
          & $\la \Sigma^{*0} \overline{\Delta^{+}}|\bar s_L \gamma_\mu b_L|B^-\ra$
          & $(2\sqrt 2, \sqrt 2, \sqrt 2)$
          \\
$\la \Sigma^{*-} \overline{\Delta^{0}}|\bar s_L \gamma_\mu b_L|B^-\ra$
          & $(2,1,1)$
          & $\la \Xi^{*0} \overline{\Sigma^{*+}} |\bar s_L \gamma_\mu b_L|B^-\ra$
          & $(4,2,2)$
          \\
$\la \Xi^{*-} \overline{\Sigma^{*0}} |\bar s_L \gamma_\mu b_L|B^-\ra$
          & $(2\sqrt 2, \sqrt 2, \sqrt 2)$
          & $\la \Omega^- \overline{\Xi^{*0}} |\bar s_L \gamma_\mu b_L|B^-\ra$
          & $(2\sqrt 3, \sqrt 3, \sqrt 3)$
          \\          
\hline $\la \Sigma^{*+} \overline{\Delta^+}|\bar s_L \gamma_\mu b_L|\overline B{}^0\ra$
          & $(2,1,1)$
          & $\la \Sigma^{*0} \overline{\Delta^0}|\bar s_L \gamma_\mu b_L|\overline B{}^0\ra$
          & $(2\sqrt 2, \sqrt 2, \sqrt 2)$
          \\
$\la \Sigma^{*-} \overline{\Delta^-} |\bar s_L \gamma_\mu b_L|\overline B{}^0\ra$
          & $(2\sqrt 3, \sqrt 3, \sqrt 3)$
          & $\la \Xi^{*0} \overline{\Sigma^{*0}} |\bar s_L \gamma_\mu b_L|\overline B{}^0\ra$
          & $(2\sqrt 2, \sqrt 2, \sqrt 2)$
          \\
$\la \Xi^{*-} \overline{\Sigma^{*-}} |\bar s_L \gamma_\mu b_L|\overline B{}^0\ra$
          & $(4,2,2)$
          & $\la \Omega^- \overline{\Xi^{*-}} |\bar s_L \gamma_\mu b_L|\overline B{}^0\ra$
          & $(2\sqrt 3, \sqrt 3, \sqrt 3)$
          \\         
\hline $\la \Sigma^{*+} \overline{\Sigma^{*+}}|\bar s_L \gamma_\mu b_L|\overline B{}^0_s\ra$
          & $(2,1,1)$
          & $\la \Sigma^{*0} \overline{\Sigma^{*0}}|\bar s_L \gamma_\mu b_L|\overline B{}^0_s\ra$
          & $(2,1,1)$
          \\
$\la \Sigma^{*-} \overline{\Sigma^{*-}} |\bar s_L \gamma_\mu b_L|\overline B{}^0_s\ra$
          & $(2,1,1)$
          & $\la \Xi^{*0} \overline{\Xi^{*0}} |\bar s_L \gamma_\mu b_L|\overline B{}^0_s\ra$
          & $(4,2,2)$
          \\
$\la \Xi^{*-} \overline{\Xi^{*-}} |\bar s_L \gamma_\mu b_L|\overline B{}^0_s\ra$
          & $(4,2,2)$
          & $\la \Omega^{-} \overline{\Omega^{-}} |\bar s_L \gamma_\mu b_L|\overline B{}^0_s\ra$
          & $(6,3,3)$
          \\                             
\end{tabular}
\end{ruledtabular}
\end{table}

The transition matrix element can be expressed as
\be
\la {\mathbf B}\bar {\mathbf B}^\prime | \bar q_L\gamma_\mu b_L |\overline B_{q'}\ra
&=& i\,\bar u_L(p_{{\rm\bf B}})\gamma_\mu v_L(p_{\overline{\rm\bf B}^\prime}) {\cal G}_{L}
+ i\,\bar u_R(p_{{\rm\bf B}})\gamma_\mu v_R(p_{\overline{\rm\bf B}^\prime}) {\cal G}_{R}
\non\\
&&
+i\,\bar u_L(p_{{\rm\bf B}}) {\bf F}_\mu v_R(p_{\overline{\rm\bf B}^\prime}) ,
\label{eq: asymptotic}
\en
where ${\bf F}_\mu$ can be expressed as 
\be
{\bf F}_\mu=a_F \sigma_{\mu\nu} q^\nu+b_F q_\mu +c_F (p_{{\rm\bf B}}+p_{\overline{\rm\bf B}^\prime})_\mu+d_F  (p_{{\rm\bf B}}-p_{\overline{\rm\bf B}^\prime})_\mu,
\en
with $q\equiv p_{B_q}-p_{{\rm\bf B}}-p_{\overline{\rm\bf B}^\prime}$ and form factors $a_F$, $b_F$, $c_F$ and $d_F$. These ${\cal G}_{L}$, ${\cal G}_{R}$ and ${\bf F}_\mu$ depends on the decaying meson and the final state baryon pair.

We use spacelike case for illustration.
Using the approach similar to those in \cite{Brodsky:1980sx, Chua:2002wn, Chua:2003it,Chua:2013zga,Chua:2016aqy,Chua:2022wmr}
the above form factors ${\cal G}_{L}$, ${\cal G}_{R}$ and ${\bf F}_\mu$ can be expressed in terms of three universal form factors, 
${\cal G}_{\parallel}$, ${\cal G}_{\overline\parallel}$ and ${\cal F}_\mu$ as following,
\be
{\cal G}_{L}
=e_{\parallel} \,{\cal G}_{\parallel},
\quad
{\cal G}_{R}
=e_{\overline\parallel} \,{\cal G}_{\overline\parallel},
\quad
{\bf F}_\mu
=e_F \,{\cal F}_{\mu},
\en
where the coefficients $e_{\parallel}$, $e_{\overline\parallel}$ and $e_F$ are given by
\be
e_{\parallel}
&=&
\bigg(\la {\mathbf B};\downarrow\uparrow\downarrow | O[q^\prime_L(1)\to q_L(1)]|{\mathbf B}^\prime; \downarrow\uparrow\downarrow\ra
\non\\
&&+\la {\mathbf B};\downarrow\uparrow\downarrow | O[q^\prime_L(3)\to q_L(3)]|{\mathbf B}^\prime; \downarrow\uparrow\downarrow\ra
\bigg),
\non\\
e_{\overline \parallel}
&=&
\la {\mathbf B};\uparrow\downarrow\uparrow | O[q^\prime_L(2)\to q_L(2)]|{\mathbf B}^\prime; \uparrow\downarrow\uparrow\ra, 
\non\\
e_F
&=&
\bigg(\la {\mathbf B};\,\downarrow\downarrow\uparrow| 
O[q^\prime_R(1)\to q_L(1)]|{\mathbf B}^\prime\,;\uparrow\downarrow\uparrow\rangle
\nonumber\\
&&+\la {\mathbf B};\,\uparrow\downarrow\downarrow|
O[q^\prime_R(3)\to q_L(3)]
|{\mathbf B}^\prime\,;\uparrow\downarrow\uparrow\rangle\bigg).
\en
Note that $q'$ is the anti-quark in $\overline B_{q'}$ meson and $q$ is the quark in the $\bar q_L \gamma_\mu b_L$ current. 
Applying $Q[q^\prime_L(1)\to q_L(1)]$ to $|{\mathbf B}';\downarrow\uparrow\downarrow\ra$ changes the 
parallel spin $q^\prime(1)|\downarrow\rangle$ part of 
$|{\mathbf B}^\prime;\downarrow\uparrow\downarrow\rangle$
to a parallel spin $q(1)|\downarrow\rangle$ part, where the flavor is changed from $q'$ to $q$, 
and likewise for the operation of $O[q^\prime_L(3)\to q_L(3)]$ on $|{\mathbf B}';\downarrow\uparrow\downarrow\ra$. 
As the operation involves only the parallel spin component, the coefficient is called $e_\parallel$ and the correspondent form factor is ${\cal G}_{\parallel}$.
Likewise $e_{\overline\parallel}$ involves only the anti-parallel spin component, while $e_F$ involves operations that flip the spin of the quark in addition to changing the flavor from $q'$ to $q$.
Note that annihilation diagram is not included in the above analysis, as the flavor flow structure is different, see Fig.~\ref{fig:TA}, 
where, as far as the flavor structure is concern, $\overline B_{q'}$ is annihilated by the current and the baryon pair is created from vacuum.
The coefficients $e_\parallel$, $e_{\overline\parallel}$ and $e_F$ for all relevant transitions considered in this work are obtained accordingly and are shown in 
Tables~\ref{tab: eBB}, \ref{tab: eBD}, \ref{tab: eDB}, \ref{tab: eDD}.

By comparing the Tables~\ref{tab: eBB}, \ref{tab: eBD}, \ref{tab: eDB}, \ref{tab: eDD} and Tables~\ref{tab: TPBB}, \ref{tab: TPBD}, \ref{tab: TPDB}, \ref{tab: TPDD}, we found the following correspondences of topological amplitudes and $(e_{\parallel}, e_{\overline\parallel}, e_F)$,
\be
T_{1\BB}&:& (e^{(1)}_{\parallel}, e^{(1)}_{\overline\parallel}, e^{(1)}_F)= (1,2,\frac{1}{2}),
\non\\
T_{2\BB}&:& (e^{(2)}_{\parallel}, e^{(2)}_{\overline\parallel}, e^{(2)}_F)= (4,-1,-\frac{5}{2}),
\non\\
T_\BD&:& (e'_{\parallel}, e'_{\overline\parallel}, e'_F)= (1,-1,\frac{1}{2}),
\non\\
T_\DB&:& (e''_{\parallel}, e''_{\overline\parallel}, e''_F)= (1,-1,\frac{1}{2}),
\non\\
T_\DD&:& (e'''_{\parallel}, e'''_{\overline\parallel}, e'''_F)= (1,\frac{1}{2},\frac{1}{2}),
\label{eq: Teee}
\en
and similar relations for $P_{i\BB,\BD,\DB,\DD}$.

In general, the topological amplitudes, $T_{i\,\BB}$, $T_\BD$ $T_\DB$ and $T_\DD$, are given in Eqs. (\ref{eq: Ti}), (\ref{eq: T BD}), (\ref{eq: T DB}) and (\ref{eq: T DD}).
It is useful to show that $T_{\BD}$, $T_{\DB}$ and $T_\DD$ have the structure of $T_{i\,\BB}$ in the asymptotic limit. 
Note that the Rarita-Schwinger vector spinor $u_\mu$ can be expressed in terms of Dirac spinors and polarization vectors as following \cite{Moroi:1995fs}
\be
u_\mu\bigg(\vec p, \pm\frac{3}{2}\bigg)&=&\epsilon_\mu(\vec p,\pm1) u\bigg(\vec p,\pm\frac{1}{2}\bigg),
\non\\
u_\mu\bigg(\vec p,\pm\frac{1}{2}\bigg)&=&
\frac{1}{\sqrt 3}
\epsilon_\mu(\vec p,\pm1) u\bigg(\vec p,\mp\frac{1}{2}\bigg)
+\sqrt{\frac{2}{3}}\,\epsilon_\mu(\vec p,0) u\bigg(\vec p,\pm\frac{1}{2}\bigg),
\en 
where $\epsilon_\mu(\vec p,\lambda)$ is the polarization vector,
\be
\epsilon_\mu(\vec p,0)=\left(\frac{|\vec p|}{m},\frac{E}{m}\hat n\right),
\quad
\epsilon_\mu(\vec p,\pm 1)=\big(0, \vec \epsilon(\vec 0,\pm 1)\big).
\en
with $\hat n\equiv \vec p/|\vec p|$ and $\vec\epsilon(\vec 0,\pm 1)\cdot\hat n=0$, and, for example, in the case of $\vec p=(0,0,p)$, we have $\hat n=\hat z$ and $\vec \epsilon(\vec 0,\pm 1)=\mp(1,\pm i,0)/\sqrt 2$.
Spinors $v^\mu(\vec p,\lambda)$ have similar relations.
When $|\vec p|\gg m$, $\epsilon_\mu(\vec p,0)$ dominates over $\epsilon_\mu(\vec p,\pm 1)$ and, 
consequently, $u^\mu(\vec p,\pm 1/2)$ and $v^\mu(\vec p,\pm 1/2)$ dominate over $u^\mu(\vec p,\pm 3/2)$ and $v^\mu(\vec p,\pm 3/2)$, respectively,
and they can be approximated as
\be
u_\mu\bigg(\vec p,\pm\frac{1}{2}\bigg)\simeq 
\sqrt{\frac{2}{3}}\,\frac{p_\mu}{m} u\bigg(\vec p,\pm\frac{1}{2}\bigg),
\quad
v_\mu\bigg(\vec p,\pm\frac{1}{2}\bigg) \simeq 
\sqrt{\frac{2}{3}}\,\frac{p_\mu}{m} v\bigg(\vec p,\pm\frac{1}{2}\bigg).
\label{eq: u v asymptotic}
\en
Using the above relations and Eqs. (\ref{eq: T BD}), (\ref{eq: T DB}) and (\ref{eq: T DD}), in the large momentum limit, we should have
 \be
T_{\BD}
&\simeq&
i\frac{G_F}{\sqrt2} V_{ub} 
\bar l_L \gamma_\mu \nu_L\,\,
\delta_{|\lambda_{\overline{\cal D}}|,1/2}
\non\\
&&
\times
\bar u(p_{{\cal D}},\lambda_{\cal B})
\sqrt{\frac{2}{3}}\,\frac{1}{m_{\overline{\cal D}}}
\Big\{
 \Big[
 (g'_1 p_{{\cal B}}\cdot p_{\overline{\cal D}}+g'_6 q\cdot p_{\overline{\cal D}}) \gamma_\mu
 +i(g'_2 p_{{\cal B}}\cdot p_{\overline{\cal D}}+g'_7 q\cdot p_{\overline{\cal D}}) \sigma_{\mu\rho} q^\rho
 \non\\
 &&\qquad
 +(g'_3 p_{{\cal B}}\cdot p_{\overline{\cal D}}+g'_8 q\cdot p_{\overline{\cal D}}) q_\mu
 +(g'_4 p_{{\cal B}}\cdot p_{\overline{\cal D}}+g'_9 q\cdot p_{\overline{\cal D}}) p_{{\cal B}\mu}
 +g'_5 p_{\overline{\cal D}\mu}
 \Big]
 \gamma_5
 \non\\
 &&\qquad
  -\Big[
 (f'_1 p_{{\cal B}}\cdot p_{\overline{\cal D}}+f'_6 q\cdot p_{\overline{\cal D}}) \gamma_\mu
 +i(f'_2 p_{{\cal B}}\cdot p_{\overline{\cal D}}+f'_7 q\cdot p_{\overline{\cal D}}) \sigma_{\mu\rho} q^\rho
 \non\\
 &&\qquad
 +( f'_3 p_{{\cal B}}\cdot p_{\overline{\cal D}}+ f'_8 q\cdot p_{\overline{\cal D}}) q_\mu
 +( f'_4 p_{{\cal B}}\cdot p_{\overline{\cal D}}+ f'_9 q\cdot p_{\overline{\cal D}}) p_{{\cal B}\mu}
 +  f'_5 p_{\overline{\cal D}\mu}
 \Big]
 \Big\}
 v(p_{\overline{\cal D}},\lambda_{\overline{\cal D}}),
 \non\\
T_{\DB}
&\simeq&
i\frac{G_F}{\sqrt2} V_{ub} 
\bar l_L \gamma_\mu \nu_L
\,\,\delta_{|\lambda_{{\cal D}}|,1/2}
\non\\
&&
\times
\bar u(p_{{\cal D}},\lambda_{\cal D})
\sqrt{\frac{2}{3}}\,\frac{1}{m_{{\cal D}}}
\Big\{
 \Big[
 (\bar g''_1 p_{\overline{\cal B}}\cdot p_{\cal D}+\bar g''_6 q\cdot p_{\cal D}) \gamma_\mu
 +i(\bar g''_2 p_{\overline{\cal B}}\cdot p_{\cal D}+\bar g''_7 q\cdot p_{\cal D})  \sigma_{\mu\rho} q^\rho
\non\\
&&\qquad
 + ( g''_3 p_{\overline{\cal B}}\cdot p_{\cal D}+ g''_8 q\cdot p_{\cal D})  q_\mu
 +  g''_5 p_{{\cal D}\mu}
 + ( g''_4 p_{\overline{\cal B}}\cdot p_{\cal D}+ g''_9 q\cdot p_{\cal D}) p_{\overline{\cal B}\mu}
 \Big]
 \gamma_5
 \non\\
 &&\qquad
 - \Big[
 ( f''_1 p_{\overline{\cal B}}\cdot p_{\cal D}+ f''_6 q\cdot p_{\cal D}) \gamma_\mu
 +i( f''_2 p_{\overline{\cal B}}\cdot p_{\cal D}+ f''_7 q\cdot p_{\cal D})  \sigma_{\mu\rho} q^\rho
\non\\
&&\qquad
 + ( f''_3 p_{\overline{\cal B}}\cdot p_{\cal D}+ f''_8 q\cdot p_{\cal D})  q_\mu
 +  f''_5 p_{{\cal D}\mu}
 + ( f''_4 p_{\overline{\cal B}}\cdot p_{\cal D}+ f''_9 q\cdot p_{\cal D}) p_{\overline{\cal B}\mu}
 \Big]
 \Big\}
 v(p_{\overline{\cal B}},\lambda_{\overline{\cal B}}),
 \non\\
 \label{eq: T BD DB asymptotic}
 \en
 and
 \be
T_{\DD}
&\simeq&
i\frac{G_F}{\sqrt2} V_{ub} 
\bar l_L \gamma_\mu \nu_L\,\,\delta_{|\lambda_{{\cal D}}|,1/2}\,\,\delta_{|\lambda_{\overline{\cal D}}|,1/2}
\non\\
&&
\times
\bar u(p_{{\cal D}},\lambda_{\cal D})\frac{2}{3} \frac{p_{\cal D}\cdot p_{\overline{\cal D}}}{m_{\cal D} m_{\overline{\cal D}}}
\Big\{
 \Big[
 g'''_1 \gamma_\mu
 +i g'''_2 \sigma_{\mu\rho} q^\rho
 +g'''_3 q_\mu
 +g'''_4 (p_{{\cal D}}+p_{\overline{\cal D}^\prime})_\mu
 \non\\
&&
+g'''_5 (p_{{\cal D}}-p_{\overline{\cal D}^\prime})_\mu
 ]\gamma_5
-[f'''_1 \gamma_\mu 
+if'''_2 \sigma_{\mu\nu} q^\nu
+f'''_3 q_\mu 
+f'''_4 (p_{{\cal D}}+p_{\overline{\cal D}^\prime})_\mu
 \non\\
&&
+f'''_5  (p_{{\cal D}}-p_{\overline{\cal D}^\prime})_\mu]\}
 \Big]
 \Big\}
 v(p_{\overline{\cal D}},\lambda_{\overline{\cal D}}).
 \label{eq: T DD asymptotic}
\en
Comparing the above equations and Eq. (\ref{eq: Ti}), we see that $T_{\BD}$, $T_{\DB}$ and $T_\DD$ indeed have the structure of $T_{i\,\BB}$ in the asymptotic limit. 
Their asymptotic form can be obtained by using Eqs. (\ref{eq: asymptotic}), (\ref{eq: Teee}) and the above equations.

\section{Formulas of decay rates for $\overline B_q\to{\bf B}{\overline {\bf B}}' l\bar\nu$ and $\overline B_q\to {\bf B}{\overline {\bf B}}'\nu\bar\nu$ decays}\label{App: formulas}

The $\overline B_q\to\bfBB' l\bar\nu$ and $\overline B_q\to \bfBB'\nu\bar\nu$ decays involve 4-body decays.
The decay rate of a 4-body decay is given by \cite{Geng:2011tr,Geng:2012qn,Cheng:1993ah}
\be
d\Gamma=\frac{|M|^2}{4(4\pi)^6 m_{B_q}^3} \beta X ds\, dt\, d\cos\theta_{\bf B} \, d\cos\theta_{\bf L} \, d\phi,
\en
where $s$ is the invariant mass squared of the lepton pair, $t$ is the invariant mass squared of the baryon pair, 
$\theta_{{\bf B}({\bf L})}$ is the angle of the baryon ${\bf B}$ (the lepton $l$ [or $\nu$]) in the baryon pair (lepton pair) rest frame with respect to the opposite direction of the lepton pair (baryon pair) total 3-momentum direction, $\phi$ is the angle between the baryon and the lepton planes, and
\be
X=\left(\frac{(m^2_{B_q}-s - t)^2}{4}-s t\right)^{1/2},
\quad
\beta=\frac{1}{t}\left [t^2-2 t (m_{\bf B}^2+m^2_{\overline{\bf B}})+(m_{\bf B}^2-m^2_{\overline{\bf B}})^2\right]^{1/2}.
\en
The ranges of $s$, $t$, $\theta_{\bf B}$, $\theta_{\bf L}$ and $\phi$ are
\be
0\leq s \leq (m_{B_q}-t^{1/2})^2,
\quad
(m_{\bf B}+m_{\overline{\bf B}})^2\leq t \leq m^2_{B_q},
\quad
0\leq \theta_{{\bf B},{\bf L}}\leq\pi,
\quad
0\leq \phi\leq 2\pi,
\en
where the masses of leptons are neglected.

The amplitude squared $|M|^2$ can be obtained by using $T_{i\,\BB}$, $T_\BD$, $T_\DB$ and $T_\DD$ as shown in 
Eqs. (\ref{eq: Ti}), (\ref{eq: T BD}), (\ref{eq: T DB}) and (\ref{eq: T DD}) with the help of FeynCalc \cite{Mertig:1990an,Shtabovenko:2016sxi,Shtabovenko:2020gxv}.
The scalar products of momenta and the contracted Levi-Civita symbol need to be expressed in terms of 
$s$, $t$, $\theta_{\bf B}$, $\theta_{\bf L}$ and $\phi$ before the integration $\int d\Gamma$ can be carried out.
The expressions have been worked out in ref. \cite{Cheng:1993ah}.
Defining
\be
P\equiv p_{\bf B}+p_{\overline {\bf B}},
\quad
Q\equiv p_{\bf B}-p_{\overline {\bf B}},
\quad
L\equiv p_{l(\nu)}+p_{\bar\nu},
\quad
N\equiv p_{l(\nu)}-p_{\bar\nu},
\en
one has \cite{Cheng:1993ah}
\be
P^2&=&t,
\quad
P\cdot Q=m_{\bf B}^2-m^2_{\overline{\bf B}},
\quad
Q^2=2(m_{\bf B}^2+m^2_{\overline{\bf B}})-t,
\non\\
L^2&=&s,
\quad
L\cdot N=0,
\quad
N^2=-s,
\en
\be
L\cdot P&=&\frac{1}{2}(m_{B_q}^2-t-s),
\quad 
L\cdot Q=\beta X \cos\theta_{\bf B}+\frac{m_{\bf B}^2-m^2_{\overline{\bf B}}}{t} L\cdot P,
\quad
N\cdot P=X\cos\theta_{\bf L},
\non\\
N\cdot Q&=&\frac{m_{\bf B}^2-m^2_{\overline{\bf B}}}{t} X \cos\theta_{\bf L}+\beta (L\cdot P)\cos\theta_{\cal B}\cos\theta_{\bf L}
-\sqrt{s t} \beta \sin\theta_{\bf B}\sin\theta_{\bf L}\cos\phi,
\en
\be
p_B\cdot P&=&\frac{1}{2}(m_{B_q}^2-s+t),
\quad
p_B\cdot Q=\frac{(m_{\bf B}^2-m^2_{\overline{\bf B}})(m_{B_q}^2-s+t)}{2t}+\beta X\cos\theta_{\bf B},
\non\\
p_B\cdot L&=&\frac{1}{2}(m_{B_q}^2+s-t),
\quad
p_B\cdot N=X\cos\theta_{\bf L},
\en
and
\be
\epsilon_{\mu\nu\rho\sigma} N^\mu P^\nu p_B^\rho Q^\sigma=\sqrt{s t}\beta X \sin\theta_{\bf B}\sin\theta_{\bf L}\sin\phi,
\en
with $\epsilon_{0123}=-1$.

In $\overline B_q\to\BD l\bar\nu (\nu\bar\nu)$, $\overline B_q\to\DB l\bar\nu (\nu\bar\nu)$ and $\overline B_q\to \DD'\l\bar\nu (\nu\bar\nu)$ decays, the calculation of $|M|^2$ involve polarization sums of Rarita-Schwinger vector spinors.
The following formulas for polarization sums [see, for example, Eq. (4.31) of ref. \cite{Moroi:1995fs}] are needed,
\be
\sum_{\lambda=-3/2}^{3/2}u_\mu(p, \lambda) \bar u_\nu(p, \lambda)
&=&-(\not\! p+ m)
\bigg(G_{\mu\nu}-\frac{1}{3} G_{\mu\sigma} G_{\nu\lambda}\gamma^\sigma\gamma^\lambda\bigg),
\non\\
\sum_{\lambda=-3/2}^{3/2}v_\mu(p, \lambda) \bar v_\nu(p, \lambda)
&=&-(\not\! p - m)
\bigg(G_{\mu\nu}-\frac{1}{3} G_{\mu\sigma} G_{\nu\lambda}\gamma^\sigma\gamma^\lambda\bigg),
\label{eq: Pmunu}
\en
where $G_{\mu\nu}$ is defined as
\be
G_{\mu\nu}\equiv g_{\mu\nu}-\frac{p_\mu p_\nu}{m^2}.
\en
Note that in the above formulas the signs of $m$ differ from those in ref. \cite{Moroi:1995fs}.
It is useful to check that in the large momentum limit, we have
\be
\sum_{\lambda=-3/2}^{3/2}u_\mu(p, \lambda) \bar u_\nu(p, \lambda)
&\simeq &(\not\! p+ m) \frac{2}{3 m} p_\mu p_\nu
=\frac{2}{3 m^2} p_\mu p_\nu \sum_{\lambda=-1/2}^{1/2}u(p, \lambda) \bar u(p, \lambda),
\non\\
\sum_{\lambda=-3/2}^{3/2}v_\mu(p, \lambda) \bar v_\nu(p, \lambda)
&\simeq&(\not\! p - m)\frac{2}{3 m^2} p_\mu p_\nu,
=\frac{2}{3 m^2} p_\mu p_\nu \sum_{\lambda=-1/2}^{1/2}v(p, \lambda) \bar v(p, \lambda),
\label{eq: Pmunu asymptotic}
\en
which agree with Eq. (\ref{eq: u v asymptotic}). 
Note that in our calculations involving Rarita-Schwinger vector spinors, only the exact polarization formulas in Eq. (\ref{eq: Pmunu}) are used. 

With these the $\overline B_q\to\bfBB' l\bar\nu$ and $\overline B_q\to \bfBB'\nu\bar\nu$ decay rates can be readily obtained once the topological amplitudes are given.



\begin{thebibliography}{99}
\bibitem{referee}
The author would like to thank the referee for pointing this out.
\end{thebibliography}
\end{document}